%% file: mastersthesis.tex
\documentclass[dvips, master, tocprelim, final]{cornell}


\let\ifcolor\iftrue
\newcommand{\comment}[1]{}

\usepackage{url}
\usepackage{graphicx}
\usepackage{ifpdf}
\usepackage{color}
\definecolor{brickred}      {cmyk}{0   , 0.89, 0.94, 0.28}

\ifpdf
  \DeclareGraphicsExtensions{.jpg,.pdf,.png}   
\else
  \DeclareGraphicsExtensions{.eps}             
\fi

\comment{
\ifpdf
\usepackage[dvips,
        plainpages=false,
        pdfpagelabels,%
        a4paper,
        hyperfootnotes=false,
        pagebackref=true,
        pdfdisplaydoctitle=true
]{hyperref} 
\else
}
    \usepackage[dvips,plainpages=false,hyperfootnotes=false,pagebackref,a4paper]{hyperref}
\ifpdf
\fi

\hypersetup{%
  pdftitle = {Using Access Data for Paper Recommendations on ArXiv.org},
  pdfsubject = {Master's Thesis},
  pdfkeywords = {thesis access data recommendation arxiv},
  pdfauthor = {Stefan Pohl}
  pdfproducer = {LaTeX with hyperref and thumbpdf}
}

\renewcommand*{\backref}[1]{}
\renewcommand*{\backrefalt}[4]{%
\ifcase #1 %
(Not cited.)%
\or
Cited on page~#2.%
\else
Cited on pages~#2.%
\fi
}

\title{Using Access Data for Paper Recommendations on ArXiv.org}
\author{Stefan Pohl}
\conferraldate{12}{2006}

\usepackage{array}
\usepackage{subfigure}

\usepackage{amsthm}

\theoremstyle{definition}
\newtheorem{definition}{Definition}[chapter]

\usepackage{algorithm2e}
\restylealgo{algoruled}

\begin{document}

\def\arxiv{arXiv\index{arXiv}}
\def\arxivorg{arXiv.org\index{arXiv}}
\def\tfidf{TF$\cdot$IDF\index{TF$\cdot$IDF}}



\input{prelim/title}

\thispagestyle{empty} 
\ifdraft This page is intentionally left blank \fi
\clearpage
\mbox{}
\clearpage

\pagestyle{plain}
\pagenumbering{roman} 

\noindent
\parindent 0cm
\input{prelim/abstract}   
\input{prelim/dedication} 
\input{prelim/ack}        
\newpage{\pagestyle{empty}\cleardoublepage}

\setcounter{tocdepth}{1}
\contentspage

\tablelistpage 
\figurelistpage 
\input{chapters/abbr}
\clearpage

\comment{
# I followed the instructions given: I placed \makeindex[ListOfAbbr] in the preamble. Right before the \printindex command, I inserted the list of my abbreviations, e.g. \index[ListOfAbbr]{K@Kelvin}, etc. When I typeset my document, the title (List of Abbreviations) appeared on a new page where I inserted the \printindex command (in bold, however, while the other headings are plain). The list of abbreviations appeared on the next page, under the heading "Index" and the list contained only the full name (i.e. the word after the @ sign in the \index command), comma, and then a page number. Am I missing something?

Because our Latex stylesheet supports multiple indexes, you can add a list of abbreviation or other things which you want to include. Therefore, the \printindex command is a little different from the default one. Our \printindex is defined as following:

    * \printindex[0]{1}{2}{3}
    * 0: refers to the name of the index; for this case, it should be ListOfAbbr.
    * 1: The heading appears in the table of contents
    * 2: The heading of the index starting page.
      3: Some textual information about the index, if necessary. 

You can also consult our sample latex file and documentation in which the command is clearly defined and documented.

Index Title is not printed:
\renewenvironment{theindex}{%
        \if@twocolumn
            \@restonecolfalse
        \else
            \@restonecoltrue
        \fi
        \bigskip
        \thispagestyle{plain}%
        \parindent\z@
        \parskip\z@ \@plus .3\p@\relax
        \let\item\@idxitem
    }{%
        \if@restonecol
            \onecolumn
        \else
            \clearpage
        \fi
    }

}



\normalspacing
\setcounter{page}{1}
\pagenumbering{arabic}
\pagestyle{cornell}


\noindent
\parindent 0cm
\input{chapters/chap1_introduction}
\input{chapters/chap2_basics}

\input{chapters/chap3_accessdata}
\input{chapters/chap5_evaluation}

\input{chapters/chap6_conclusion}

\appendix
\input{chapters/app1}



\renewcommand{\bibname}{Bibliography}

\input{mastersthesis.bbl}
\bibliographystyle{unsrt}
\clearpage




\end{document}

%% file: prelim/title.tex
%


%

\newcommand{\deckblatt}[8]{
  \begin{titlepage}
  \vspace*{-35mm}
  \hspace{#1}               
  \begin{minipage}[t]{#2}       
      #3
    \begin{center}
    \vspace{10mm}
      #6
        \\   
    \vspace{10mm}
    #7
    \end{center}
  \end{minipage} \\         
  \vfill
  \hspace{#1}               
  \begin{minipage}[t]{#2}       
    \begin{center}
    #8
    \end{center}
  \end{minipage}            
  \end{titlepage}
}


\newcommand{\titelseite}[3]{%
  \deckblatt
    {3mm}
    {120mm}
    {
      \vspace*{15mm}
	\hspace{-2.7cm}
\begin{tabular}{@{}cc@{}c@{}c@{}}
  \begin{tabular}{@{}l@{}}
\ifcolor \includegraphics[height=0.875in]{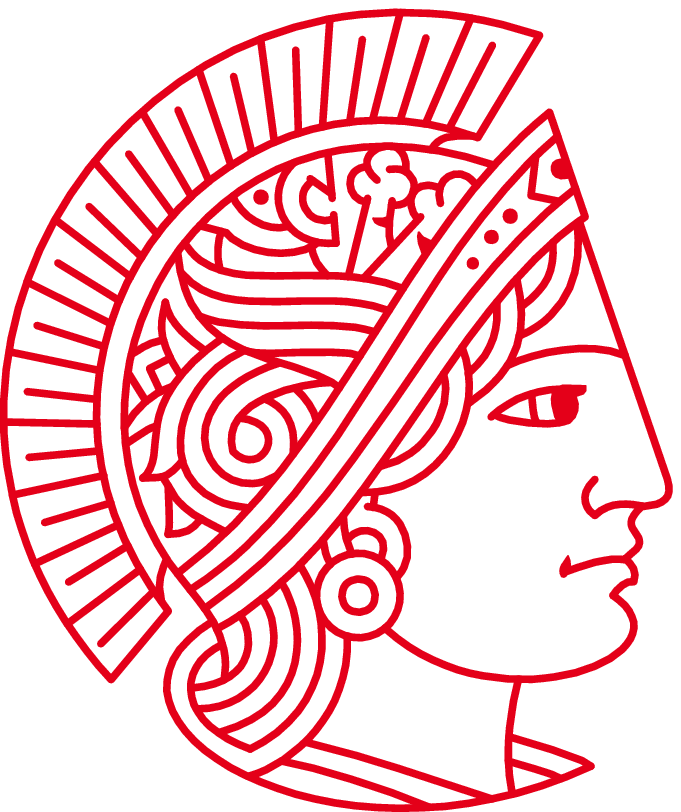}
\else    \includegraphics[height=0.875in]{figs/athene_sw.eps}
\fi
  \end{tabular}
&
\begin{tabular}{@{}l@{}}
\parbox[0.875in]{5.8cm}
  {\renewcommand{\baselinestretch}{1.15}\begin{flushleft}
     \fontsize{15}{15} \selectfont {\fontencoding{OT1}\fontfamily{ppl}\selectfont Technische Universit\"at Darmstadt\\
     Fachbereich Informatik}
   \end{flushleft}} \vspace{2.5mm}
\end{tabular}
&
\mbox{\hspace{1.5cm}}
&
  \begin{tabular}{@{}l@{}}
\ifcolor \includegraphics[height=0.875in]{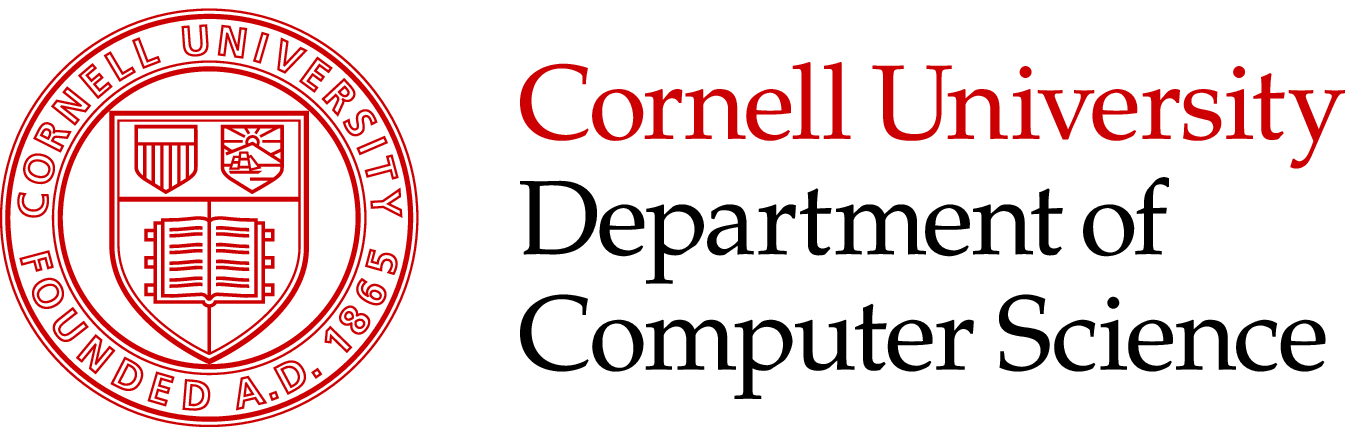}
\else    \includegraphics[height=0.875in]{figs/cs_3line_blk_pc.eps}
\fi
  \end{tabular} 

\end{tabular}
    }
    {185mm} 
    {68mm}
    {#1}
    {#2}
    {#3}
}


\titelseite{
  {\large \sc Master's Thesis} \\ [5mm]
  {\LARGE \bf Using Access Data for Paper Recommendations on ArXiv.org} \\ [5mm]
\comment{Recommending citations, Using usage data }  {von} \\
  {\large Stefan Pohl}
}{
  Pr\"ufer: \\
  Prof.~Dr. Thomas Hofmann \\ [5mm]
  Betreuer: \\
  Assoc.~Prof.~Dr. Thorsten Joachims \\ [15mm] 

  Fachgebiet Intelligente Systeme \\
  Fachbereich Informatik \\
  Technische Universit\"at Darmstadt 
}{
  Ithaca, New York, im Dezember 2006
}

\comment{
\begin{titlepage}

\begin{center}

  \large
  {\bf \sc Master's Thesis}
 
  {\huge\bf <Diplom-Titel>}
  
  von
  
  {\bf <Vorname> <Name>}\\
  Matrikelnummer <Nr>
  
  \vspace{2cm}

  \normalsize
  Technische Universität Darmstadt\\
  Fachbereich Informatik\\
  Fachgebiet Graphisch-Interaktive Systeme\\
  Fraunhoferstraße 5\\
  64283 Darmstadt

  \mbox{}
Fachgebiet Intelligente Systeme
  
  Betreuer: Asst.~Prof. Thorsten Joachims
            Ulrike
  
  Prüfer: Prof.~Dr. Thomas Hofmann

\end{center}
\vfill\mbox{}

\end{titlepage}

}

%% file: prelim/abstract.tex
%
%

\comment{500 words, not more than 700}

\begin{kurzreferat}
Diese Arbeit untersucht, wie n\"utzlich Zugriffsprotokolle als Informationsquelle zum Auffinden von verwandten, wissenschaftlichen Ver\"offentlichungen sind.
Dies wird anhand von {\em arXiv.org} gezeigt, {\em der} 
Instanz zur Vorver\"offentlichung sogenannter preprints in verschiedenen Bereichen der Physik.
Verglichen mit Zitatinformationen haben Zugriffsdaten den Vorteil, dass sie sofort verf\"ugbar sind und nicht erst manuell oder automatisch extrahiert werden m\"ussen.
Deshalb liegt ein Schwerpunkt dieser Arbeit auf der Frage, inwieweit das Verhalten von Nutzern als Ersatz f\"ur explizite Meta-Daten dienen kann, welche potentiell teuer oder \"uberhaupt nicht verf\"ugbar sind. Hierf\"ur werden zugriffs-, inhalts- und zitatbasierte Verwandtschaftsmasse in verschiedenen Szenarien miteinander verglichen.
Als abschliessendes Ergebnis wurde ein Empfehlungssystem erstellt, welches Wissen\-schaftlern dabei helfen kann, weitere relevante Literatur zu finden, ohne aktiv danach suchen zu m\"ussen.
\end{kurzreferat}

\begin{abstract}
This thesis investigates in the use of access log data as a source of information for identifying related scientific papers. This is done for {\em arXiv.org}, the authority for publication of e-prints in several fields of physics.
Compared to citation information, access logs have the advantage of being immediately available, without manual or automatic extraction of the citation graph.
Because of that, a main focus is on the question, how far user behavior can serve as a replacement for explicit meta-data, which potentially might be expensive or completely unavailable. Therefore, we compare access, content, and citation-based measures of relatedness on different recommendation tasks.
As a final result, an online recommendation system has been built that can help scientists to find further relevant literature, without having to search for them actively.
\end{abstract}

%% file: prelim/dedication.tex
%
%

\thispagestyle{empty} 
\begin{dedication}
{\it To Agata Daria --- for unconditional love, comprehension and\\ 
support throughout the course of this thesis.
}

\comment{
This thesis is dedicated to my wife or husband or parents whose encouragement
have meant to me so much during the pursuit of my graduate degree and the
composition of the thesis.
}

\end{dedication}

%% file: prelim/ack.tex
%
%
\begin{acknowledgements}

{
I owe an immense debt of gratitude to my advisors, Thomas Hofmann and Thorsten Joachims, who provided me with the experience to work in a scientific environment on a challenging task.
Both of them constantly supported my work during the whole process.

Besides Thorsten, I would also like to thank Filip Radlinski for all the fruitful discussions and comments, which always have been stimulating and constructive.
Being mentors in the fullest sense of the word, they provided me with comprehensive insight into the scientific mode of operation.

Cornell University is an experience in itself. Its excellence and diversity, in terms of culture and fields of study, is ubiquitous.
My visit to Cornell was an enriching experience, for which I am very grateful and which I do not want to miss.

Also, I want to mention Paul Ginsparg and Simeon Warner from arXiv.org, having made a wealth of data available for the experiments, in an uncomplicated manner.
Not only that their data forms the basis of this thesis, also their domain knowledge has been invaluable.

Finally, special thanks go to my parents and Agata Daria, who always have loved and believed in me, for their support, trust and encouragement.
To each of them, I extend my deepest appreciation.
}

\end{acknowledgements}

%% file: chapters/abbr.tex

\abbrlist 
\begin{center}
{\singlespacingplus
\begin{tabular*}{0.95\textwidth}{l@{\hspace{1.5cm}}l}
ADS    & Astrophysics Data System \\
AP     & Average Precision \\
\arxiv & pronounced ``archive'', as if the ``X'' were the Greek letter $\chi$ \\
CF     & Collaborative Filtering \\
DAG    & Directed Acyclic Graph \\
HEP    & High-Energy Physics \\
HITS   & Hypertext-Induced Topic Search \\
HTML   & Hypertext Markup Language \\
HTTP   & Hypertext Transfer Protocol \\
IR     & Information Retrieval \\
LANL   & Los Alamos National Laboratory \\
LRU    & Least Recently Used \\
LSA    & Latent Semantic Analysis\\
LSI    & Latent Semantic Indexing (application of LSA in IR) \\
MAP    & Mean Average Precision \\
NAT    & Network Address Translation \\
OAI    & Open Archives Initiative \\
PDF    & Portable Document Format \\
PR     & PageRank \\
SLAC   & Stanford Linear Accelerator Center \\
\TeX   & Typographic language and program \\
\tfidf & Term-Frequency $\cdot$ Inverted-Document-Frequency \\

\end{tabular*}
}
\end{center}

%% file: chapters/chap1_introduction.tex
\cleardoublepage

\chapter{Introduction}
\label{sec:introduction}

\comment{general stuff what this is good for, history}

The amount of available information continually grows with an increasing rate, so that the search for a specific information need takes more and more time.
Search engines help to find online available information given a few keywords, but users have to search for them actively and express their demand explicitly.
Thinking of what could exist and by which means it could be retrieved (e.g.\ the keywords to use, or even the search engine to choose) is a time-consuming, straining task.

Recommendation system complement search algorithms by trying to actively push informations to users, which might be useful to them and otherwise eventually overseen. Depending on which information is already available and how expensive it is to acquire new data, normally a mixture of explicit and implicit informations about users and the objects to propose is used.

Because explicit information provided by users is expensive in terms of work imposed on users (rating, assessment, preference), the exploitation of available implicit information is generally to prefer. As an example, the solely aim of {\em Web Usage Mining}, as part of Web Mining, is to apply data mining techniques to the copious available, implicit informations contained in web logs.

\section{Motivation} 

Besides descriptive information about a paper, todays digital libraries also show up relations to some other papers. Relatedness is usually based on either textual similarity or the links, induced by citations.
Similar documents based on text are inherently always available. But normally only obvious, superficial relationships with other papers are found.
Looking for related documents on the basis of citation data might be able to find non-obvious relations to other papers, but here we suffer from the problem that citations are rare and not immediately available. Even for papers, published already for years, there are on average only few citations for them. But, there are even more papers, which never are cited and thus are not covered, i.e.\ will never show up in recommendations given by such systems.
Another problem is to make citation data available. One has to extract citations out of papers and resolve them, which is an instance of a DB record matching problem. Research is occupied with the difficulty of this problem in itself already for decades!
Access data is particularly promising because it doesn't necessarily have this limitations. Not only that it is available for free and easy to collect, it is also likely that all papers in a collection are accessed. The question is, if those accesses contain implicitly enough valuable information about relationships between papers to be useful.
Also, it might last some time to collect enough access data, so that it starts to become useful, but in domains where the availability of explicit information takes even more time, is completely absent or expensive to generate, this might be an interesting alternate source of information. For instance, todays intranet environments typically lack of explicit link information.

Therefore, the aim of this thesis is to investigate in the usefulness of access data for the exemplary problem to recommend related scientific papers. This is done in the scholarly physics community by analyzing data of \arxivorg.
We discuss key aspects that should be considered for building a recommendation system on the basis of access data. By examining data of the considered application domain, we find patterns of the underlying processes and properties of scientific writing.
Finally, to leverage the usage of access data, an online available system has been built that showcases the usefulness of access data and can serve physicists as a complementary tool to filter relevant literature, out of the growing amount of documents.

In this chapter, some background information is provided. This should help to understand the context of the problem and the type of data, we will use throughout this thesis.
In Chapter~\ref{sec:basics} state-of-the-art approaches are introduced to face the problem. These include text- and link-based (usage of citations) methods.
Chapter~\ref{sec:accessdata} is dedicated to explain processing steps to make access data usable and a measure, derived from the implicit informations contained in access logs.
Subsequently, Chapter~\ref{sec:evaluation} discusses the results achieved with the proposed methods, evaluating them on different scenarios for the prediction of citations.
Finally, Chapter~\ref{sec:conclusions} concludes with remarks and envisages possible future work.

\section{\arxiv}

The \arxiv\footnote{\url{http://arxiv.org}}, introduced by Paul Ginsparg, is an online repository for self-archived, so-called e-prints of scientific papers covering different fields of physics, computer science, mathematics and biology. Authors upload their preprints (before peer-review) or postprints (after peer-review) in source format (mostly \TeX), which are automatically converted into postscript or PDF files.
In \arxiv's model, documents consist of meta-data and the document itself in source format. Furthermore, meta-data include title, author list, abstract, submission date and optionally a reference to a journal and categories.
In~\cite{OConnell02:arxivhistory}, O'Connell gives a good overview over the historic development of \arxiv.

Since its foundation, it showcased the possibility of distributing scientific documents freely over the internet, which led to the current revolution in scientific publishing, known as open access movement.
As of September 2006, \arxiv\ consisted of more than 380,000 papers. So it's not astonishing that it is the largest centralized Open Access~\cite{Kurtz2004:OAI} archive available today. Systems like CiteSeer\footnote{\url{http://citeseer.ist.psu.edu}} are still bigger, but they crawl the web for papers, instead of letting the authors archive and maintain their papers themselves. In \arxiv\, a system of endorsements of other authors leads intrinsically to a better data quality.

Initially already started in 1991, today almost all papers in many fields of physics are placed into \arxiv. This makes the scientific community in physics special, because on top of a nearly complete document collection, the evolution and also the access patterns of users are digitally traced already for years.
This makes \arxivorg\ a perfect testbed for evaluations of the potential of access data. Although, the processing steps which we will describe have been very time-consuming, this can easily be justified by further research that might be conducted on its basis.

\section{Recommendation systems}

Recommender try to predict items, a user might be interested in. They represent one instance of information filtering, retrieving relevant information out of a large information space. For that, they use knowledge about a user's profile and the items to propose. Often such informations are provided in turn by other users, either explicitly or implicitly.
There has been a lot of research and also successful commercial recommender systems have been built, e.g.\ for the fields: books~\cite{Linden03_Amazon}, news~\cite{resnick94_grouplens_news}, movies~\cite{Miller03_movielens}, music~\cite{Shardanand95_social}, even jokes~\cite{goldberg01_eigentaste}. A comprehensive survey is provided in~\cite{Schafer99_RecommenderSystems}.
Mostly, they are implemented with a collaborative filtering algorithm.

{\em Collaborative Filtering (CF)\index{Collaborative Filtering}}
is the prediction of a small subset of items (filtering) for a specific user, that is derived out of the taste information of many other users (collaborating). The assumption, underlying of CF approaches, is that the behavior of entities (mostly users) in the past is representative for current ones.
Formally, CF systems can be described with an underlying binary user-item matrix $M_{m \times n}$. Every row $M_{i,*}$ corresponds to a user and describes which items he has purchased or is interested in, and each column $M_{*,j}$ corresponds to an item and marks which users have been interested in it.
The first CF systems calculated k-nearest neighbors, and proposed items that these users also preferred. Such user-based algorithms have their limitations, since they don't scale very well with the number of users and all calculations have to be done online. Item-based algorithms precompute in a first step relationships between pairs of items offline, and combine in a second, online step those, a user is currently interested in. Due to the better scalability of this approach, recommendations have also become feasible for large datasets like \arxiv.
Nowadays, (item-based) CF is applied in most recommendation systems and has proven to help users in all mentioned domains.

\section{Related work}
\label{sec:introduction:related}

There have been a lot of effort in the investigation of recommendation systems. Also for the domain of research papers, there have been many attempts to use the inherent properties available here, like text or citations to learn something about patterns in research communities.

Citation analysis is a large part of what today is called {\em bibliometrics}\index{Bibliometrics}. The usage of citations in itself already reveals much insight about the connected papers. E.~Garfield is a well-known representative for citation analysis, having invented the most applied measure for assessing scientific impact~\cite{Garfield79:citationindexing}.

To research papers, the first application of recommenders has been done by McNee et~al.\ in 2002 for the field of computer science~\cite{McNee2002}. They applied different recommendation algorithms and evaluated their quality offline and also online via user ratings. The experiments show that no algorithm is the best for multiple usage scenarios. In a system called {\em TechLens}~\cite{McNee2005:Techlens} the authors try to provide a basis for their future research in the field of providing in itself configurable information retrieval methods to meet the needs of different kinds of applications.
Furthermore, the KDD~Cup 2003 consisted of one task to predict citations for a small subset of \arxiv\ out of full-text and citation data~\cite{KddCup2003}.

There also have been some attempts to investigate in the potential of access data.
Brody et~al.\ showed that download data can serve as an early predictor of later citation impact~\cite{Brody2005:webusage}. Some basic work for this has been done in preliminary analysis of web logs in the {\em Open Citation Project}~\cite{Brody00_opcit}.
Kurtz et~al.\ examine article readership information of the {\em NASA Astrophysics Data System~(ADS)} and already suggest a two-dimensional view of access and citation counts for assessing an individual's scientific productivity~\cite{Kurtz05_readcite}. This suggests that a part of the information contained in access data is sensitive to a different kind of research usage of publications.

Woodruff~et~al.~\cite{Woodruff2000} combine text and citations into a model to make reading recommendations, which evaluate significantly better than the other models on their own. In contrast, the combination of text with collaborative methods, but also the usage of access-based measures in a real application are rare to find.

An early exception is WebWatcher~\cite{joachims97_web}, a tour guide accompanying users browsing the web and recommending related web pages.


%% file: chapters/chap2_basics.tex
\chapter{Basics}
\label{sec:basics}

\newcommand{\lvec}[1]{{\overrightarrow{#1}}}

Due to the vast amount of data that is subject to be recommended, we have chosen to implement an item-based recommendation system. For that, in a first step we need to extract relationships between items (papers in our case), which reflect the kind of relevance, proposed recommendations shall have. The big advantage is that they can be performed once offline, so their computation is not time-critical. The approach resembles a divide-and-conquer strategy, which can often be seen to succeed.
Because the way, how to aggregate such pre-calculated measures in a second step, is not obvious, we defer those consideration to Chapter~\ref{sec:evaluation}, where further setups are provided to avoid aggregation.
This chapter introduces basic relatedness measures which reflect state-of-the-art approaches and are used in most recommendation systems deployed in digital scientific libraries like CiteSeer. Later, they will serve as a baseline, the utility of access data will be compared against.



\section{Text-based methods}

The usage of text-based similarity for revealing relationships between scientific papers seems to be the most obvious. The assumption is that scientists are interested and will more likely refer to other papers which have similar content.
Textual methods have been widely used in Information Retrieval (IR). The prevailing approach is to see a document as a {\em bag-of-words}, calculating weights for them, for instance with equation~\ref{eqn:tfidf}. Although there is evidence that the word order and with it some contextual information is important on the word and sentence level, the bag-of-words model even seems to be sufficient for representing topicality of paragraphs~\cite{Landauer97_bagofwords}.
An easy weighting scheme is given by Salton~\cite{Salton89_IR}. The \tfidf\ weight $w_{ij}$ for a term $t_j$ in a document $d_i$ is given by
\begin{equation}
w_{ij} = t\!f_{ij} \cdot \log \frac{N}{n}
\label{eqn:tfidf}
\end{equation}
where $t\!f_{ij}$ is the frequency of term $t_i$ in document $d_j$ and $^n\!/\!_N$ is the fraction of documents, the term occurs in at least once. The formula reflects the simple assumption that a term describes a document the better, the more often it occurs in the document (term frequency), but the less often it occurs in other documents (inverse document frequency).
There exist a lot of variants and extensions to this elementary formulation, e.g.\ to normalize the inversed document-frequency differently or with another base than two for the logarithm.
To represent whole documents, the vector-space model allows to use simple vector algebra for estimating the similarity between documents. Here, a document is a vector consisting of all possible words, resp.\ their \tfidf\ weights. Thus, document similarity between two documents $d_i$ and $d_j$ can be defined as the angle between their feature vectors $\vec{d_i}$ and $\vec{d_j}$:
\begin{equation}
\textrm{rel}(d_i, d_j)\ {\buildrel\rm def\over=}\ \cos(\vec{d_i}, \vec{d_j}) =
\frac{\vec{d_i} \cdot \vec{d_j}}
{\|\vec{d_i}\|_2 \  \|\vec{d_j}\|_2}
\label{eqn:cossim}
\end{equation}

Having a model, how to compute similarities is just the first step, the probably more important one is to choose the right data to apply it on. The available textual parts of a document consists of title, abstract and full-text. We opted not to use the title in itself, because it consists a very limited amount of words to describe different topics of the whole document. Instead, we calculated similarities using either concatenations of title and abstract (we refer to this as meta-data) or title, abstract and the full text (referred to as full-text).

Experience has shown that the rendering of documents in source format (\TeX) to a final representation like PDF and subsequently the extraction of type-setted text is the most passable way to retrieve a textual representation of a document. This gives us an equivalent to what readers see. On the other hand, we lose all markup information, normally contained in \TeX\ documents. Unfortunately, if we would like to retrieve those, a complete \TeX\ system would have to be extended because of the extensive use of author specific, freely defined macros. However, scientific papers have a very similar textual structure, so that for our purposes the indirect procedure is to favor.

To calculate similarities between $n=350,000$ documents implies an order of $O(n^2)$ pairwise comparisons. Such a calculation becomes easily prohibitive, should the 349,999 similarity calculations for one paper last multiple seconds. Even worse, also with use of a sparse feature vector representation for every document, not all documents can be held in memory.
However, such an implementation would be useful for an exact determination of the used similarity formula, but, since it shall only serve as a baseline, we have chosen to use an existing standard implementation for our purposes.

\begin{figure}
  \centering
    \includegraphics[width=12cm]{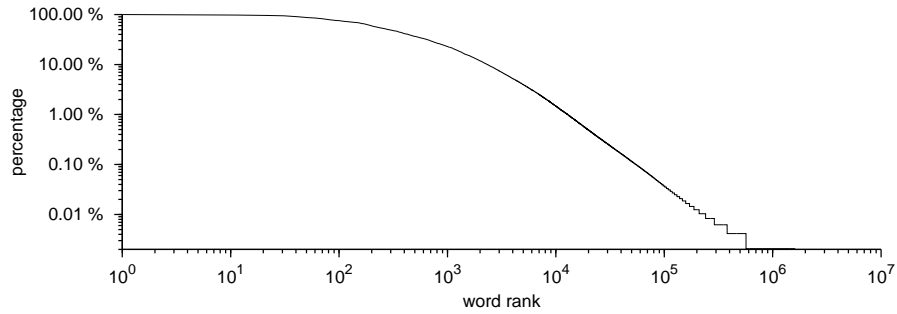}
  \caption[Ranked document frequencies of words]{Ranked document frequencies of words, before stop-word removal. After removal the curve reflects a typical power law distribution. }
  \label{fig:wordfreq}
\end{figure}

We made intensive use of the freely available, open-source Java implementation of a search engine, {\em Lucene}.\footnote{\url{http://lucene.apache.org}} It offers classes for word-parsing, indexing, querying and is easily extendable. The $\approx100$ in every document occurring words, as depicted in Figure~\ref{fig:wordfreq}, suggest the need for a customized stop-word list, which not only contain english filler words, but also artefacts like variable names of formulas, digits, etc.
The way we extract text imposes the need for de-hyphenation of words at end of lines. This has been done only for words, for which the de-hyphenated version already exists at least three times in the dictionary over all found words. Not to give the textual model an advantage in the prediction task, which we consider during this thesis, we located and removed the reference section by textual heuristics.\footnote{For those, see Table~\ref{tbl:regex}.}
For the generation of similarities, we use implicitly the adjusted \tfidf-formula implemented in Lucene for calculation of score values between a query and the indexed documents. For this, we extract 1,000 words with the highest \tfidf\ weight out of each paper and formulate a query containing those.
However, since we ignore within the query the exact \tfidf\ weights of the document to query, we might not obtain exact cosine-similarities as described by Equation~\ref{eqn:tfidf}, but we achieve an equivalent to what a user might find, using a search engine manually. For our experiments this is preferable.
The size of the final index, built over roughly 350,000 papers, takes 3~GB disk space.

\section{Citation-based recommendations}

Almost all scholarly documents contain a reference section, listing other papers text passages refer to. That way, a relationship between documents has already been established explicitly by the author. The characteristics of this relation is not so obvious, because it might be either credit or acknowledgement to the influence of previous work, or also refutation of others statements.
However, a common denominator would be that a citation is a relevance judgement of the author of a paper, i.e.\ references express some kind of relatedness.
In information retrieval, the inclusion of graph-based measures lead to a significant improvement over pure text based systems. The best known example for this is probably the success of Google, using PageRank as an importance measure derived from the web graph.

Because papers refer by means of citations to other papers, which in turn have their own set of references, citations can be resolved into a so-called {\em citation graph}. Such a graph $G = (V,E)$ consists of a set of {\em vertices}~$V$ (the papers) and a set of {\em edges}~$E$ (the references), connecting the vertices. The edges $(u,v) \in E$ are directed, building non-symmetric relationships between vertices $u, v \in V$.
For clarity reasons, throughout this thesis we will use the term ``reference'' for papers referred to in a paper, and the term ``citation'' for a reference from a paper. In terms of the citation graph this means, references are the outgoing edges of a paper, while the citations are the incoming ones, which in turn refer uniquely to the citing papers.

As any graph, a citation graph can be uniquely described in matrix form by means of an {\em adjacency matrix} $C_{n \times n}$. An element $C_{i,j}$ of the matrix is $1$ if paper $d_i$ has a reference to paper $d_j$, and $0$ otherwise. While a row vector $\lvec{C_{i,*}}$ refers to all references of a paper~$d_i$, a column vector $\lvec{C_{*,j}}$ contains all citations to paper~$d_j$.
The usage of an adjacency matrix also maps the domain of paper recommendations directly into the framework of collaborative filtering. Here, a paper would represent a `user', giving its relevance judgement with a reference to papers, which would represent `items'.
\comment{
\begin{equation}
C = 
\left(
\begin{array}{ccc}
adj(0,0) & \dots & adj(0, N) \\
\vdots & adj(i,j) & \vdots \\
adj(N, 0) & \dots & adj(N, N) \\
\end{array}
\right)
\end{equation}
where
\begin{equation}
adj(i,j) = \left\lbrace
\begin{array}{rl}
1, & \textrm{if there is an edge }D_i \to D_j \\
0, & \textrm{otherwise.}  \\
\end{array}
\right.
\end{equation}
}

The underlying process inherently leads to properties of citations and they again to specifics of the graph and matrix structure, distinguishing them from arbitrary graphs.
First of all, the graph is very sparse due to the average number of references in every paper, compared to the amount of papers potentially to refer to.
Then, a paper can only reference past work and being referenced by papers published in future. This leads to the following inequality where $t(d_i)$ is the publication date of an arbitrary document $d_i$:
\begin{equation}
\forall d_i: \quad
  \max_{\forall j: (i,j) \in E}{t(d_j)}
  < t(d_i) <
  \min_{\forall k: (k,i) \in E}{t(d_k)}
\label{eqn:ref_time}
\end{equation}
This leads to a strictly {\em directed acyclic graph~(DAG)}, which is continuously extending itself over time, introducing more edges to past vertices with every new vertex. It is likely that the homogeneity of such a graph is strongly biased by time.

Before the digital age, it was the field of activity of librarians to resolve and index references. Later, such informations have been increasingly used for bibliometrics, basing measures on them for the evaluation of research.
With the digital availability of documents also automatic citation indexing became possible, which allows for the generation of bigger and more complete citation graphs in a much cheaper way~\cite{giles98_citeseer}.

\begin{table}
\scriptsize \centering
\comment{
\begin{verbatim}
[1] A. Vilenkin and E. P. S. Shellard, Cosmic Strings and Other Topological Defects
    (Cambridge University Press, Cambridge, England, 2000).
[2] M. B. Hindmarsh and T. W. B. Kibble, Rep. Prog. Phys. 58, 477 (1995).
[3] M. Sakellariadou, Cosmic strings, [arXiv:hep-th/0602276].
[4] T. W. B. Kibble, J. Phys. A 9, 387 (1976).
[5] R. Jeannerot, J. Rocher and M. Sakellariadou, Phys. Rev. D 68, 103514 (2003).
[6] F. R. Bouchet, P. Peter, A. Riazuelo and M. Sakellariadou, Phys. Rev. D 65, 021301 (2002);
    L. Pogosian, M. Wyman and I. Wasserman, J. of Cosm. and Astrop. Phys. 09, 008 (2004);
    M. Wyman, L. Pogosian and I. Wasserman, Phys. Rev. D 72 (2005) 023513.
[7] J. Rocher and M. Sakellariadou, JCAP 0503,004 (2005);
    J. Rocher and M. Sakellariadou, Phys. Rev. Lett.94, 011303 (2005);
    J. Rocher and M. Sakellariadou, D-term inflation in non-minimal supergravity
   [arXiv:hep-th/0607226]
[8] S. Sarangi, S.-H. H. Tye, Phys. Lett. B 536, 185 (2002).
[9] T. Damour and A. Vilenkin, Phys. Rev. D 71, 063510 (2005).
[10] M. Sakellariadou, Phys. Rev. D 42 , 354 (1990).
[11] M. Snajdr and V. Frolov, Class. Quant. Grav. 20 (2003) 1303.
[12] M. Snajdr, V. Frolov and J.-P. de Villiers, Class. Quant. Grav. 19 (2002) 5987.
[13] A.M. Polyakov, Phys. Lett. B 103 (1981)207.
[14] C.W. Misner, K.S. Thorne and J.A. Wheeler, Gravitation (W.H. Freeman, San Francisco, 1973).
[15] H. Thirring and J. Lense, J. Phys. Z 19 (1918) 156.
\end{verbatim}
}
\begin{minipage}{0.7\textwidth}
\begin{verbatim}
Bondi, H. 1952, MNRAS, 112, 195
Brown, G.E. 1995, ApJ, 440, 270
Burrows, A., & Woosley, S. 1986, ApJ, 308, 680
Cannon, R.C. 1993, MNRAS, 263, 817
Cannon, R.C., Eggleton, P.P., Zytkow, A.N., P. 1992, ApJ, 386, 206
Chevalier, R.A. 1989, ApJ, 346, 847 (C89)
Chevalier, R.A. 1993, ApJ, 411,L33
Colgate, S.A. 1971, ApJ, 163, 221
Colgate, S.A., Herant, M., & Benz, W. 1993, Phys. Rep., 227, 157 (CHB)
Cox, A.N., Vauclair, S., & Zahn, J.P. 1983, Astrophysical Processes
     in Upper Main Sequence
Stars, (CH-1290 Sauverny : Geneva Observatory)
Davies, R.E., & Pringle, J. 1980, MNRAS, 191, 599
Davies, M.B. & Benz, W., 1995, MNRAS, submitted
\end{verbatim}
\end{minipage}
\caption[Excerpt of a typical reference section in physics papers]{Excerpt of a typical reference section in physics papers.}
\label{fig:ref_section}
\end{table}
Unfortunately, the automatic extraction of references is much harder in physics and would need a lot of expert and background knowledge. Table~\ref{fig:ref_section} gives an impression of how a typical reference section of a physics paper looks like.
Luckily, the Stanford Linear Accelerator Center~(SLAC) makes already a long time an effort with the SPIRES database\footnote{\url{http://www.slac.stanford.edu/spires}} to manually keep track of references for papers of High-Energy Physics~(HEP). It is run by the late 1960's and was 1991 one of the first web sites available. Next to their own indexing and keys used, they also maintain cross-references to \arxiv.
This gives us a valuable source of information for \arxiv\ papers. Although \arxiv\ consists of multiple fields and SLAC/SPIRES covers primarily only HEP papers we still obtain a high coverage. As of July 2006, we have citation information for 192,963 out of roughly 350,000 papers. Even though, we have complete reference lists for those papers, for now, we have chosen to ignore such references of them which point to papers not contained in \arxiv. However, Figure~\ref{fig:cocit:freq1} indicates that a considerable amount of references is left. This is due to the almost complete set of documents in fields like High-Energy Physics. The decreasing amount of submitted papers is evidence for that.

\begin{figure}
{\centering
  \includegraphics[width=0.8\textwidth]{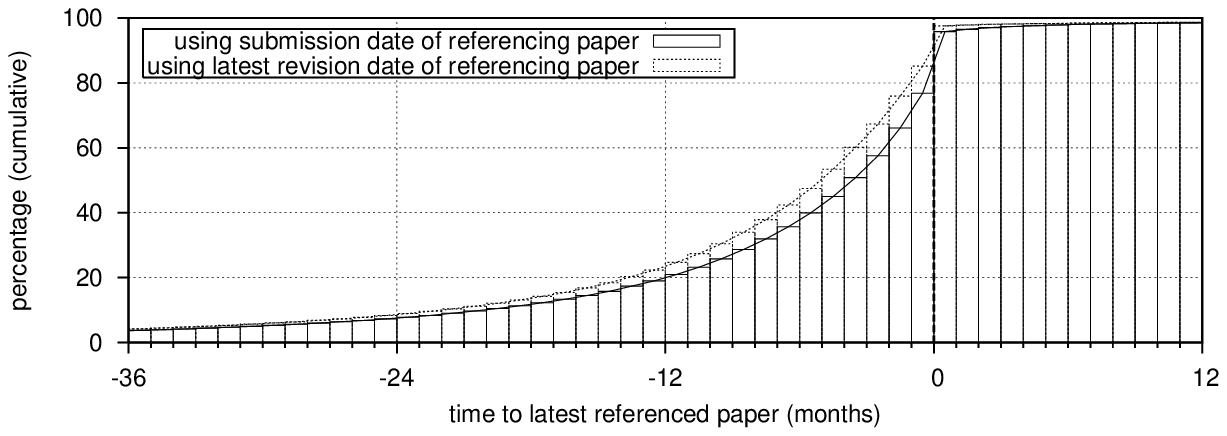}
  \caption[Time between paper publication and latest reference in it]{Time between paper publication and latest reference in it.
The steepness of the curve gives information on the currency, how fast published preprints can be incorporated into new papers. It is remarkable that 40\% of the papers refer to another paper, having been published at most 3 months ago.
}
  \label{fig:paper:lastreftime}
}
\end{figure}

When it comes to real data, due to noise and artefacts often general assumptions are contradicted. Figure~\ref{fig:paper:lastreftime} shows that already such a trivial supposition like Inequation~(\ref{eqn:ref_time}) doesn't hold anymore.
For roughly $2$\% of the documents, the most recent referenced paper has been published in future. This is because the submission to \arxiv\ does not necessarily coincide with the real initial publication, which is unfortunately not generally available.\footnote{Even though for some submissions there is a textual note in the comments field, it is unstructured and thus not easily accessible.} Also papers can be updated such that new references might be added.
However, this example shows the importance to test every assumption on real data to account for the fraction by which the data violates them.


\subsection{Co-citation and co-reference}

A citation is an explicit expression that a cited paper has relevance and is important for a topic. In this sense, the fact that papers are cited together forms some kind of relationship between them.
This so-called {\em co-citation} of papers has been used in bibliometrics for long time and has shown to be able to establish patterns of linkage~\cite{Small73_cocitation}. This can be explained with the circumstance that independently from a paper itself, a co-citation for this paper refers to a judgement of a third party.
Although the reference list of a paper is normally diverse, consisting of papers covering different subtopics, they are still related in a broader sense, indirectly over the referring paper. Also, its measurement over multiple citing papers is likely to account for this flaw by interpolation, giving more often co-occuring papers a higher similarity.
\begin{samepage}
\begin{definition}[Co-citation]
Given the adjacency matrix $C$ of the citation graph $G = (V,E)$, {\em co-citation} between two documents $d_i$ and $d_j$ is defined as the absolute number of papers which cite both documents $d_i$ and $d_j$, i.e.,
\begin{eqnarray}
\textrm{rel}(d_i, d_j)\ {\buildrel\rm def\over=}\ \textrm{co-cit}(d_i, d_j)
 & = & | \lbrace d_k : \exists (k,i) \in E \land \exists (k,j) \in E \rbrace |  \nonumber \\
 & = & \lvec{C_{*,i}} \cdotp \lvec{C_{*,j}} = \sum_k{C_{k,i}C_{k,j}} \nonumber \\
 & = & 
			\left\{
			\begin{array}{ll}
			(C^TC)_{i,j} & \textrm{if $i \ne j$} \\
			0            & \textrm{otherwise.} 
			\end{array}
			\right.
\end{eqnarray}
\label{def:cocitation}
\end{definition}
\end{samepage}

\begin{figure}[b]
{\centering
  \includegraphics[width=0.55\textwidth]{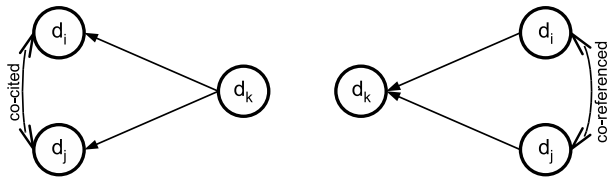}
  \caption{Co-citation and co-reference}
  \label{fig:cocit:cocitation}
}
\end{figure}

The counterpart to co-citation is {\em co-reference, a.k.a.\ bibliographic coupling}~(see Figure~\ref{fig:cocit:cocitation}). Apart from bibliometrics, it is also used in linguistics to describe the relation between two strings, which point to the same entity.
\begin{definition}[Co-reference]
Analogously to definition~\ref{def:cocitation}, given the adjacency matrix $C$ of the citation graph $G = (V,E)$, {\em co-reference} between two documents $d_i$ and $d_j$ is defined as the absolute number of papers that are co-referred to by $d_i$ and $d_j$, i.e.\ the number of references shared between paper $d_i$ and $d_j$,
\begin{samepage}
\begin{eqnarray}
\textrm{rel}(d_i, d_j)\ {\buildrel\rm def\over=}\ \textrm{co-ref}(d_i, d_j)
 & = & | \lbrace d_k : \exists (i,k) \in E \land \exists (j,k) \in E \rbrace |  \nonumber \\
 & = & \lvec{C_{i,*}} \cdotp \lvec{C_{j,*}} = \sum_k{C_{i,k}C_{j,k}} \nonumber \\
 & = & 
			\left\lbrace
			\begin{array}{ll}
			(CC^T)_{i,j} & \textrm{if $i \ne j$} \\
			0            & \textrm{otherwise.} 
			\end{array}
			\right.
\end{eqnarray}
\end{samepage}
\end{definition}

Co-reference has the advantage that it can be also applied on papers which have not been cited yet, which means they don't have co-cited papers. Thus, for broad application of co-citation it is important to examine its coverage.

The maximal possible co-citation value for a pair of papers is given by the minimum of their individual numbers of citations. This upper bound could be used to normalize the inherently unbounded co-citation values. While making co-cited papers of often cited papers more comparable, this could lead to an overestimation for rarely cited papers.
Similarly, the upper bound for co-reference is the smaller number of references of both of them (Figure~\ref{fig:cocit:freq1}):
\begin{figure}
{\centering
\includegraphics[width=0.8\textwidth]{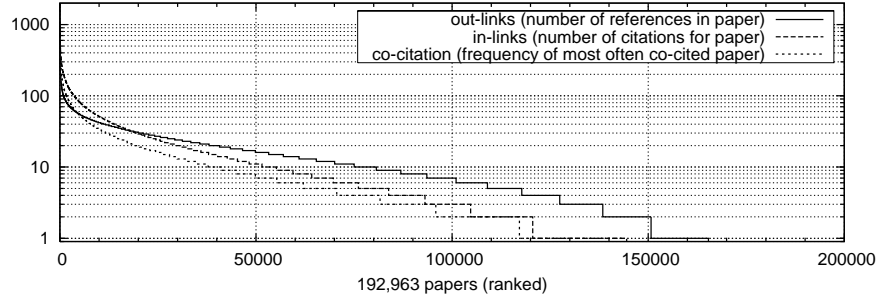}
  \caption[Frequencies of references, citations and the most often co-cited paper]{Frequencies of references, citations and the most often co-cited paper over all papers, ranked independently.}
  \label{fig:cocit:freq1}
}
\end{figure}
\begin{eqnarray}
\textrm{co-cit}(d_i, d_j) & \le & \min( \| \lvec{C_{*,i}} \|_1, \| \lvec{C_{*,j}} \|_1 ) \\
\textrm{co-ref}(d_i, d_j) & \le & \min( \| \lvec{C_{i,*}} \|_1, \| \lvec{C_{j,*}} \|_1 )
\end{eqnarray}

Another noticeable difference is that co-citations are able to adjust over time because of the possibility to become cited still long time after initial publication. Relationships derived from co-reference are either fixed, or if new revisions of \mbox{papers} are possible in the hand of one set of authors, whose choice in turn might be biased.

\begin{figure}
{\centering
\includegraphics[width=0.7\textwidth]{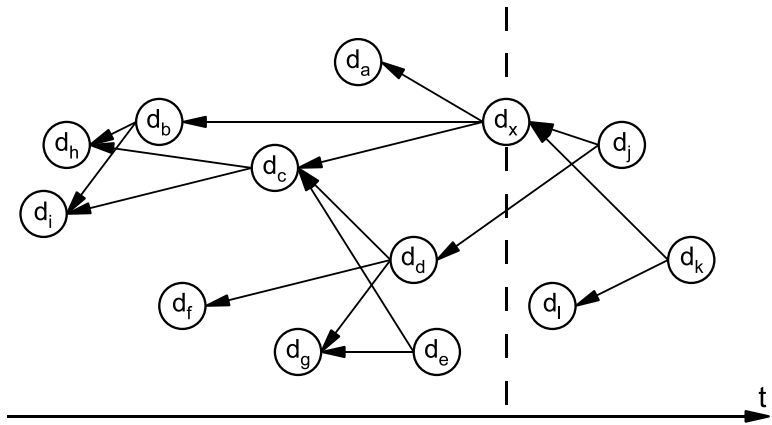}
\caption{Exemplary citation graph}
  \label{fig:citgraph}
}
\end{figure}

Co-citation and co-reference are symmetric pair-wise measures, such that for recommendation purposes one of the input documents can become fixed, while the set of other papers ordered by their score gives us ranked recommendations.
For a given scenario, an essential question is where to apply the measures.
Figure~\ref{fig:citgraph} shows an exemplary citation graph, in which $d_x$ is considered a present paper. Co-citation for $d_x$ would retrieve papers $d_d$ and $d_l$, since they are co-cited by $d_j$ and $d_k$. For this case, co-reference would also yield to $d_d$, and $d_e$ because they share the reference to $d_c$ with $d_x$.
But, depending on the scenario, i.e.\ the amount of knowledge available at a point in time, some possibilities can not be considered. For instance, co-citation cannot be applied to $d_x$, given that we don't know future citations of $d_x$ yet.
Since the references of a paper lie in the past, there is a chance that for those, we already have citations.
With such an application, we receive $d_g$ and $d_f$ co-cited, and $d_b$ co-referenced with $d_c$.
A problem with this approach is that we now have multiple papers and a ranked list of co-cited papers for each of those. As mention earlier, the choice of an aggregation function is crucial in a collaborative filtering implementation.
To examine all possibilities we will also consider co-reference applied to the references of papers and compare the recommendations achieved.

\section{Importance measures} \label{sec:importance}

Given the citation graph, the question arises, how the structure of the graph can reveal properties of its vertices. For information retrieval, some query-independent {\em importance measures} have been investigated, which are able to improve retrieval performance, if they are introduced into the retrieval function.
The assumption for this is that important vertices, as derived by means of the graph structure, should be presented at a higher rank than less important ones.
We give an overview over proposed importance measures and apply them to the citation graph.
Out of bibliometrics or citation analyzes, basic importance measures have emerged. Among those are the well established citation impact of a paper, but also measures to assess the reputation of authors or journals~\cite{Garfield79:citationindexing}.

{\em Citation impact} is probably the most known quantity which simply specifies the number of citations for a scientific article, resp.\ paper, i.e.\ the number of in-links in the citation graph.
This very local measure has some problems which have been criticized in the past. For example, comparability can only be achieved between papers with the same age and it doesn't distinguish between the papers which cite them. A citation from another paper with a high impact is the same valuable as a self-citation of the same author. On the other hand, its simplicity led to a diffusiveness, another measure would have to compete with before it could be accepted by a large audience. However, there is no reason, not to use more sophisticated algorithms for the estimation of importance for internal usage.

The next subsections introduce the two most widely used importance measures: {\em PageRank} and {\em HITS}.
We have chosen to implement them to prepare their usage in further work. For instance, it would be interesting to see if importance measures can help to improve the performance for our specific prediction task.

Further proposed algorithms are the {\em Hilltop}-algorithm~\cite{Bharat2001_Hilltop}, which ranks documents based on authority scores, which in turn depend on the search query terms and {\em TrustRank}~\cite{Gyongyi04_trustrank}, combating Web~Spam with the propagation of the notion of trust starting from a seed set of trusted vertices, or {\em SpamRank}~\cite{Benczur05_spamrank}, penalizing vertices with biased distributions over their in-links.

\subsection{PageRank}

Originally developed at Stanford University by Larry Page, PageRank is today integral part of the web search engine Google~\cite{Page98_pagerank}. Basically, it calculates a \mbox{numerical} weight for every vertex, reflecting its relative importance. This is done with the following recursive equation:
\begin{equation}
PR(D_i) = (1-d) + d\sum_{\forall j : (j,i) \in E}{\frac{PR(j)}{C_j}}
\label{eqn:pr1}
\end{equation}
$C_j$ is the number of out-links of vertex $j$ and with $d$ a so-called damping factor is introduced.
The equation reflects the ideas of the {\em Random-Surfer}-model, in which a surfer randomly chooses with probability~$d$ to follow links, instead of jumping to another initial vertex (with probability $1-d$). Hence, $d$ is between $0$ and $1$.\footnote{In Brin et~al.~\cite{Brin98_google}, a damping factor of 0.85 is suggested.}
The iterative calculation of PageRank with Equation~\ref{eqn:pr1} leads to weights that sum up to the number of vertices. To obtain a probability distribution, instead the following formulation can be used directly:
\begin{equation}
PR(D_i) = (1-d)\frac{1}{n} + d\sum_{\forall j : (j,i) \in E}{\frac{PR(j)}{C_j}}
\  ;\qquad
\sum_{i=1}^n{PR(i)} = 1
\label{eqn:pr}
\end{equation}

\cite{Page98_pagerank} states that ``we [Page et~al.] found on tests of the Stanford web that PageRank is a better predictor of future citation counts than citation counts themselves.'' This justifies that PageRank, incorporating multiple levels of indirection, might be better than simple citation impact measures. On the other hand, as strong improvements as Page et~al.\ have seen would probably not be possible because of the slightly different properties of the citation graph.

\begin{figure}[h]
{\centering
  \includegraphics{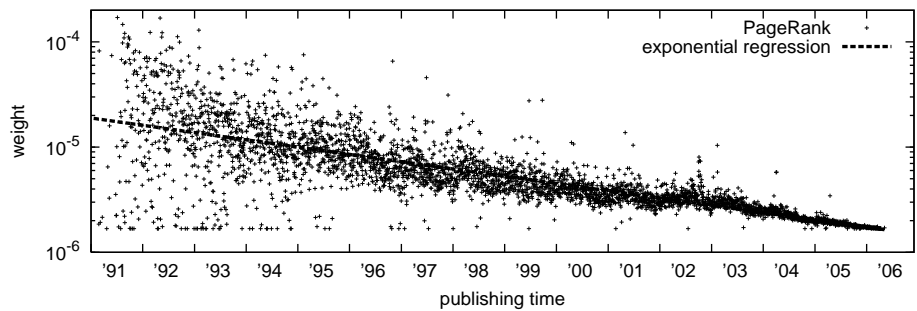}
  \caption{PageRank scores over time}
  \label{fig:cit:pr}
}
\end{figure}

We calculated PageRank on the citation graph and receive an interesting distribution over time (see Figure~\ref{fig:cit:pr}), which can be explained by the properties of citation data. A comparison of papers of different publication date by PageRank is prohibitive, if not some kind of normalization is performed.

\subsection{HITS}

{\em HITS} (Hypertext-Induced Topic Search), another link-based ranking method, was introduced by Jon Kleinberg~\cite{Kleinberg99_hits}. Basically, it not only estimates one importance measure like PageRank, but simultaneously two of them: {\em hubs} and {\em authorities}. Authorities are such vertices in a graph, many other vertices with a high hub-value point to. Hubs, in turn, are defined as such vertices, which point to strong authorities. Each vertex is assigned an authority score $x_i$, and a hub score $y_i$. The inherent circularity can be broken by iteratively solving
\begin{equation}
x_i'=\sum_{j:(j,i) \in E}y_j, \quad y_i'=\sum_{j:(i,j) \in E}x_j, \qquad x_i= x_i'/\|x'\|_2, \quad y_i=y_i'/\|y'\|_2
\end{equation}
For not having to calculate HITS over the whole internet, Kleinberg proposed also heuristics to generate a subgraph out of the results of a standard web search, that is still connected enough to calculate reasonable importance measures. For our amount of data, such a complexity reduction is not needed.

Motivated by the fact that there are different kinds of scientific papers, we opted also to implement HITS. Rather than simply measuring the impact of a paper by the number of references to it, we are more interested in its authority, resembling to be referred to by good hubs. Documents with high hub values should resemble to survey papers, tutorials or introductory articles, which refer to important papers.

\begin{figure}
\centering
  \includegraphics[viewport=0 0 360 232, width=11cm]{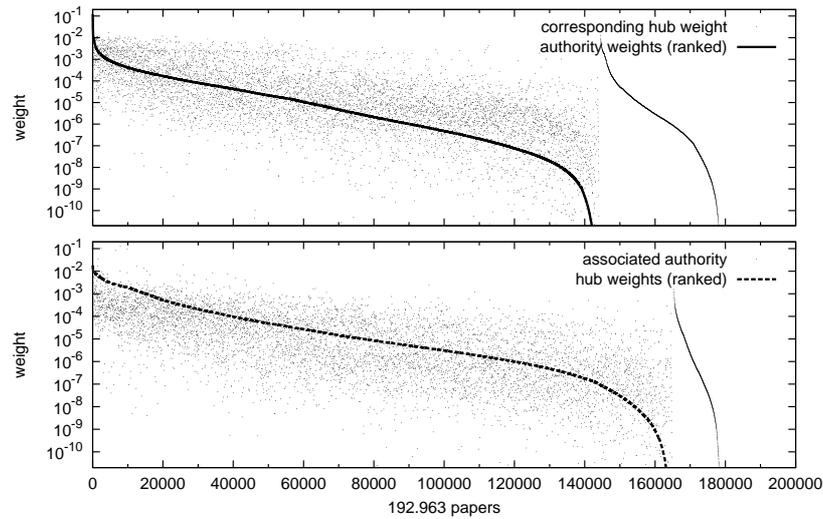}
  \caption[Corresponding hub and authority weights (ranked)]{Distribution of authority (resp. hub) weights over ranked papers, and their corresponding hub (resp. authority) weights.}
  \label{fig:cit:hits_dist}
\end{figure}

We calculated hub and authority values for all documents with available citation data. Primarily, the results in Figure~\ref{fig:cit:hits_dist} show that hub and authority weights are correlated. For the domain of research papers, this might be explained by the fact that often cited papers (authorities) also cite many other authorities and thus, serve as a hub.
Interesting are also high density areas in the graphs: The top 30,000 authorities tend also to have a high hub weight (around $2\cdot{}10^{-3}$).
Since there exists a bias in the data towards having more references than getting cited, the hub weights tend to exceed the authority weights. This becomes more significant for papers with low authority values.
 It is also shown that there are less non-zero authorities than hubs. This can be explained by the fact that nearly every paper references others, while they might not be cited and thus have an authority of 0.
\comment{
 There are 27621 papers without references (pointing to other \arXiv papers).
 48450 don't get citations.
}Even though for some papers the weight of authorities (resp.\ hubs) is zero, their corresponding hub (resp.\ authority) weights still form a smooth distribution.

\begin{figure}[b]
{\centering
  \includegraphics{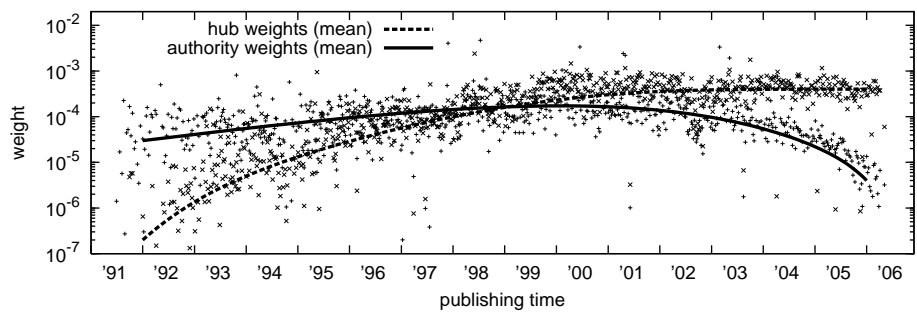}
  \caption{Hub and authority scores over time}
  \label{fig:cit:hits_mean}
}
\end{figure}

Like PageRank, also HITS shows certain irregularities depending on the publishing date (Figure~\ref{fig:cit:hits_mean}). The references of recent papers are more likely to exist already in \arxiv. Thus, the mean of the hub weights is strictly increasing, beginning lowest for the oldest papers published to \arxiv. Interestingly, the curve seems to flatten out from around 2002, which indicates that from 2002 \arxiv\ already contained a corpus of scientific literature such that the fraction of references not directed to papers of \arxiv\ became constant.
Authority weights, similar to PageRank show the opposite behavior: most recent papers denote a strong decrease in their weight due to missing citations.
Furthermore, the authority of papers also declines the older they are from around 1999. This can be explained by the steadily increasing amount of published papers to \arxiv, whose references in turn form a distribution over the age of previous work. For physics, the mean of such distributions lies typically within a decade (see Figure~\ref{fig:citations:reference_dist2005}).

\subsection{Convergence}
Brin et~al. argue, that the convergence of PageRank depends on the structure of the considered graph~\cite{Page98_pagerank}. They refer to Motwani Raghavan for mathematical details~\cite{Motwani95_randomwalks}. Basically, the elimination of dangling vertices (documents without references) ensures that PageRank converges.
HITS even converges for any graph, without the need for adjustments~\cite{Kleinberg99_hits}.

\begin{figure}
{\centering
  \includegraphics[width=11cm]{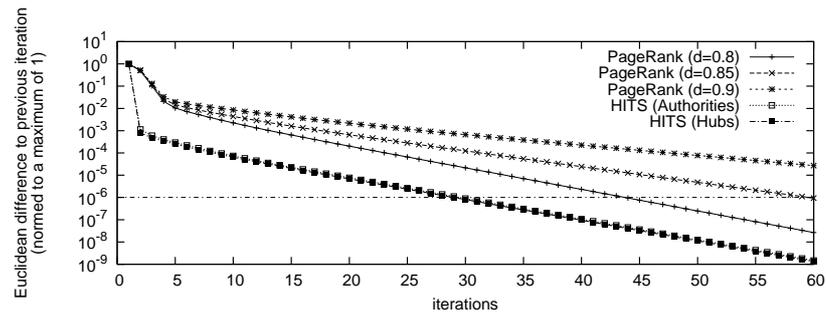}
  \caption{Convergence of PageRank and HITS}
  \label{fig:cocit:prhits}
}
\end{figure}

Figure~\ref{fig:cocit:prhits} shows the convergence behavior of PageRank and HITS on top of the citation graph. It reveals, depending on the damping factor $d$, how many iterations are needed for a specific level of precision.

As we have seen, there are many obstacles for the direct usage of importance measures. The strong bias over time has to be corrected, before they can be used comparing papers with different publication date. Otherwise, we would end up with recommendations of very old papers. We could have done so, but the main focus of this thesis is not on the improvement of an existing recommendation system by incorporation of importance measures, but to build an initial one on the basis of access data and its performance compared to currently used measures. This might be addressed in future work.

\comment{
\section{Relatedness}

, while all others ranked yields to a ranking of related documents
$r() : d_i \to {d_{r_1}, d_{r_2}, \dots, d_{r_N}}$
$r() = sim(d_i, d_{r_1}) \ge sim(d_i, d_{r_2}) \ge \dots \ge sim(d_i, d_{r_N})$

(show a simple graph with timeline)
We have chosen to use the explicit citation data provided through the authors to consider as the ground truth.

\begin{definition}[Non-Symmetric relatedness]
sjahfajf
\begin{equation}
\textrm{related}(d_i) = \{ d_j : \exists (i,j) \in E \}
\end{equation}
\end{definition}

relatedness = 
related( $d_i$, $d_j$) = 1, iff $d_i$ cited $d_j$

not symmetric rel($d_i$, $d_j$) $\ne$ rel($d_j$, $d_i$)

take time-aspect into account:
Another possibility is to define a set of related papers for every paper
related( $d_i$, $d_j$) = ${d_k: d_i, d_j are cocited between t_1 and t_2}$

\begin{definition}[Symmetric Relatedness]
$t_1$ is the ...with $t_1 > t(d_i)$ \\
Two papers $d_i$ and $d_j$ are related, if they are co-cited between a time-interval 
\begin{eqnarray}
\textrm{related}(d_i) & = & \{ d_j : \textrm{co-cit}(d_i, d_j) \ne 0 \}
\end{eqnarray}
\label{def:rel_cocit}
\end{definition}

Evaluation function / paper -> ranking, set -> ranking \\

AP=

MAP=

}

\clearpage
\section{Normalization}\label{sec:basics:normalization}

Normalization is typically conducted to enforce certain mathematical properties (e.g.\ the probabilities of a distribution to sum up to $1$, or limiting values to an interval).
Another kind of normalization is to reverse the influence of unwanted artefacts, i.e.\ removal of systematic errors. For example, it can be often observed that people try to enforce an even distribution for a measure, such that it performs well and equally for all instances of the data.

As we have already seen in section~\ref{sec:importance}, the usage of citation data leads to uneven distributions over time. More recent documents haven't had much time to be cited. Also, papers with high in-degree tend to have high co-citation with other documents. Although absolute values do not influence the recommendation ranking given one paper, it has disadvantages for the comparability given multiple papers. Hence, it plays a role in our setup, because we combine rankings based on scores.
On the other hand, it is likely that papers having lots of references are not very selective\comment{picky}, expressing weaker relationships between the involved papers.
To allow for those assumptions can improve the quality of a recommender, if the right normalizations for the given data are applied and the assumptions fit to the properties of the data.

Karypis et~al.\ pointed out that such kind of normalization issues are a general problem dealing with recommendation systems. To face those, they incorporated basic column- and row-wise normalizations into the CF~framework and showed that their selection is application dependent~\cite{Karypis04}.
We follow their approach and apply the following normalizations using the L2-norm:

{\bf Column normalization} allows for the equalization of different numbers of citations per paper and refers to column-wise normalization of the CF~matrix.

{\bf Row normalization} reflects its complement and allows for the normalization of the amount of references a paper contains.

Obviously, the proposed changes in the input data have a higher impact for CF~matrices with a high variance in the number of non-null elements of column- and row-vectors.

\begin{samepage}
\begin{equation}
\lvec{C_{*,i}}' = \frac{\lvec{C_{*,i}}}{\|\lvec{C_{*,i}}\|_2}
\  ; \quad
\lvec{C_{j,*}}' = \frac{\lvec{C_{j,*}}}{\|\lvec{C_{j,*}}\|_2}
\end{equation}
\end{samepage}

Interestingly, McNee et~al.\ show that cosine-similarity used in item-to-item~CF performs significantly better than pure co-citation~\cite{McNee2002}. It turns out, instead of being a different approach, this simply reflects a normalization concern. Cosine similarity not only can be transformed into normalization of co-citation values~(\ref{eqn:norm1}), but also to the proposed column-wise normalization serving as input for the calculation of co-citation (\ref{eqn:norm2}):
\begin{samepage}
\begin{eqnarray}
\textrm{cos}(\vec{d_i}, \vec{d_j})
 & = & \frac{\lvec{C_{*,i}} \cdotp \lvec{C_{*,j}}} 
            {\|\lvec{C_{*,i}}\|_2 \  \|\lvec{C_{*,j}}\|_2}
   = \frac{\textrm{co-cit}(d_i, d_j)}
            {\|\lvec{C_{*,i}}\|_2 \  \|\lvec{C_{*,j}}\|_2} \label{eqn:norm1} \\
 & = & \frac{\lvec{C_{*,i}}}{\|\lvec{C_{*,i}}\|_2} \cdotp
       \frac{\lvec{C_{*,j}}}{\|\lvec{C_{*,j}}\|_2}
   = \lvec{C_{*,i}}^{'} \cdotp \lvec{C_{*,j}}^{'}
   = \textrm{co-cit}(d_i^{'}, d_j^{'}) \label{eqn:norm2}
\end{eqnarray}
\end{samepage}

However, the values that we receive doesn't reflect anymore co-citation values as stated in definition~\ref{def:cocitation}. As we will see, normalization as proposed here is able to significantly improve the examined prediction task.

\section{Other approaches}

For both kinds of data further methods can be used and other informations incorporated. But as more sophisticated methods imply a higher computational complexity and are less reproducible, we opted for simpler methods.

{\em Latent Semantic Indexing (LSI)} is in contrast to collaborative filtering a content-based technique to extract topics out of documents. A singular value decomposition is used to find an approximation for the original term-document matrix with a lower rank~\cite{Hofmann99plsa}. The underlying model belongs to the family of mixture models, in which documents consist of a distribution over different topics, while each topic in turn is represented by a distribution of terms.
For bibliographic purposes scientific papers normally already are tagged with keywords. Either they could be incorporated for the generation of topics, or, where missing, they could be generated via LSI and being used for further improvements of the final ranking.

Ziegler et~al.\ show that topic diversification can greatly increase the usefulness of recommendations~\cite{Ziegler05divers}.
Co-reference has been applied on the citation graph and is thus on document level, but as mentioned, in linguistics similar attempts have been undertaken on the word level. With {\em co-word} analyzes relationships between words can be built.

The citation graph is one further dimension supplementing content information, but we still ignore some of the meta-data that is available. Documents are written by multiple authors, which in turn are affiliated with faculties or research establishments. However, most available algorithms do not differentiate between  different types of vertices in such a generalized graph and are not directly applicable.

%% file: chapters/chap3_accessdata.tex

\chapter{Usage of access log data}\comment{Exploiting, Utility}
\label{sec:accessdata}

This chapter considers access data as potential source of information.
Since it has not been intended a~priori to serve for this purpose, some additional steps have to be applied before one can make use of it.
The first two sections are dedicated to those steps, then an analog of co-citation is introduced which will be used for evaluations through the rest of this thesis.

In the authoring process a prerequisite for the ability to cite another paper is to read it, at least to know about its existence. Having only one centralized, freely accessible archive of such papers, it is unlikely that papers still distribute in a peer-to-peer manner circumventing \arxiv, so that accesses to them can not be observed.
The results of Brody et~al.\ confirm that there is a correlation between downloads and later citations~\cite{Brody2005:webusage}.
On the other hand, references in papers also lead to downloads of those. The trivial explanation for following a reference is further interest in its topic.
As downloads influence citations and citations influence downloads it is likely that there would be some correlation between the two.
On the other hand, another part of downloads seem to be sensitive to a different kind of research usage of publications. Kurtz et~al.\ showed its orthogonality to citations on the basis of readership information~\cite{Kurtz05_readcite}.

In \arxiv, we not only have one type of download, instead, there is a distinction between accesses to summary pages of each paper and downloads of the paper itself. The operators endeavor to direct links and search results first to the summary, mainly to save computational resources and bandwidth. However, this way user have the chance to make their relevance and importance judgement already on the basis of the summary which helps to keep the download data clean of undiscriminated downloads. We should see in the evaluation that measures based on downloads perform better than such based on views of the summary.

\begin{figure}
\centering
  \includegraphics[width=0.8\textwidth]{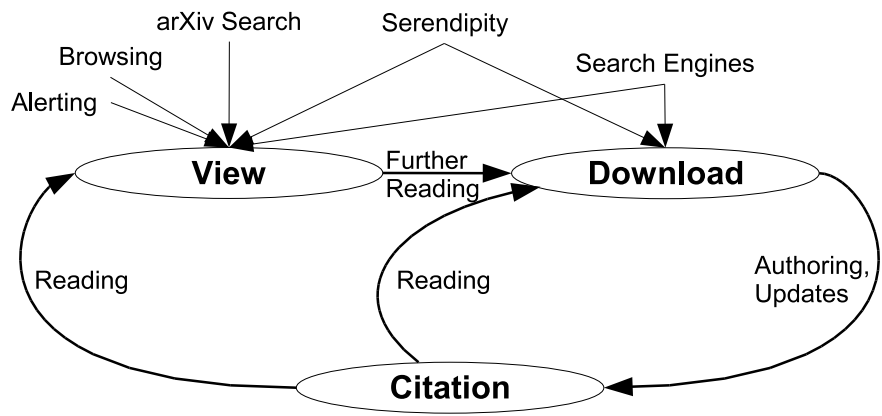}
\caption[The relation of views, downloads and citations]{The relation of views, downloads and citations. Customized version, following first proposal in \cite{Brody00_opcit}.}
\label{fig:accessdata}
\end{figure}

The downside is that we do not have a closed loop between citations and accesses. As Figure~\ref{fig:accessdata} indicates, there are a lot other influences, being responsible for accesses.
Part of them are biased and to some extent induced by \arxiv\ itself. Alerting services, but also browsing through the web pages favors accesses to papers of the same (actual) month.
Search engines, internal and external (e.g.\ Google), have the capability to direct users to papers not easily reachable by means of browsing, considering many of their (mostly textual) properties. Problematic with external engines might be that they only show a textual snippet and the title of the document, and directly link to the full-text. Not being provided with the abstract, users are probably more likely to download the full-text before making a relevancy decision. On the other hand, search engines are able to attract new users and direct them to related informations in \arxiv\ so that we can make use of the implicit information contained in such accesses.
Finally, there is still some kind of serendipity left in access data, that can't be explained by objective reasons.

Being able to make use of access data would have several advantages.
Access data is immediately available. Authoring processes take their time and classically entail peer reviews, till they finally eventually are published. It might last years in some areas of research until a first citation for a paper is able to appear.
This way, the implicit information contained in access data might be better on the prediction of actual and recent work.
Although access data is much noisier, due to its large coverage it might still be useful as a source of information for papers, for which citation data is not available.
Since it doesn't represent authored relationships, but joint implicit judgements of all users, it has the ability to discover unknown relationship (for instance to different fields of research). On the other hand, it is also biased by authored citation data,
can be inflated by automated web crawlers, short-changed by intermediate caches, abused from authors by hits to their own papers, and it can not easily discriminate between normal browsing and item-specific reading. Still, it seems to have some signal value in it too, partly correlated with and partly independent of citation data.

One can imagine a typical scenario of a researcher who searches for literature in his actual field of research.
So, to extract useful information out of access data, a straight-forward approach is to assume that users rather access papers over a period of time, related to their information need, than unrelated ones. 
For this purpose, we need to conduct preprocessing steps which preserve the type of information we are interested in and eliminate as much noise as possible.

\section{Preprocessing}

The source for access data are operational systems, mostly web-server, which underlie steadily modifications. Data formats change over time, already with usage of different implementations and versions. Also, it is likely that there have been changes in what is stored. Making use of distribution like installing mirrors in different countries worsens the situation. Apart from daylight saving, then also different time zones come into play.
In short, one has to deal with everything, known to be the most time-consuming step in every data warehouse project.

All this leads to the necessity of a preceding preprocessing step. But it not only has to deal with the unification of different formats and representations, also some kinds of filtering should be applied to minimize the amount of data having to be processed at later stages. Performance reasons prohibit sophisticated filtering methods here, so that filtering decisions should be made on a local per request level.

The most important unwanted artefacts in the data originate from search engines crawling the website, or automatic scripts. Most of those can be recognized by their user agents, but some of them are ambiguous or they hide themselves.
Besides that, it is also important that an implementation doesn't have susceptibilities against malformed log entries, which occasionally happen due to attacks intending to cause buffer overflows.

In our case, \arxiv\ experienced several changes. With its formation, the main site was hosted at the {\em Los Alamos National Laboratory~(LANL)}. With the transition of Paul Ginsparg to Cornell also the main site moved and LANL became one of the dozen mirrors.
The total amount of data that we have been provided with consists of 741 million accesses in total~(156~GB, or 13~GB compressed). This massive dataset consists of logs of LANL and Cornell. Three different data formats had to be unified, two timezones and their clock changes were translated into a universal timestamp.
To profit from the experience of log analyzing software, we made use of the lists of known crawlers and mirroring tools in {\em AWStats}.\footnote{\url{http://www.awstats.net}} We had to manually extend the lists with \arxiv-specific automated scripts and two dozen crawler of mirrors. The biggest impact of the search engines has had Google (see Figure~\ref{fig:googlebot}).
The filtering of access data involved significant amount of work for analyzes to identify unwanted artifacts. As a result, a much less peaked distribution could be obtained (Figure~\ref{fig:accesses}).

Because log entries are written after the completion of the request, but timestamped with initiation of the request, log files are generally unordered. This effect is not only significant for broken connections, but also for long lasting downloads of slowly connected clients.
Because many people probably still have bookmarks to the original LANL main site and they basically belong to the same basic population, we have chosen to merge the two datasets.
At least the latter fact justifies the need to sort the access logs.\footnote{The usage of standard tools like UNIX~{\bf sort} was not appropriate, since it uses temporary files even for the simpler task of just merging big pre-sorted files into one. Also the usage of compressed input files is not supported directly without the usage of {\em named pipes}. Those problems lead to an own implementation of merge sort, working completely in memory, using only compressed external storage and taking advantage of multi-core, resp.\ multiple processors for compression tasks.}

Although it has its advantages also to filter out unwanted requests early, we have chosen to defer further processing to later steps which are able to make decisions considering the context of a request.


\section{Session extraction}\label{sec:sessionextraction}

\begin{figure}
{\centering
  \includegraphics{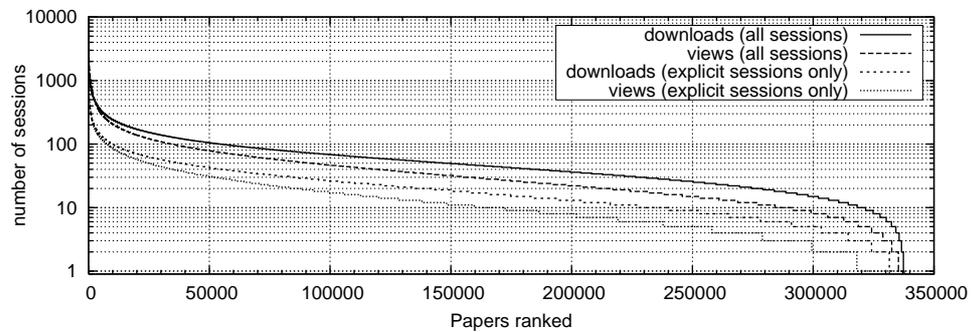}
  \caption[Number of sessions papers occur in]{Number of sessions papers occur in.}
  \label{fig:paper:sessfreq}
}
\end{figure}

\begin{figure}
{\centering
  \includegraphics{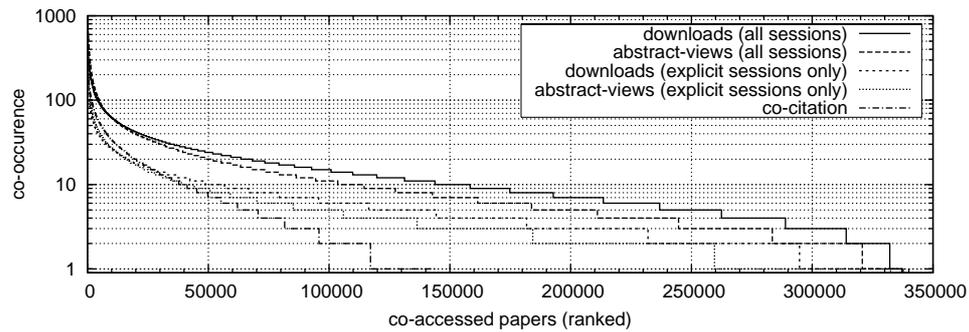}
  \caption[Frequency of co-accessed paper with highest rank over all papers]{Frequency of co-accessed paper with highest rank over all papers.
For over 260,000 papers there exist one paper that was at least 5 times co-downloaded, if also generated sessions are incorporated. This serves as a good base for using download data for recommendations.
It also shows that there are downloads, that can't be an effect of a previous look at the abstract. Those are direct downloads.
Furthermore, the coverage of papers by sessions can be seen as an upper bound for the highest possible co-access frequency (see Figure \ref{fig:paper:sessfreq}).
}
  \label{fig:codl:freq1}
}
\end{figure}

For e-commerce transactions, it is essential to be able to track users, such that different requests can be associated to the same person. Because the original HTTP protocol~(1.0) was not intended to be aware of connections, it has been extended later to support {\em sessions}.\footnote{The explicit support of so-called {\em session cookies} freed web applications to use workarounds like {\em URL~rewriting} or {\em hidden form fields}.}
Explicit session support allows nowadays a very accurate tracking of users. But to make also use of older access data, we still have to apply some heuristics to derive sessions out of a-priori independent requests. Figures~\ref{fig:paper:sessfreq}--\ref{fig:codl:freq1} will show the differences in coverage that we can achieve with the incorporation of non-explicit sessions.
In the early years of the internet, users still could be tracked uniquely by their IP address. Nowadays, due to the limited amount of IP addresses and performance optimizations, we have to cope with proxies, {\em Network Address Translation~(NAT)} and dynamic IP addresses. However, for a short period of time the supposition still might hold that an IP address is assigned to one user unambiguously. Another information available to distinguish between users, even though they share the same IP, is the used user-agent and its version. Orthogonal to this, two kinds of approaches have been proposed to recreate sessions out of access data.

{\bf Time-oriented} approaches consider the time aspect of requests. The assumption is that requests of the same user are typically clustered over time. Although web server are able to track the activity of users, they still suffer from the problem not always to know when a session is finished logically. Because of that and not to run out of memory maintaining too much state information they use the heuristic to declare a session as finished when a time-out of usually 30~minutes of inactivity occurs. 
Adjusting a time-out allows to retrieve relatedness in a conceptual narrower or broader way. It also embraces the likelihood that users underlie a concept drift, i.e.\ they search for papers of the same topic during a short period, while longer periods consist of searches for different topics.

{\bf Navigation-oriented} approaches use the background knowledge about the structure of a web page to identify `possible' and `impossible' transitions from one website to another.
This assumes that users are only able to navigate through a website following the provided links, not choosing new start points for their navigation.
Such an approach is questionable if one also want to take advantage of the many users coming in over search engines, going directly to a sub-webpage.
In our domain, there exist global scientific indexes like {\em Google Scholar} or {\em CiteSeer}, many users search by. And we are especially interested in such user, wanting to satisfy an information need by searching for related documents and therefore only following links to related documents in \arxiv.

We opted for the first approach because of its simplicity and the fact that only one parameter has to be adjusted. The other methods would also have resulted in reengineering of the whole website historically for every point in time.

There are two possibilities to realize session extraction based on a heuristically determined maximum time between requests belonging to the same session. The first is to sort the data by users, resp.\ sessions, grouping the requests of a session together. Depending on the type of sort algorithm used, this normally assumes that the decision if any two requests belong to the same session can be made independently of other requests, only based on the information in those two requests.
Not to restrict us by the with this approach associated loss of context information, we have chosen to simulate the original sequence of requests over time. This has the advantage that no sorting on external data and only one linear run over the access data is required to extract all sessions. There only has to be enough main memory to maintain the state information needed to track all currently active sessions. ``Expiring'' sessions still underlie several processing steps, before they are written out to disk in a application specific manner. To recognize expiring sessions efficiently, an enhanced least-recently used~(LRU) queue is used. As a side effect, at all times the number of currently active sessions is known, which gives interesting, retrospective insight into the access statistics of \arxivorg~(see Figures~\ref{fig:sess:concur30}--\ref{fig:sess:concur3}).

The generation of sessions serves as the basis for a created framework, in which multiple independent steps form a processing chain. Each step can be reused and configured independently by external configuration.
There are steps, allowed to manipulate sessions and such, which decide if a session should be further processed or rejected. We implemented filtering on the basis of time stamps or the extent of sessions (duration, number of accesses), as well as filter based on HTTP attributes, categorizing the domain specific type of request (search queries, author lookups, downloads, \dots), searching for valid \arxiv\ identifiers or unification of identical (consecutive) accesses. They are organized in the filter chain in order of increasing computational complexity, still respecting local orderings imposed by preconditions of those steps.

Brody et~al. have shown that the publishing of a paper leads to a rush of accesses to a paper~\cite{Brody00_opcit}.
Main reasons for this are the various alerting services announcing new publications. Furthermore, access data also suffers from a kind of presentation bias, induced by the navigation on the website. Apart from being alerted, users are likely to navigate through different lists of new (containing the same day), recent (covering the recent week) or current (same month) papers.
Since we are not interested in such artificially self-induced accesses, we investigated in different ways to limit their effect, but still to profit from early accesses. Ignoring all accesses to a paper in the first 31~days after its publication is a very strong regularization. Instead we chose to apply the algorithm, described in Function~{\em\ref{alg:filtering}}, on page~\pageref{alg:filtering}. Here, we still allow for pairs of accesses of a just published paper and such not being on the same weekly, or monthly (alert) list.

\begin{figure}
\newcommand{\incfig}[1]{\includegraphics[width=0.8\textwidth]{#1}}
{\centering
\subfigure{\incfig{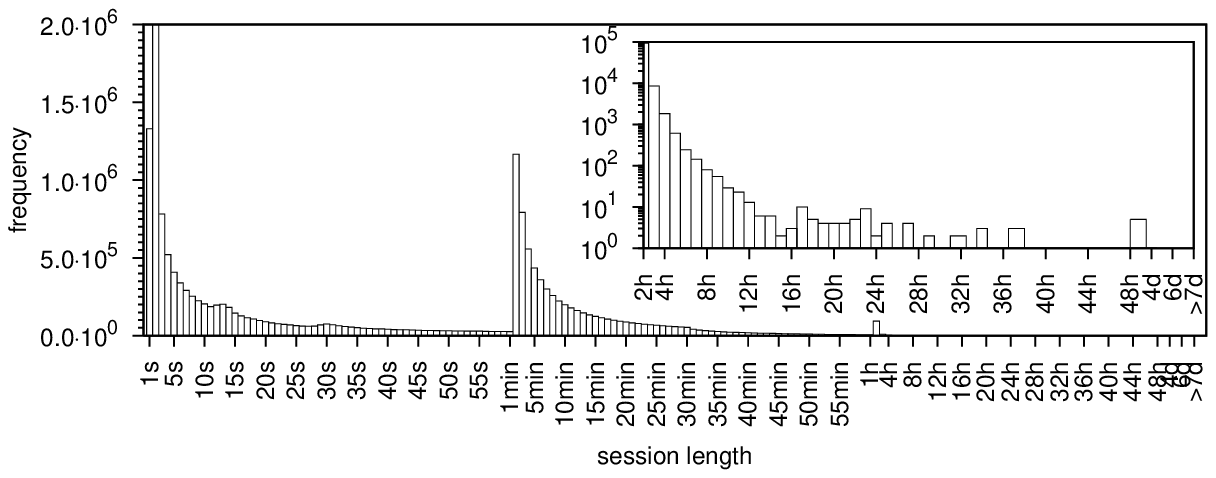}}
\\
\subfigure{\incfig{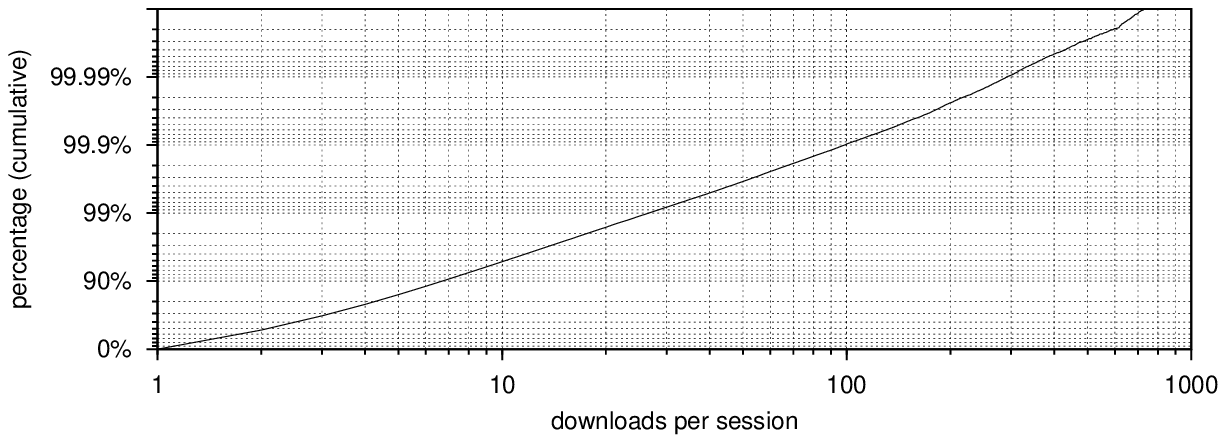}}
  \caption[Session lengths]{Session lengths. Generated with a time-out of 30 minutes.
}
  \label{fig:sess:length}
}
\end{figure}
\begin{figure}
{\centering
  \includegraphics[width=8cm]{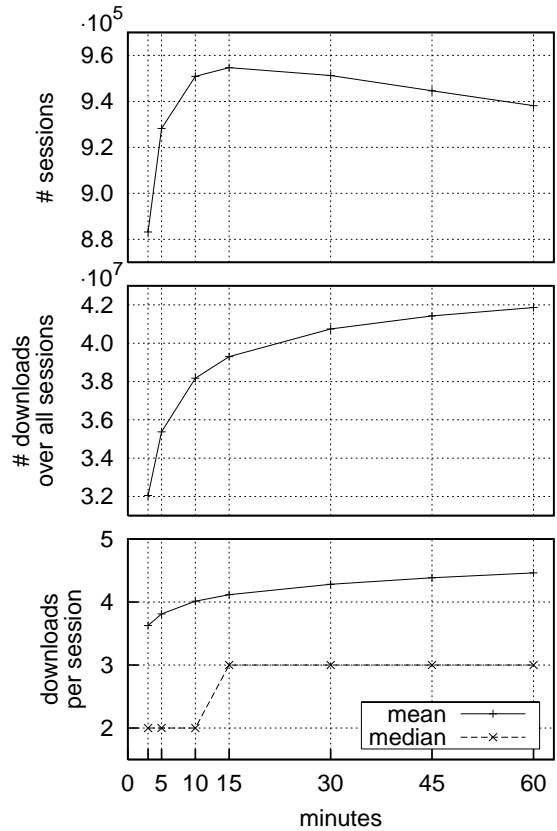}
  \caption[Number of sessions over the time-out used]{Number of sessions over the time-out used.
With a larger time, allowed between consecutive requests, the number of sessions slightly decreases because more sessions become merged. With a shorter time-out, a lot sessions become too small for allowing measurement of dependencies between accesses in a session~($< 2$) and are thus filtered.
The effect of filtering can be seen purely in the number of downloads made over all sessions.
}
  \label{fig:paper:sessgap}
}
\end{figure}

Figure~\ref{fig:sess:length} gives some insight into the properties of the sessions which we receive after the preprocessing steps. The lengths of sessions in terms of time, but also in the number of downloads of \arxiv\ papers follow a power law distribution. Except for some outliers, the length is bounded to 16~hours and a few hundred downloads. With consideration of proxies, the sessions generated might still reflect human behavior, which suggests the effectivity of the filtering applied. Further control can be exercised with adjustment of the time-out used to generate the sessions.
Figure~\ref{fig:paper:sessgap} indicates that shorter time-outs are able to break long sessions apart which might be helpful for the treatment of proxies.

\section{Co-access measures}

Analogously to co-citation, in which a paper expresses explicitly the relatedness to other papers by the choice of its references, we can make use of sessions to derive relatedness between papers accessed in a session. This assumes that users are selective and their behavior reflects some kind of discrimination between papers accessed to those not accessed, hence relatedness between elements of those two groups.
The ideal scenario for this is a scientist who searches for literature in a specific topic, accessing more likely related documents than unrelated ones.
Following the basic collaborative filtering approach, we regard every session $S_i$ as a set of votes for documents, each originating from a different `user'.
We receive a user-item matrix $A_{m \times n}$, in which $A_{i,j}=1$, if session $S_i$ accessed document $d_j$, and $0$ otherwise.
\begin{definition}
Co-access between two papers measures the co-occurrence of accesses to those papers over sessions, i.e.\ the number of sessions both papers are accessed in. Thus, relatedness between two papers is defined as:
\begin{samepage}
\begin{eqnarray}
\textrm{rel}(d_i, d_j)\ {\buildrel\rm def\over=}\ \textrm{co-acc}(d_i, d_j)
 & = & | \{ S_k : d_i, d_j \in S_k \} |  \nonumber \\
 & = & \lvec{A_{*,i}} \cdotp \lvec{A_{*,j}} = \sum_k{A_{k,i}A_{k,j}} \nonumber \\
 & = & 
			\left\{
			\begin{array}{ll}
			(A^TA)_{i,j} & \textrm{if $i \ne j$} \\
			0            & \textrm{otherwise.} 
			\end{array}
			\right.
\end{eqnarray}
\end{samepage}
\label{def:coaccess}
\end{definition}
Normally, one can distinguish between different types of accesses that are performed on a website.
Next to the navigation through HTML pages, user also download files or even purchase items. Increasing cost for different possible user actions correlates with the interest a user have: The purchase of an item in a session of an e-commerce site is probably the strongest statement of interest. On the other hand, with increasing bandwidths and cheaper internet access, the difference in cost between accesses to a HTML page and downloads of a whole document diminishes.
For \arxiv, as an open content provider, users most often access overview pages (presenting meta-data of a paper like its abstract) and downloads of the full-text.
We investigate the question to what extent the usage of download information is preferable over plain accesses. Therefore, we distinguish between co-download and co-view as two instantiations of co-access, which are built on different types of accesses.


\section{Implementation issues}

For large amounts of data like access logs, it becomes easily prohibitive to retain all base data in memory to work on it efficiently. Even worse, it is already impossible to maintain intermediate results in memory. Using better coding schemes leads only to a fraction of the required memory with the cost of more computation.
Therefore, a feasible implementation should work on an external representation of the base data and use the available memory to store and update a subset of frequently changing intermediate results.

Summarizing, an implementation for the calculation of co-downloads has to fulfill the following two key requirements:
\begin{samepage}
\begin{enumerate}
\itemsep -.08in
\item ability to work efficiently on external data, thus being scalable to the growing amount of available access data
\item make use of the sparsity of papers, being accessed together
\end{enumerate}
\end{samepage}
\comment{
Co-access between two papers simply measures the co-occurrence of accesses to those papers over the sessions, i.e. the number of sessions both papers are accessed in. 

The first requirement can possibly be relaxed for specific datasets, because there should be a limited time frame for the used access data, so that it is still useful for deriving 
concept drift over time (e.g. it might be enough to count co-accesses for a paper in the first 6 months after publishing
New papers that are very related to older ones should be allowed to catch up in their co-access measures with other related papers. Hence, there's no limit to the size of access data that has to be taken into account.
Nevertheless, there should be done some pre-processing to filter out artifacts that introduce noise. Robots access papers consecutive, that likely have nothing to do with each other. But there are also a lot of similar artificial accesses, e.g.\ induced by monthly mailing lists, announcing the new papers. Scientist might follow the links to papers presented in such lists, that are not mandatory related. Brody et~al. propose for \arxiv\ to ignore log data in the first 7~days after publishing~\cite{Brody2005:webusage}.
Cleaning of noise increases sparsity and therewith the efficiency (less pairs of papers have to be counted).

Counting
The incrementation of variables can be performed a magnitude faster than summation or multiplication on most architectures.
This frees time of the cpu, that can be used in the trade-off between memory footprint and computational complexity, most algorithms underlie.
}

Our implementation assumes session data as input. Every session consists of a number of accesses to papers, i.e.\ either downloads or accesses to the document summary. In a first run over the data, the session data is transformed into a binary representation, encoding accesses to papers as plain numbers (for faster reading and parsing of external data). Then, multiple linear runs are performed over the session data in such a way that for each run all co-occurrences with other papers of as most as possible papers are counted. Since this number is unknown beforehand, running out of memory leads to an exponential decrease in the number of papers that are tried to count in another run. An improved version might determine the number of co-occurring papers for every paper as a side effect of a previous run and estimate the number of papers that can be counted in the next run in advance.
For every paper a hash table is maintained to memorize co-accesses with other papers.
Finally, by iteration over every hash table the $N$ most often co-occurring papers are collected efficiently using a heap data structure.

\comment{
\begin{figure}

{\bf for} each vertex {\bf do} \\

\caption{Calculating top-N co-accessed papers}
\end{figure}

Further remarks:

}

%% file: chapters/chap5_evaluation.tex

\chapter{Evaluation}
\label{sec:evaluation}

In this chapter, first the main setting and the application of the measures used are described.
On that basis, first results are discussed. Then main parameter are examined in their influence and further experiments are conducted to examine the quality of recommendations over time after publication.
Finally, we introduce a research prototype of a recommendation system for \arxivorg, which makes the outcomes tangible for human users and can serve as a basis for further research.

\section{Metrics \& experimental design}

Recommendation systems are intended to filter relevant items out of the set of all possible items. Ideally, the set of proposed items~$P$ is congruent with the set of relevant items~$R$. In information retrieval, out of the perspectives of these sets, retrieval performance has been defined in terms of precision and recall:
\begin{equation}
precision=\frac{|R| \cap |P|}{|P|} ,\qquad  recall=\frac{|R| \cap |P|}{|R|}
\end{equation}
Here, it is assumed that a fixed set of documents is returned, without any ordering imposed on them. Because nowadays the number of returned results normally exceeds what humans are able to consider, a ranking of the items by their expected relevance is used. Rankings can be measured by computing precision and recall at different cut-off points $i$ of the ranking. This can be depicted in a precision-recall-graph, in which the area under the curve is equivalent to what is called {\em average precision} $AP$
\begin{equation}
 AP=\sum_{i=1..n}{(precision_i \cdotp rel_i)} / |R| ,
\end{equation}
where $rel_i$ is 1, if the $i$th item in the ranking is relevant, and 0 otherwise. {\em Mean average precision (MAP)} averages over a set of queries issued to a system to measure its overall retrieval effectiveness and makes relative comparisons possible.
However, for simplified interpretation we use the recall achieved over the top ranks in our main experiment.

For the evaluation of the different measures we choose the following setting, which has been formalized in Setting~\ref{alg:setting1}. 
Since we want a representative evaluation for the performance of a system deployed and used today, we chose to evaluate on recent papers. Not to suffer from possible seasonal bias (see Figure~\ref{fig:accesses}), we consider all papers $d_k$ that have been submitted in 2005 as our test set (line~2), ignoring 2006. The training set, out of which the system is built, consists of the whole corpus of papers published beforehand and for which citation data is available. This means, we assume the actual point in time to be January 1, 2005.
\newcommand{\tgtbg}{t_{gt}^{begin}}
\newcommand{\tgten}{t_{gt}^{end}}
\newcommand{\tevbg}{t_{eval}^{begin}}
\newcommand{\teven}{t_{eval}^{end}}
\newcommand{\alg}{{\textrm{al}}}
\newcommand{\GETS}{\leftarrow}
\begin{algorithm}
\renewcommand{\algorithmcfname}{Setting}
\scriptsize
\SetLine \linesnumbered \dontprintsemicolon
\KwIn{points in time: 
$ \tevbg < \teven $,
citation graph vertices~$V$,
evaluation~ranks~$N \in \{1,3,5,10,\dots , 100\}$
}
\KwOut{Recall for ($\alg, N$)-tupel}
\ForEach{recommendation algorithm \alg}{
  \ForEach{document $d_k: \tevbg \le t(d_k) \le \teven$ }{
    $R_{d_k} \GETS \{ d_i: (k,i) \in E  \land t(d_i) < \tevbg \}$\;
    \If{$|R_{d_k}| \ge 2$}{
      \ForEach{reference $d_j \in R_{d_k}$}{
        $I \GETS R_{d_k} \setminus \{d_j\}$\;
        $O \GETS \textrm{AggregateRankings}_{d_i \in I}( \{ O(d_i) \GETS al(d_i) \} )$\;
        $O \GETS O \setminus I$\;
        $O \GETS O \setminus \{d_l \notin V \}$\;
        \ForAll{ranks $N$}{
        $(\alg, d_k, d_j, N) \GETS \textrm{recall}(O, d_j, N)$\tcp{ is $d_j$ in top-$N$ results}\;
        }
      }
      \lForAll{$d_j$}{$(\alg, d_k, N) \GETS \textrm{mean}_{d_j}(\alg, d_k, d_j, N)$}
    }
  }
  \lForAll{$d_k$}{$(\alg, N) \GETS \textrm{mean}_{d_k}(\alg, d_k, N)$}\;
}
\caption{Recall for held-out references}
\label{alg:setting1}
\end{algorithm}

In this setting, we use the kind of relatedness which results from the co-citation relationship between references of a paper. As Figure~\ref{fig:paper:lastreftime} indicates, 80\% of papers have at least one reference in the recent 12~months. Because we can only recommend papers before 2005 we have chosen to ignore all references to papers submitted later (line~3), amounting typically only to 10--25\% of all of them. This also allows us to estimate the performance of a recommendation system, only occasionally updated.

We evaluate by holding out one reference of a paper in 2005, make a recommendation out of the remaining ones and search for the held out reference in the top~$N$ entries.
We assure that every paper has at least two references (line~4), or skip the considered paper otherwise. Instead of randomly choosing one reference and trying to predict it, given the rest of references, we opted to do that with every reference (line~5). This makes more usage of the available data and leads to \mbox{fractioned}, more precise recall values for every examined paper $d_k$ and given rank~$N$.
The setting proposed so far is an instance of a realistic CF problem, in which the input consists of a set of references of a paper, for which we want to find further related documents. Since our relatedness functions are only capable to build a ranking for one input document, we apply them on every input document and aggregate the resulting rankings (line~7). For the purpose to generate a new ranking, we generally used the sum of the scores in the separate rankings. This turns out to be the best choice out of usually used aggregation functions (see Section~\ref{sec:eval:agg}).
Although {\em higher-order} models have been proposed~\cite{Karypis04}, this still reflects the prevalent approach in CF. Because we apply recommendations for each item of the set independently, we might also recommend items of the given set. Also, due to the use of relations derived from citation data as our golden truth, we penalize measures with higher coverage. To minimize this, we filter the recommendations for documents, given as input to the recommender (line~8), and such, not contained in the citation graph (line~9).

For every examined recommendation algorithm $al$, document $d_k$, each of its references $d_j$ and every rank~$N$, we obtain a binary recall, which states if the document $d_j$ could be found in the first $N$ recommendation entries (line~11). We take the mean over all references $d_j$ of each document $d_k$ and finally over all documents $d_k$ themselves.

Although in the CF framework it is fairly clear on which properties of the given input data algorithms should be applied, due to that we have access to many properties of a document itself and to those for every of its references as well, we obtain manifold possibilities. Co-reference, using past data for the estimation of relatedness, as well as textual similarity can also be applied on the references' hosting document, referred to as $d_k$. In the results, we will distinguish between those cases.

Another approach would have been to follow a leave-one-out strategy, testing on recent papers given all previous papers that time. This would imply recalculation of the measures for a lot of points in time to pretend that it was used and built the time the paper was published. Otherwise, the system could recommend very related, but future papers in top ranks, which in turn leads to lower ranks for papers, which we would like to see in the evaluation. However, although recalculations for every paper are not feasible, it still can be done a bound number of times (e.g.\ on a monthly basis). We follow this strategy to determine the timeliness of recommendations based on access data in a further experiment. Another approach is to factor the influence of later papers out of the measures, calculated once for all papers. Such an approach is a difficult task in itself, susceptible to too optimistic estimation due to overseen effects.
After all, our setting gives us a clean separation between training and test set and is easier to control.
\comment{
\begin{table}[htb]
\centering
  \begin{tabular}[h]{l|llll}
      & top 1 & top 5 & \dots & top~k \\
\hline
$d_1$ &  0.3  & 0.4   & \dots & 0 \\
$d_2$ &  0    & 0.1   & \dots & 0.7 \\
\dots & & & & \\
$d_m$ &  0    & 0.1   & \dots & 0.7 \\
\hline
mean  & 0.2   & 0.6   & \dots & 0.8 \\
  \end{tabular}
  \caption{Evaluation matrix}
  \label{tab:eval}
\end{table}
}

\section{Results} 

\begin{figure}[h]
\newcommand{\fnwd}{0.4\textwidth}
\newcommand{\fnwdb}{0.8\textwidth}
{\centering
\setlength{\tabcolsep}{5pt} 
\setlength{\subfigcapskip}{5pt}
\begin{tabular}{cc}
\subfigure[Text-based measures]{\includegraphics[width=\fnwd]{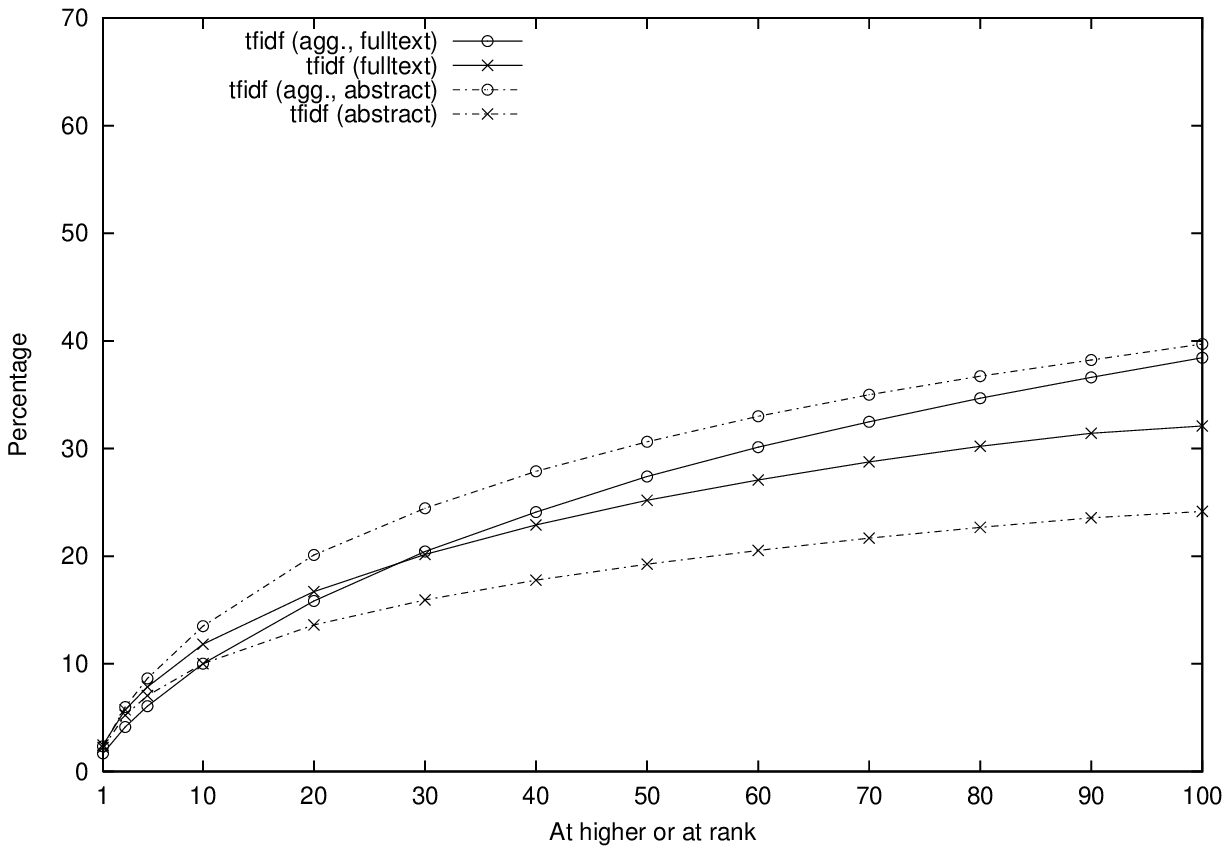}\label{fig:eval:text}} &
\subfigure[Citation-based measures]{\includegraphics[width=\fnwd]{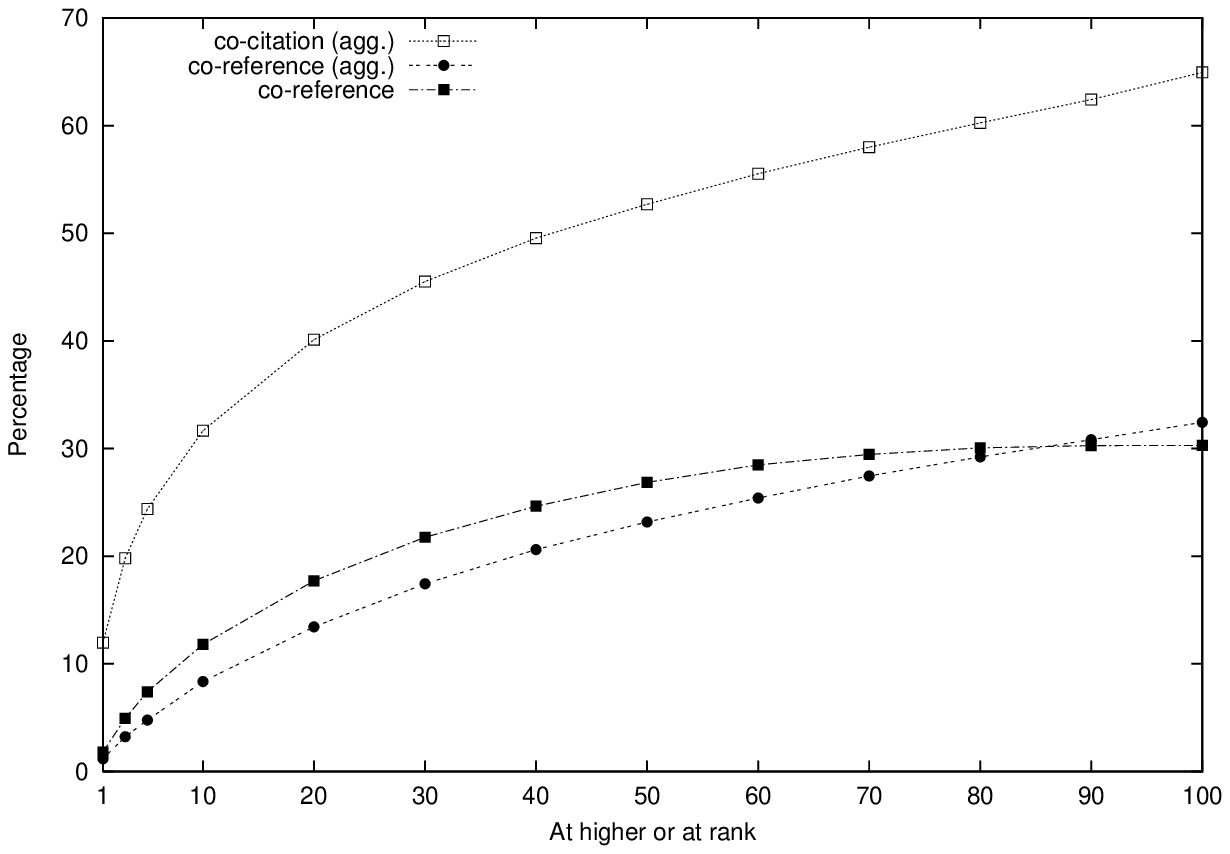}\label{fig:eval:cit}} \\
\multicolumn{2}{c}{
\subfigure[Access-based measures and best others]{\includegraphics[width=\fnwdb]{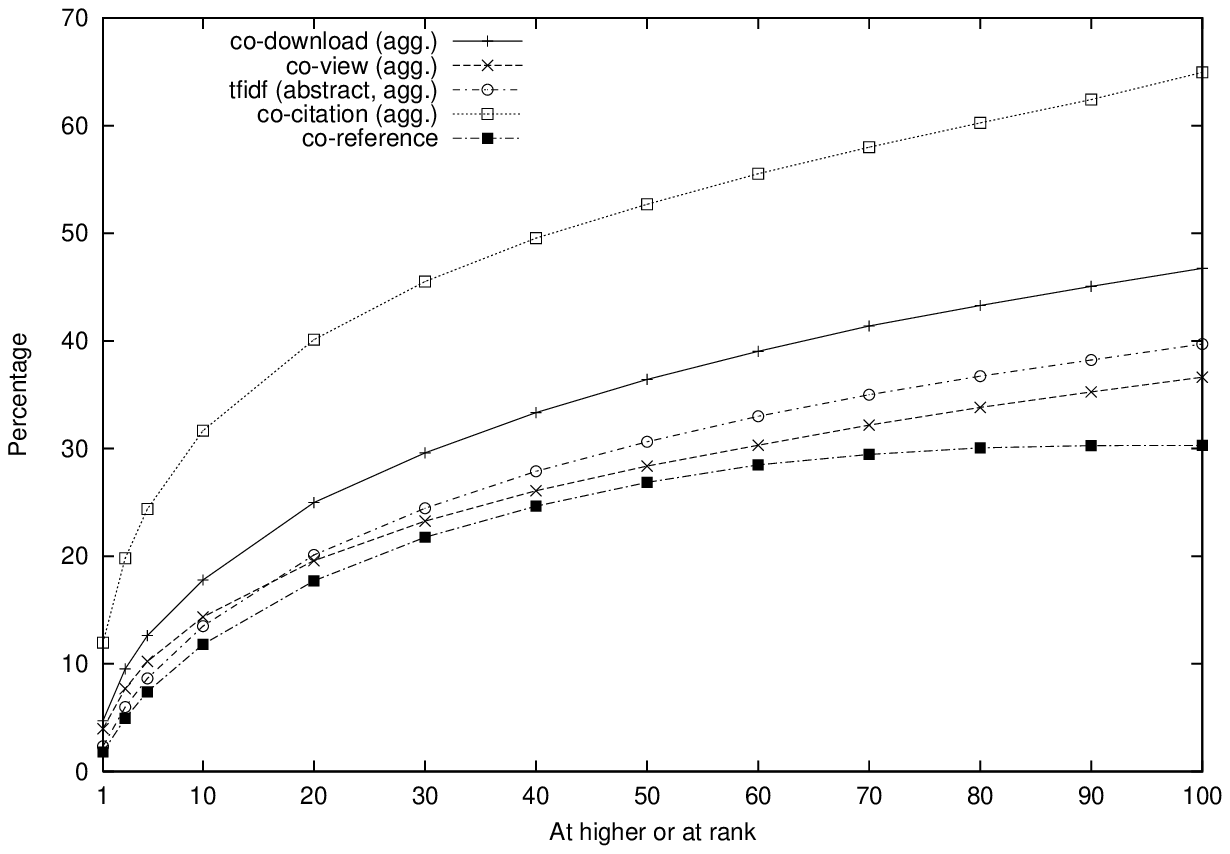}\label{fig:eval:dl}}
}
\end{tabular}
  \caption[Evaluation of text-, citation- and access-based measures]{Evaluation of text-, citation- and access-based measures. Percentage of held-out references proposed in the top 1--100 recommendations.}
  \label{fig:eval}
}
\end{figure}

For clarity, we first investigate in different ways to use text- and citation-based measures on their own. Then, because we are particularly interested in the comparison of access-based measures to the others, we use the best candidates in a final comparison to access-based measures (Figure~\ref{fig:eval}).

We considered to apply textual measures on the paper in question in itself and all its references, and either on their abstract or whole full-text. Using references, which refers to the classic CF approach, yields in turn to an aggregation of rankings.
We see that for the search of similar documents using only textual features of the evaluated document, its full-text should be used. For this, Figure~\ref{fig:measures:freq100} gives the explanation that abstracts are capable to point out few very similar papers, but are less able to retrieve further related documents in latter ranks. This ``cherry-picking'' behavior is probably due to that abstracts describe indeed very significantly the topics of a paper, but their limited amount of words doesn't allow to capture as many further relationships as the full-text in lower ranks. However, for both the curve soon flattens out.
Interestingly, using the references of a paper reverses the previous statement. Here, abstracts seem to be better to capture the different topics of the hosting document, recommending more related documents than the full-text is able to. On the other hand we can achieve the highest recall with full-texts, if we are considering more than the first 100~hits.

The evaluation of citation-based measures gives a clear indication for the strength of co-citation as a measure of relatedness. The selection of references used by an author seems to be strongly correlated with the choices of previous authors.
Co-reference, on the other hand, shows qualitatively a similar behavior to textual similarity. Its coverage is limited, so that related papers can only be found in the first 50 recommendations. Using more input data in form of the references of a paper doesn't lead to an improvement. But, finding similar papers for every reference still can find a lot of held-out papers. This indicates that authors are likely to cite multiple, topically similar papers together, instead of choosing just one of them. This is a common behavior, especially in introductory or related work sections.

For the considered setting, access-based measures can not compete with co-citation due to all the additional influences, making it more noisy.
However, co-download is better than all other measures including textual similarity, which is a benchmark for access data in terms of coverage.  Even co-view is superior in the top 10 results to our best textual measure.
As assumed, we also can observe a difference between the view of a web page and a probably still more costly download. Nowadays, because of the availability of broad-band networks, the cost of a download for a user tends to be nearly as negligible as a click to a web page. We assume the directing to the overview page instead directly to a download allows users to decide about the relevance of a paper already then.

For completeness, in Figure~\ref{fig:eval_dist}, we present the frequency of documents $d_k$ as a distribution over the recall given by $(\alg, d_k, N)$-tupel, calculated in line~14, Setting~\ref{alg:setting1}.
Here, with increasing $N$, the bar for none found references ($n=0$) decreases, such that for those documents more references are found. These usually form a normal distribution, of which the mode moves further to 1.
An exception is co-citation, in that case, it seems that either all or none of the references are found. This expresses the significance associated with explicit link data, if available.
Because for a large amount of papers only two or three references are left, the fractions of possibly found references become limited to multiples of $^{1}\!/_{2}$ or $^{1}\!/_{3}$. The distributions feature accordant peaks.

\section{Effect of normalization}

\begin{figure}
  \newcommand{\fnwd}{0.31\textwidth}
  \newcommand{\fnwdx}{\hspace{0.5cm}}
  \centering
\subfigure[Co-citation (agg.)]{\includegraphics[width=\fnwd]{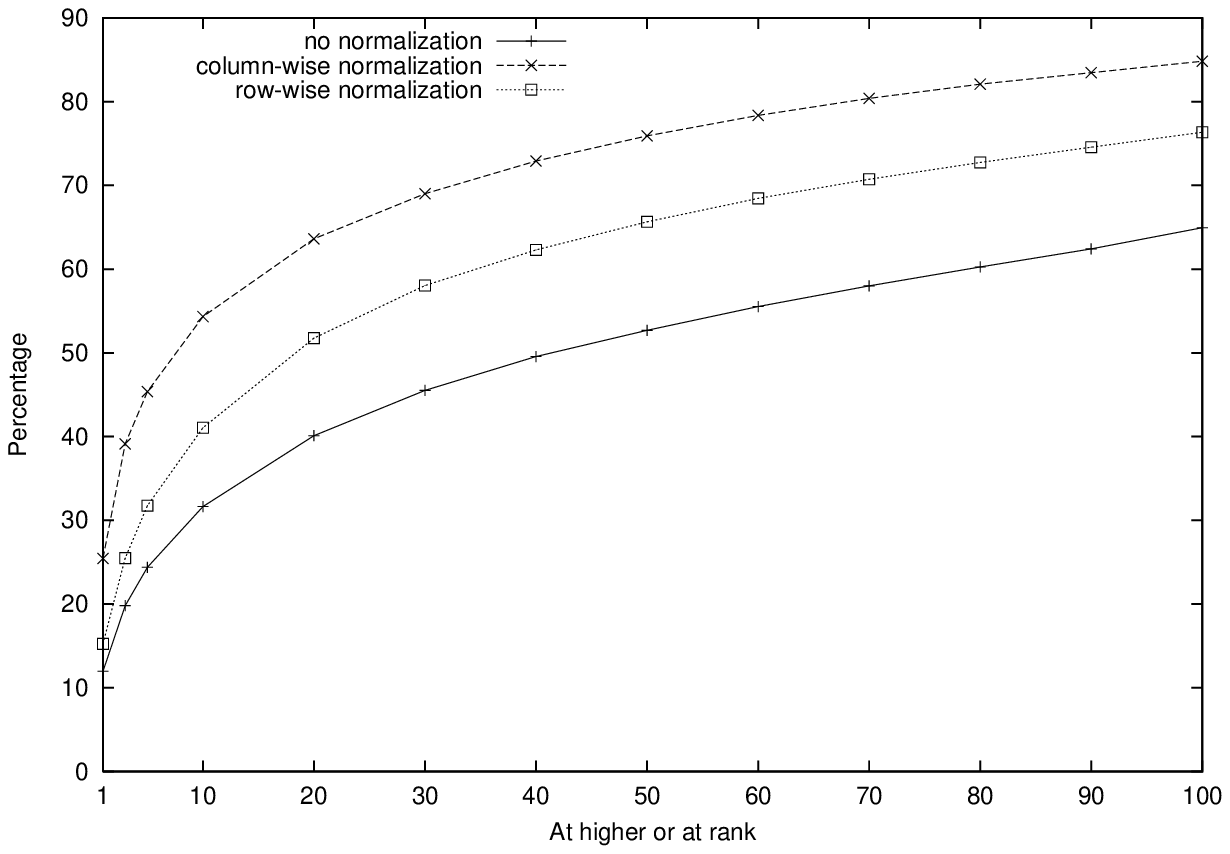}\label{fig:eval:norm:cocit}}\quad \subfigure[Co-reference (agg.)]{\includegraphics[width=\fnwd]{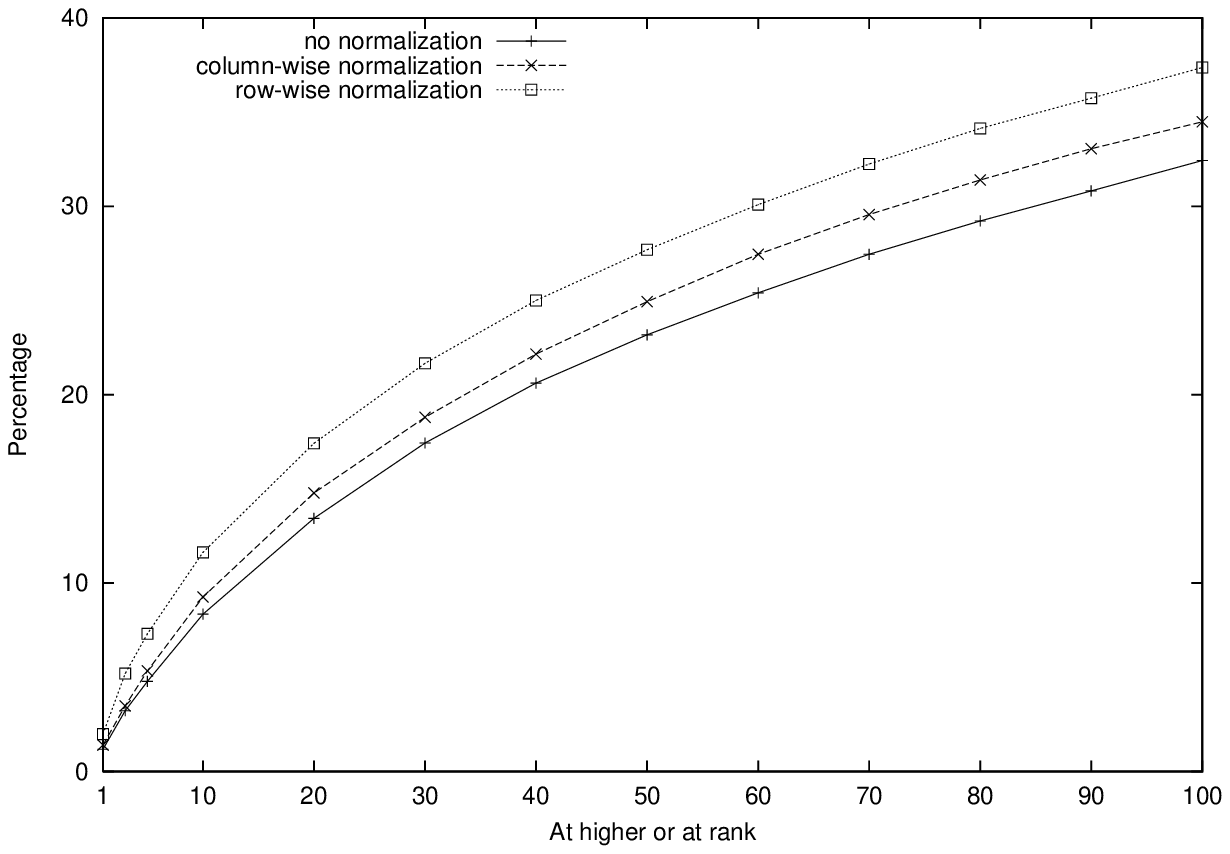}\label{fig:eval:norm:coref}}\quad
\subfigure[Co-reference]{\includegraphics[width=\fnwd]{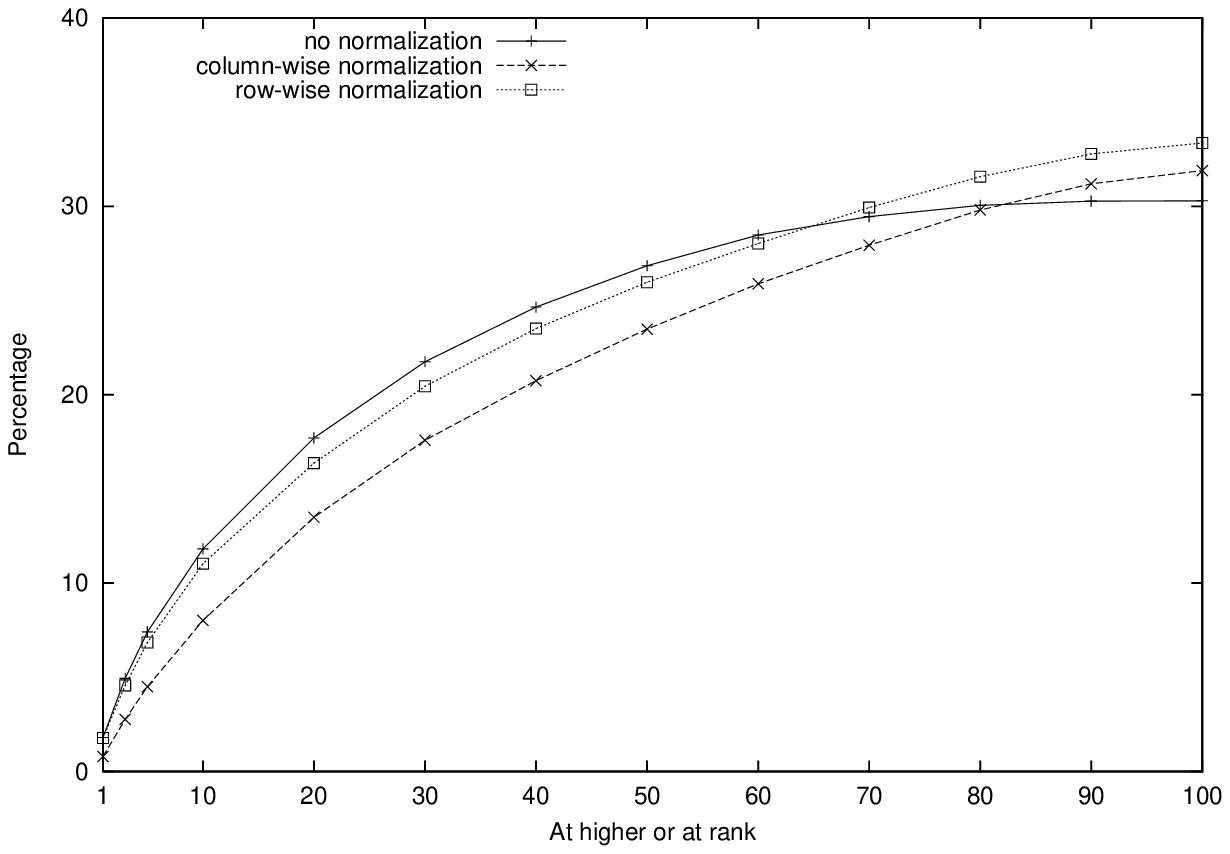}\label{fig:eval:norm:coref1}}
\\
\subfigure[Co-download (w/o~`rush')]{\fnwdx\includegraphics[width=\fnwd]{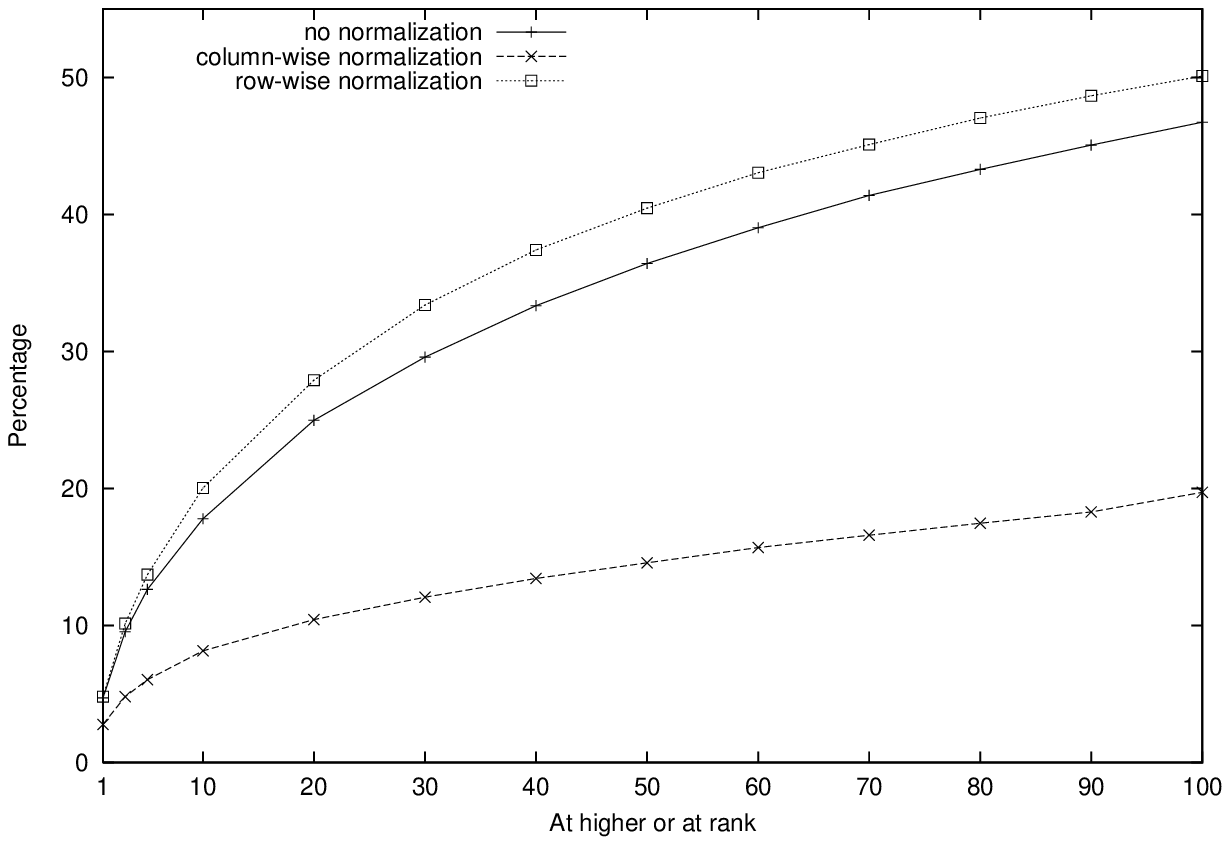}\fnwdx\label{fig:eval:norm:codlwo}}
\subfigure[Co-download (w/~`rush')]{\fnwdx\includegraphics[width=\fnwd]{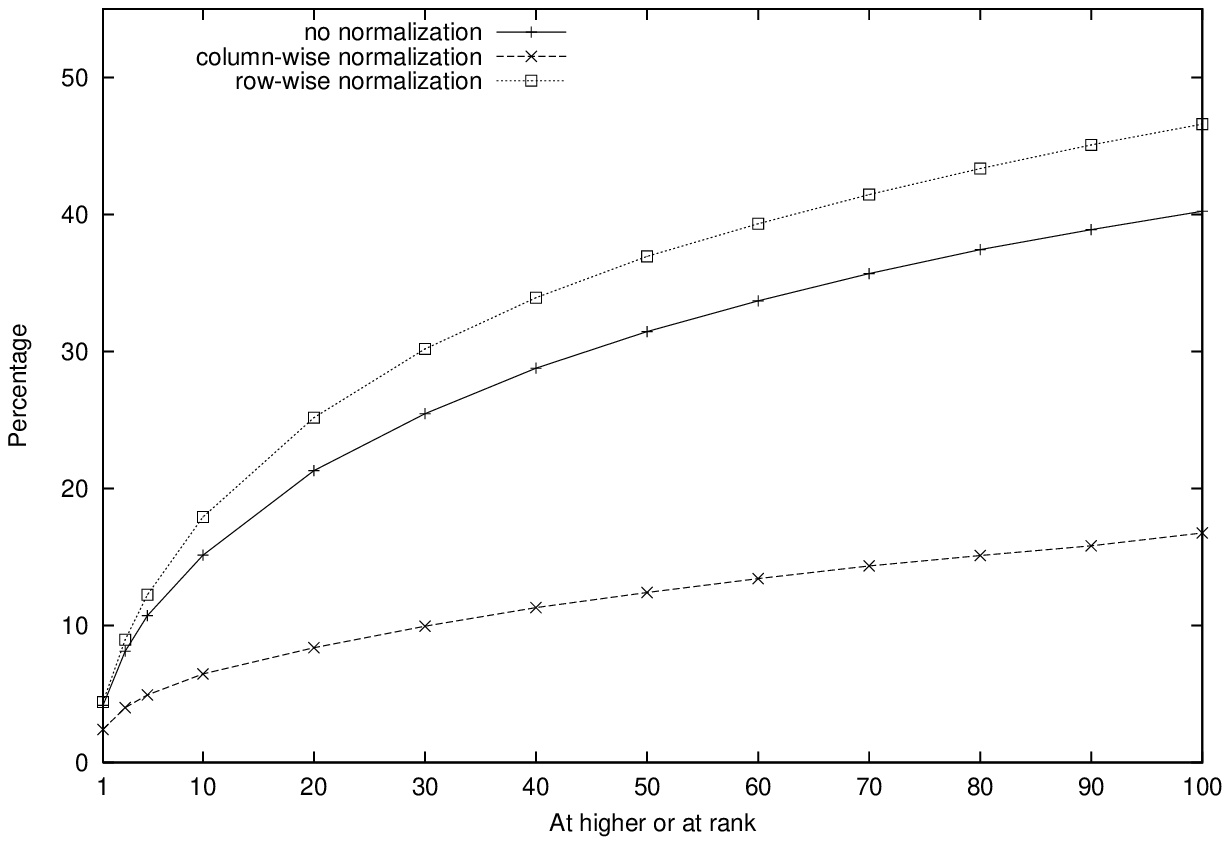}\fnwdx\label{fig:eval:norm:codlw}}
\\
\subfigure[Co-view (w/o~`rush')]{\fnwdx\includegraphics[width=\fnwd]{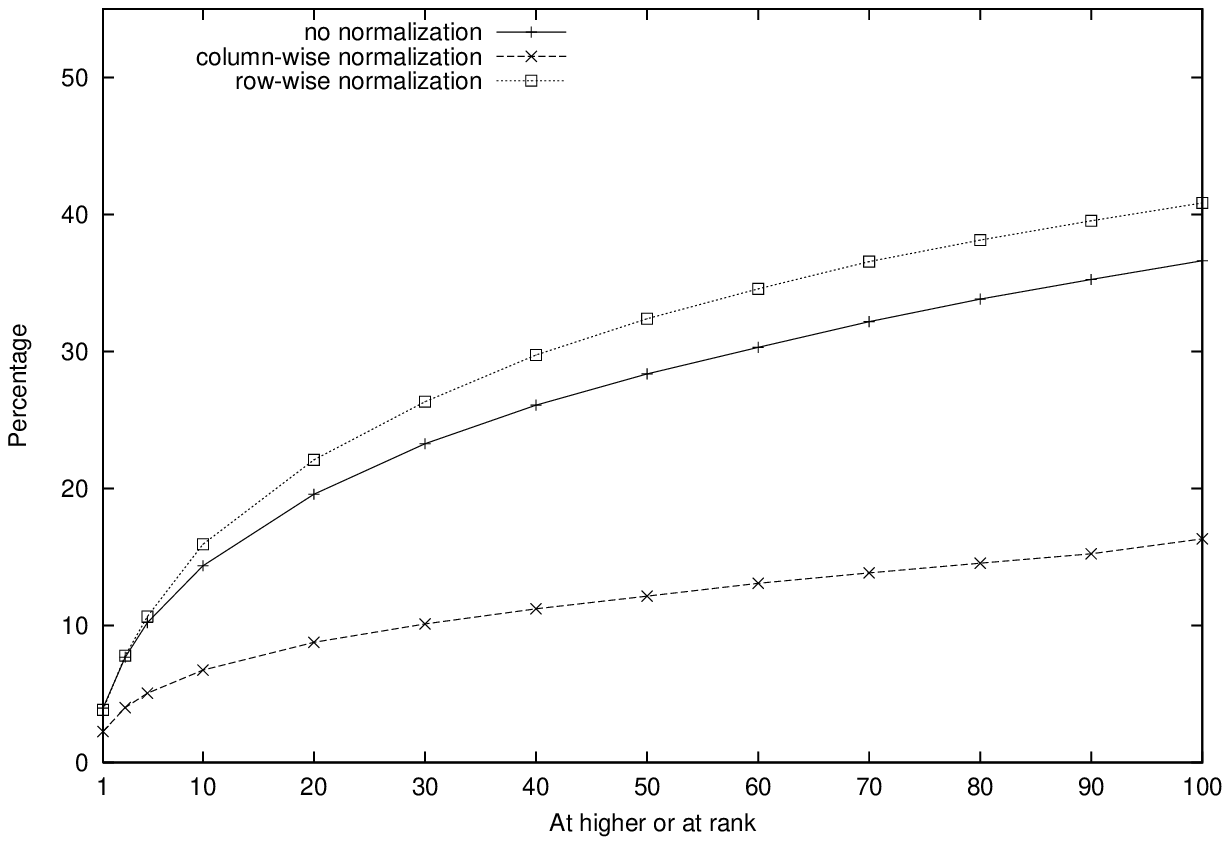}\fnwdx\label{fig:eval:norm:coviewwo}}
\subfigure[Co-view (w/~`rush')]{\fnwdx\includegraphics[width=\fnwd]{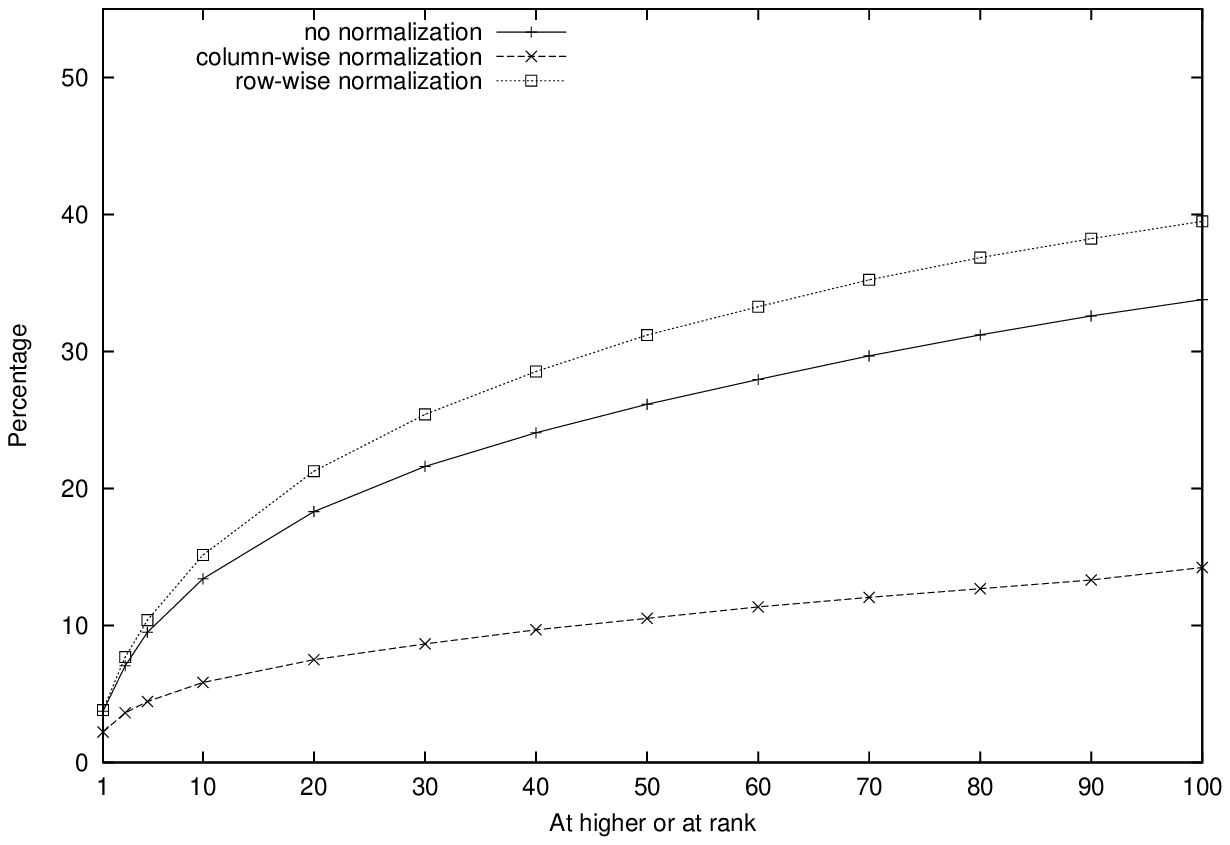}\fnwdx\label{fig:eval:norm:covieww}}
  \caption{Effect of normalization}
  \label{fig:eval:norm}
\end{figure}

We implemented the normalizations, proposed in section~\ref{sec:basics:normalization}, for citation and download data.
Figure~\ref{fig:eval:norm} summarizes the results.

For co-citation, a big improvement can be achieved by row as well as column normalization (Figure~\ref{fig:eval:norm:cocit}). This justifies the assumption that to find related papers it is better to be cited by papers which only cite few other papers. The higher selectivity of such papers expresses a stronger relationship between its references.
Over time the number of citations steadily increases, thus leading to higher citation counts for older papers. This makes it more likely for them to be in co-citation lists for other papers, thus having higher rankings. As the results indicate, authors are more selective in their choice of references. Penalizing popular, old papers improves recommendation quality for our setting.

For co-reference the roles change such that column normalization indeed leads to an improvement, penalizing often cited papers, but since we calculate co-reference as the dot product between rows, row normalization is a better choice (Figure~\ref{fig:eval:norm:coref}). It penalizes long reference lists, favoring short ones, which are more likely to be selective in their choice of references. That means, sharing references is not enough, similar papers also shouldn't have still other references. Although we are recommending related papers to the references of a paper, we achieve the same \mbox{performance} with normalization as for using co-reference on the paper in question.
The aggregation of co-reference rankings and the involved availability of more, diverse input data seems to play a big role because normalization of co-reference applied on only one paper is counterproductive (Figure~\ref{fig:eval:norm:coref1}). Here, it only increases the recall at lower ranks ($>65$), but worsens the recommendation quality for the top results.

Access data exhibits a lot of artefacts like the mentioned `rush' to new published papers. This is caused by several lists of recently published papers, which drastically restrict the choice of a user. If we do not ignore the influence of such accesses, we achieve strongly biased recommendations towards papers of the same month. Figures~\ref{fig:eval:norm:codlwo}--(g) show the two different kinds of access data, we are considering, once without and once with the `rush' to new papers.
Already without any normalization, the exclusion of such biased accesses makes a big difference.
Row normalization penalizes sessions with many accesses. Artificial sessions like those not originating from humans, but from search engines or robots would be disqualified. Besides that, row normalization is able to cover a big fraction of the unwanted accesses after the publication of a paper, since the improvements are much higher on the basis of the data including the rush of accesses to publication and is nearly able to catch up with the results achieved otherwise. Curiosity probably causes users to click on more links, given a list of new, yet unknown papers.
Furthermore, this type of normalization becomes essential to support the stability of co-accesses.\footnote{A single session accessing $m$ papers is able to increment $m(m-1)$ co-accesses!}
Column normalization, on the other hand, drastically hurts the achieved recall, which is contrary to all what yet could be observed in citation data.
Access data doesn't suffer from the problem that older papers always have an advantage in terms of the achievable amount of references to it. Here, it is the case that accesses are distributed more uniformly, but with column normalization extremely seldom accessed papers become favored. Popularity derived from access data seems to be a better indicator for its relevance than such, derived from citation data.
This confirms the call of Kurtz et~al.\ for looking at an individual's scientific productivity from a two-dimensional perspective, consisting of citations and accesses~\cite{Kurtz05_readcite}.

Summarizing, normalization can greatly improve the quality of recommendations and reduce the influence of noise in the underlying data. On the other hand, it also can lead to a decrease in performance.

\section{Choice of parameters}
\label{sec:eval:agg}

In the proposed methods, there is a number of free parameters to adjust. Since the exhaustive evaluation of the combinatory of all parameters, which potentially could influence the final results, would make this thesis unduly large, we decide on providing comparisons of key parameters after setting minor parameters to reasonable values.

\begin{figure}
\newcommand{\fnwd}{0.48\textwidth}
{\centering \setlength{\subfigcapskip}{5pt}
\setlength{\tabcolsep}{1pt} 
\begin{tabular*}{\textwidth}{cc}
\subfigure[\label{fig:eval:gap}]
{\includegraphics[width=\fnwd]{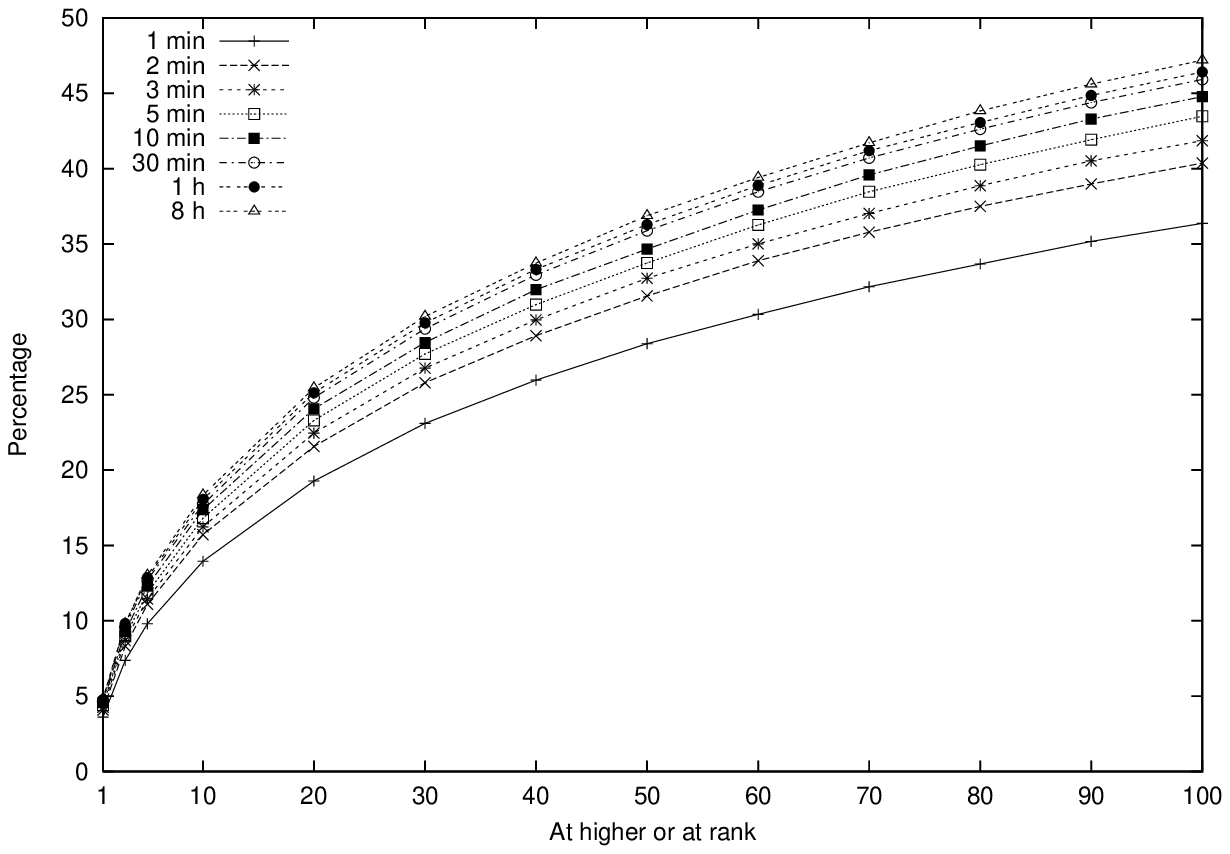}}
&
\subfigure[\label{fig:eval:agg}]
{\includegraphics[width=\fnwd]{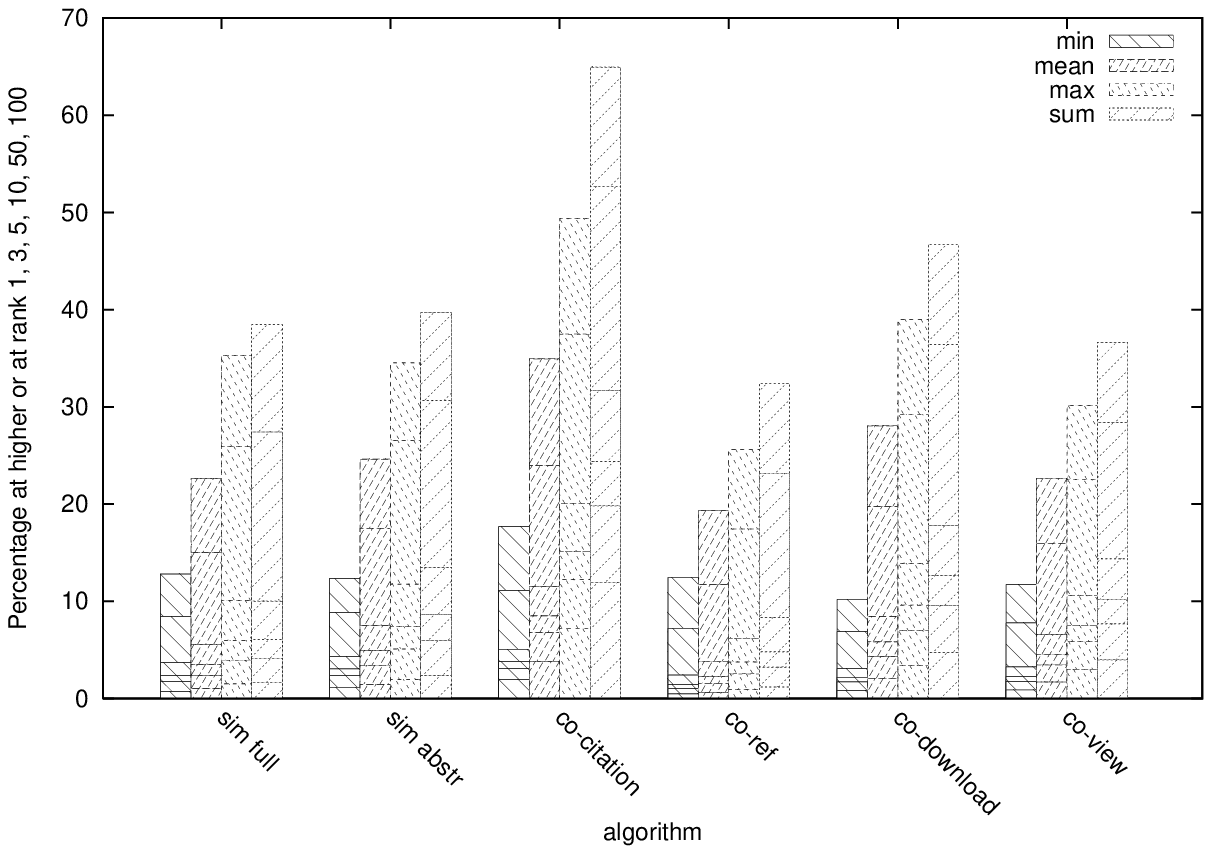}}
\end{tabular*}
  \caption[Influence of time-out and aggregation functions]{Influence of time-out~(a), used for session generation, and aggregation functions~(b), for top 1, 3, 5, 10, 50 and 100.}
  \label{fig:eval:gap-agg}
}
\end{figure}

For session generation, we have to decide about a suitable time-out. The assumption is to trade off between a long time-out, which leads to longer sessions and therewith to a higher coverage, and a short one to consider the concept drift. A user might be interested in different topics over time, such that accesses in a short time interval are highly related to each other, but less to very later accesses.
We used different time-outs from one minute to eight hours and evaluated the thereof derived co-download measure like before to find an optimal cut-off point. Figure~\ref{fig:eval:gap} shows an unexpected result, namely that even with a time-out of eight hours, which practically time-outs only over night, we still see improvement.
The concept drift of users on \arxivorg\ seems to be very slow, such that even a much higher time constraint could be used. Unfortunately, our choice to use a simulating session generator doesn't scale to longer time-outs.
However, since the influence for results in the top ranks is minimal and we don't want to run into problems with proxies, we opted always to use a time-out of 30 minutes.

Furthermore, a crucial decision in every item-based CF system is the choice of the right aggregation function, which comes into play in the second step of an item-based CF framework, performed online.
We consider the standard aggregation functions {\em min}, {\em mean}, {\em max} and {\em sum}, for which the results are presented in Figure~\ref{fig:eval:agg}.
In all cases, the ordering of the functions imposed by their performance stays the same, even though the relative improvement is different.
The superior role of {\em sum} to combine different rankings allows for that documents being related to more than one reference of the document in question should be presented at first. Obviously, this helps to promote the ``right'' documents.

Of course, one can imagine much more sophisticated ways to combine different rankings. These include to learn a combination function out of the properties of the papers involved. For this, direct properties as well as such derived from the same or different type of measure, like authority weights or PageRank, could  be incorporated.
But to find for each measure the right aggregation function is a time-consuming task, we will use a different setup in Section~\ref{sec:eval:further} to examine the performance of the types of data without the need for aggregations.

\comment{
\clearpage
\section{Properties of recommended papers}

how are the average PR, Auth, Hub, TimeStamp, etc. properties of papers proposed in top-10?

\section{The growth of an archive}

Find out how the utility over time improves... How much access data has to be collected, how recent references can be predicted? \\
How much time, collecting access data, is sufficient for good recommendations?
\\
Are we getting better, taking more access data into account???? ! {\bf This can be uniquely found out on this massive data set! 
}

over time: more absolute papers,
accelerated publishing, after catch up, constant submission.

more complete, more users, more valuable relatedness information

\begin{figure}
{\centering
  \includegraphics[width=8cm]{chapters/figs/eval_time92-05/eval_time92-05_cit}
  \caption{Recommendations quality over time}
  \label{fig:eval:92-05}
}
\end{figure}

evaltime92-05dl.eps

So the same for

the more papers in archive the more problematic it becomes for top-10, but the richer the connectionism is also.

dl-data:
more completeness, more users attracted

}

\section{Further experiments}\label{sec:eval:further}

In this section, we conduct further experiments, implying a new experimental setup. Of special interest is the comparison between the best citation- and access-based measures, namely co-citation and co-download.
We examine the questions, how recent related papers can be recommended, and what the performance of the measures is over the age of given papers.

\subsection{Recommendations for recent publications}

Papers, which have just been published, do neither have citations, nor accesses. Of course, assuming they have not been published also somewhere else, in this stage accesses are the precondition for future citations. But as we have seen in Section~\ref{sec:sessionextraction}, very early accesses are strongly biased towards co-accesses of new publications. Thus, it is interesting to see, how recent the measures are able to give recommendation and of what quality they are.

To make a more reliable comparison, not depending on used aggregation functions and the amount of references in a paper, we only apply the measures on the papers in question themselves and use another scheme to evaluate for related papers. This is formalized in Setting~\ref{alg:setting2}.

In contrast to our previous setup, we do not evaluate on past references, but on papers between $\tevbg$ and $\teven$, for which we assume to be recently published. Because we only want to find very recent, related papers, also being published later than $\tevbg$, we derive the golden truth about such relationships out of co-citations in the future. For this, we used those papers published in 2006 ($\tgtbg$=1/1/2006 and $\tgten$=6/1/2006). To track the performance over a reasonable time period, we went two years back in time and evaluated on all papers in 2003 ($\tevbg$=1/1/2003 and $\teven$=11/31/2003).

In a first step, for every paper $d_i$ a set of related papers is pre-calculated out of the union of all co-cited and recently published papers $d_j$, which have been co-cited by papers $d_k$, published in 2006 (line~2).
Additionally, for a fixed amount of points in time, all measures are pre-calculated, as if they would have been calculated that time.

Then, for each recommendation algorithm $\textrm{al}$, each document $d_k$ published in the evaluation time interval and each point in time $t_x$, iterating from the publication date of $d_k$ at least two years to the beginning of the ground truth time interval, we build the set of related papers $T$ for $d_k$ out of such papers, which have been pre-calculated and also already been published (line~4--8). If this set is not empty, $\Delta t$ reflects the time that have passed from the publication of $d_k$ to the actual point in time $t_x$ (line~9--10). Now, the recommendation algorithm runs to find related papers for $d_k$, given the knowledge at time $t_x$ (line~11). Since we are only interested in recent papers, we filter such before $\tevbg$ and also not contained in the citation graph (line~12--13). To assess the quality of the retrieved ranking, we use {\em average precision} between the recommendations $O$ and the set of related papers $T$ (line~14).
Finally, we take the mean over all papers $d_k$, on which we were evaluating on to receive the mean average precision values for each algorithm over time (line~18).
\begin{algorithm}
\renewcommand{\algorithmcfname}{Setting}
\scriptsize
\SetLine \linesnumbered \dontprintsemicolon
\KwIn{points in time: 
$\tevbg < \teven \ll \tgtbg < \tgten $,
citation graph vertices~$V$
}
\KwOut{Mean Average Precision for ($\alg , \Delta t$)-tupel}
\ForEach{document $d_i$}{
// precalculate sets of related papers
\[
\textrm{related}(d_i) \GETS \bigcup_{d_k \ s.t.\ (k,i) \in E \atop\land \tgtbg \le t(d_k) \le \tgten }
\{ d_j : (k,j) \in E\ \ \land\ \ \tevbg \le t(d_j) \le \tgtbg \}
 \setminus \{ d_i \} \nonumber
\]\;
}
\ForEach{recommendation algorithm \alg}{
\ForEach{document $d_k: \tevbg \le t(d_k) \le \teven$ }{
  \ForEach{point in time $t_x: t(d_k) \le t_x \le \tgtbg
$}{
    $T \GETS \textrm{related}(d_k)$\;
    $T \GETS T \setminus \{d_i: \comment{d_i \in T \land} t(d_i) > t_x \} $\;
    \If{$T \ne \emptyset$}{
      $\Delta t \GETS t_x - t(d_k)$\;
      $O \GETS \alg(d_k, t_x)$\;
      $O \GETS O \setminus \{d_i: t(d_i) < \tevbg \}$\;
      $O \GETS O \setminus \{d_l \notin V \} $\;
      $(\alg, d_k, \Delta t) \GETS AP(O, T)$\tcp{ average precision}\;
    }
  }
}
\lForAll{$\Delta t$}{$(\alg, \Delta t) \GETS \textrm{mean}_{d_k}(\alg, d_k, \Delta t)$}\;
}
\caption{MAP of recommendations for recent publications}
\label{alg:setting2}
\end{algorithm}

\begin{figure}
{\centering
\includegraphics[width=0.8\textwidth]{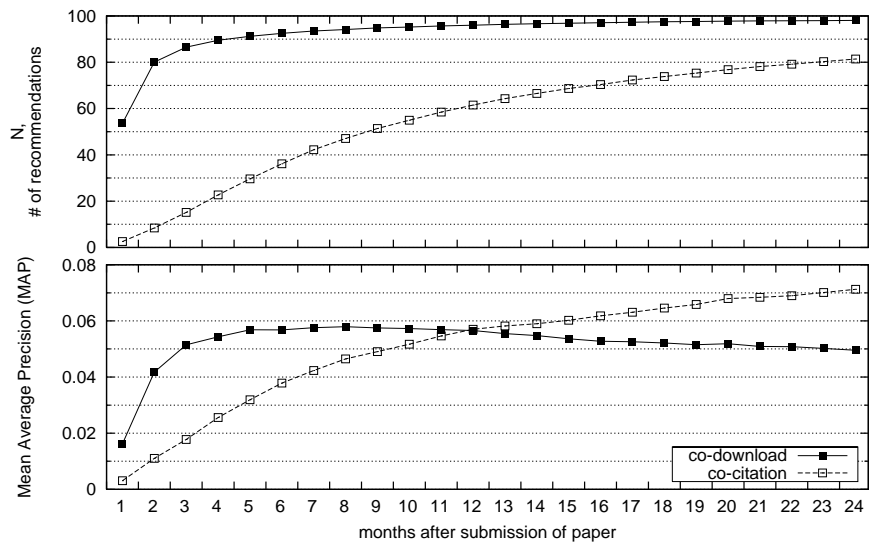}
  \caption[MAP and number of recommendations after publication]{Mean average precision and number of recommendations over time after publication. The maximal number of recommendations is limited to $N=100$.}
  \label{fig:eval:setting2}
}
\end{figure}

The results, shown in Figure~\ref{fig:eval:setting2}, indicate that co-download can compete with co-citation.
Already in the first month, co-download is able to show relationship to an average of more than 50 papers. Furthermore, in the next few months it rapidly converges to the preassigned maximum of 100 recommendations. Co-citation shows a much slower convergence. At all, it is remarkable that co-citation is already able to recommend anything in the first months, requiring that a paper already will be cited only one month after its publication! Besides the limitations of the experiment, we are sure that this fact is not generalizable and a specific of \arxivorg.
The ratio between the achieved MAP values and the number of recommendations is much higher for co-citation, speaking for more significance in citation data, resp.\ more noise in access data.

\begin{figure}
{\centering
\includegraphics[width=0.8\textwidth]{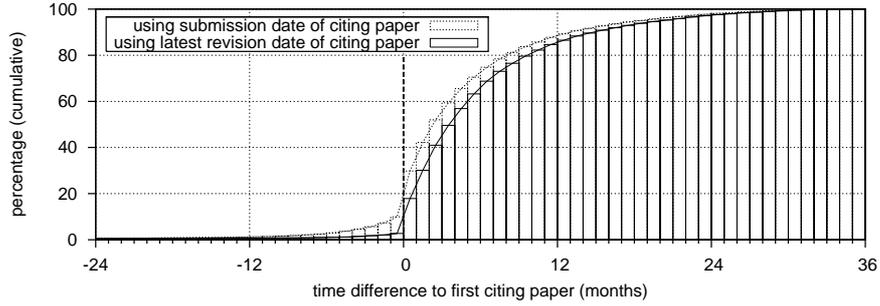}
  \caption[Time between paper publishing and first citation for it]{Time between paper publishing and first citation for it.}
  \label{fig:paper:firstcitetime}
}
\end{figure}

Also for this experiment, several limitations apply.
The biggest problem is that we assume that citation data is instantly available. In \arxiv, citations are still indexed manually, so that this kind of data is expensive and thus likely to be generated in batches.
Anyhow, assuming there would be an automatic citation indexing mechanism, the length of a publish-cite cycle is untypically short.
Figure~\ref{fig:paper:firstcitetime} shows that in \arxiv\ at least 20\% of the papers, which are cited at all, already get their first citation in the first month after publication.
Together with Figure~\ref{fig:citations:reference_dist2005}, this expresses that physics is one of the ``hot'' sciences, building new work primarily on the foundations made in the last 5 years.
Additionally, a big portion of papers also adjust their reference lists in new versions to include recently published work. Nearly $^1\!/_3$ of all papers have been updated. So, the truth lies in between the curves, but we can only determine the exact distribution with the knowledge, when every edge in the citation graph has been introduced.
Also in this experiment the higher coverage of access data has not been rewarded.
To generalize to arbitrary archives, the curves for co-citation should be shifted by a time interval to a later point in time, and co-access further rewarded for being able to give recommendations even for papers that have not been cited.


\subsection{Recommendations over age of papers} \label{sec:eval:age}

In the main setting, we could not apply co-citations on papers themselves because we assumed that they have just been published. Instead we used the knowledge about the papers that we already have had, i.e.\ its references.
However, to measure performance, we aggregated over references which follow a distribution over time (see Figure~\ref{fig:citations:reference_dist2005}).
References which have been published long time ago are likely to have more citations and thus it is easier to find related papers for them.
To examine the influence of the age of a paper, on which the measures are applied, we chose to implement Setting~\ref{alg:setting3}.

Instead of finding a held-out reference given the others, we turn Setting~\ref{alg:setting1} around such that given a reference of a paper, we try to find all others. Additionally, for different ages of a paper given, we calculate MAP values between the recommendations based on it and the list of further references, we are searching. Considering age given a fixed point in time $t_0$ frees us to compute measures multiple times.

\begin{algorithm}
\renewcommand{\algorithmcfname}{Setting}
\scriptsize
\SetLine \linesnumbered \dontprintsemicolon
\newcommand{\trec}{t_{recent}}
\KwIn{points in time: 
$ \dots < t_1 < t_0 = \tgtbg < \tgten $,
citation graph vertices~$V$
}
\KwOut{Mean Average Precision for ($\alg , \Delta t$)-tupel}
\ForEach{recommendation algorithm \alg}{
\ForEach{document $d_k: \tgtbg \le t(d_k) \le \tgten$ }{
  $R_{d_k} \GETS \{ d_i: (k,i) \in E \land t(d_i) \le t_0 \}$\;
  \ForEach{point in time $t_x, x \ge 0$}{
    $\Delta t \GETS t_x - t_0$\;
    $R_{d_k, t_x} \GETS \{d_i \in R_{d_k}: t_{x+1} \le t(d_i) \le t_x \}$\;
    \ForEach{reference $d_i \in R_{d_k, t_x}$}{
      $T \GETS \{d_j \in R_{d_k}: t_x \le t(d_j)\} \setminus \{d_i\}$\;
      \If{$T \ne \emptyset$}{
        $O \GETS \alg(d_i, t_0)$\;
        $O \GETS O \setminus \{d_l \notin V \} $\;
        $(\alg, d_k, \Delta t, d_i) \GETS AP(O, T)$\tcp{ average precision}\;
      }
    }
    $(\alg, d_k, \Delta t) \GETS \textrm{mean}_{d_i}(\alg, d_k, \Delta t, d_i)$\;
  }
}
\lForAll{$\Delta t$}{$(\alg, \Delta t) \GETS \textrm{mean}_{d_k}(\alg, d_k, \Delta t)$}\;
}
\caption{MAP of recommendations over age of paper}
\label{alg:setting3}
\end{algorithm}

We iterate for every recommendation algorithm $al$ over all documents $d_k$, which have been published between a given time interval from $t_0=\tgtbg$ to $\tgten$ (line~1--2). We have chosen to use 1/1/2005--6/1/2005.
Furthermore, we will consider only those references $R_{d_k}$ of $d_k$, which have been published before $t_0$ (line~3). Then, for every point in time $t_x$ a subset $R_{d_k, t_x}$ of $R_{d_k}$ is generated with all references being published in the momentary considered time interval (line~6).
For every reference $d_i$ in $R_{d_k, t_x}$, the set of references we are searching for is generated by exclusion of $d_i$ from those papers of $R_{d_k}$ which have been published later than $t_x$ (line~8).
We generate with algorithm $al$ recommendations for $d_i$, from which we exclude such not contained in the citation graph (line~10--11).
For every algorithm, document $d_k$, age and reference $d_i$ of $d_k$, having that age, we compute the average precision of the recommendations (line~12).
Finally, we average over the references for every age and again, over all considered documents $d_k$ (line~15,18).
In this experiment, we consistently used the date of the latest update of a paper instead of its submission date. This excludes such papers from the evaluation of which the reference list might have been updated to include papers published very recently.

\begin{figure}
{\centering
  \includegraphics[width=0.8\textwidth]{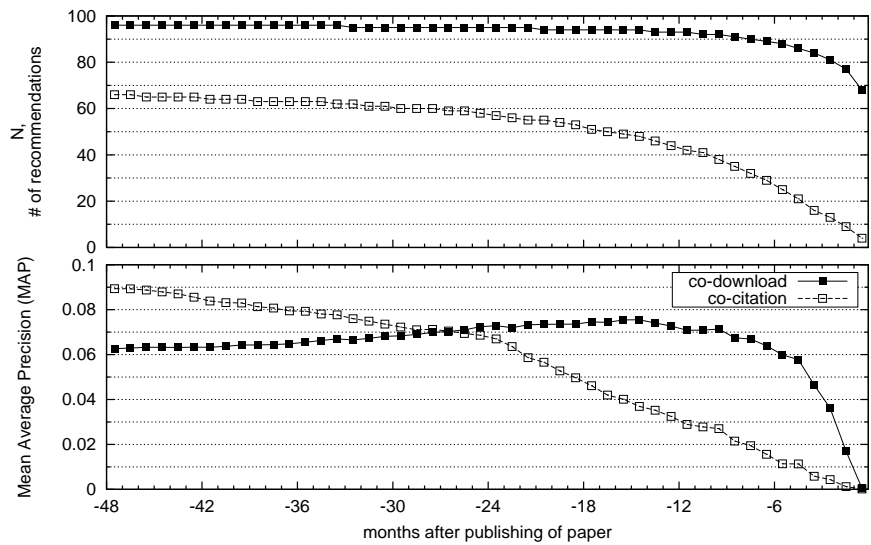}
  \caption{Mean average precision over the age of a given paper.}
  \label{fig:eval:setting3}
}
\end{figure}

The results, depicted in Figure~\ref{fig:eval:setting3}, possess an inverted behavior to what we have already seen in Setting~\ref{alg:setting2}.
Recently published papers have nearly no citations, thus, the coverage of co-citation is very limited. Because of that, relevant papers can not be found.
By contrast, co-access is able to provide a large number of recommendations already in the first month after publication. Its mean average precision exceed those of co-citation for the first months until co-citation is able to catch up.
From that point in time, we have a constant performance for both measures. This equilibrium shows similar performance ratios as seen in Setting~\ref{alg:setting1}.
This means, the older the paper given to the recommender is, the better is co-citation suited to find related papers. Co-access is able to propose more recommendations, those earlier and is slightly better for recent publications to find related papers.

This experiment is very similar to Setting~\ref{alg:setting2}. Both trace the time difference between publication of a paper and a chosen point in time, for which MAP values are computed. The main difference is how we generated the set of related papers. In this setting we allow a much smaller set to be related, namely all references of a paper. In Setting~\ref{alg:setting2}, it was the union over the references of all documents containing the paper in question. This explains why the coverage of co-citation converges slower than in Setting~\ref{alg:setting2}.

The usage of the submission date for this experiment led for co-citation to high MAP values, already for papers published in very recent months. A more detailed tracking over time, when every reference has been added to a paper, would make this treatment unnecessary and lead to more accurate results.

\clearpage
\section{Recommendation system prototype}

As a final result of this thesis, the implementations have been supplemented with a front-end in form of a web application.\footnote{Freely available under \url{http://search.arxiv.org:8090/}.}
Here, very similar to the setting in the evaluations, users can provide a set of references to \arxiv\ papers (e.g.\ those of their actual work) and receive recommendations to other related papers.

\begin{figure}
{\centering
  \includegraphics[bb=0 0 1036 841, width=0.78\textwidth]{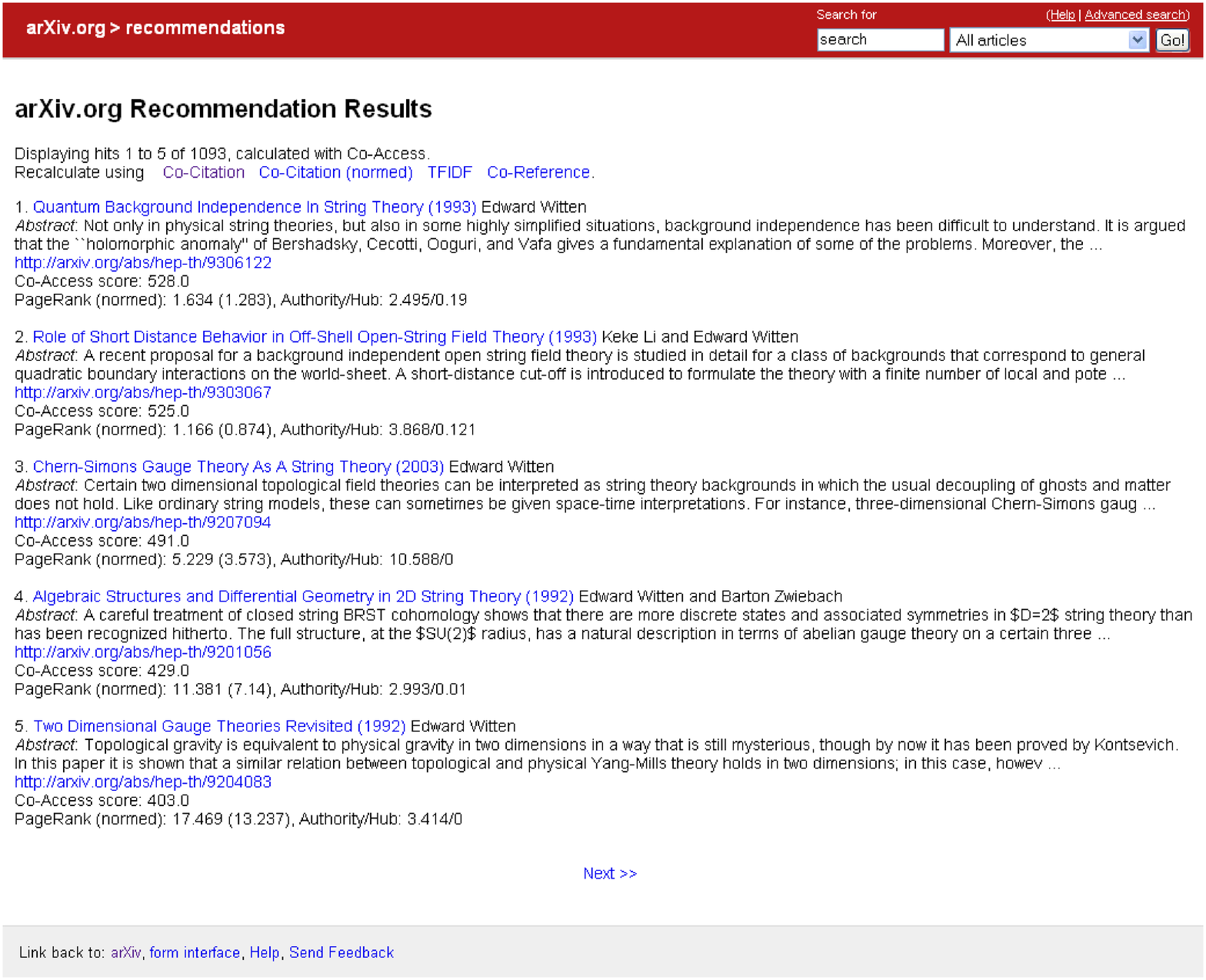}
  \caption{ArXiv recommendation system}
  \label{fig:arxivrec}
}
\end{figure}

In the introductory web page, a text field is provided, in which arbitrary information can be inserted. The only requirement is that it contains \arxiv\ identifiers in the common form ``category/ddddddd''. This way, it is possible for an author to simply paste e.g.\ the BibTeX file contents of a paper, he is actually writing into the provided form to get further pointers to related work. Just to see the system in action, it is also possible to choose a random paper out of \arxiv.
In the next step, the parsed references are resolved and shown to the user, who either agrees on them or can adjust them still.
The final step consists of the calculation of recommendations, per default, first on the basis of co-accesses. The results are presented as a ranked list of \arxiv\ papers, showing all properties discussed during this thesis~(see Figure~\ref{fig:arxivrec}).

For now, the ability to recommend papers is limited to the underlying static set of papers, which have been available by July 2006. An online system, keeping up to date with \arxiv, would require the installation of processes to push incremental data and implementations incrementally updating the measures. Thus, it leads to tight coupling with \arxiv, entailing continuous administrative, time-consuming work. Anyhow, with the existing implementations it is possible to occasionally update the data basis through complete recalculation.

\comment{
\section{Wrap Up}

As expected, the experiments with the system show that normalization leads to results not favoring such with higher publicity (high PageRank, Authority, in-links), but related and more relevant to include it in a reference section.

}

%% file: chapters/chap6_conclusion.tex

\chapter{Conclusions}
\label{sec:conclusions}

It has been shown that access data is able to identify related papers. Already a simple measure like co-access is able to compete with recommendations based on textual similarity.
Additionally, it reaches a reasonable amount of predictive power only 2 months after publication of a paper. Therefore, it is particularly suited to complement citation data as a source of information, if citations are less timely, not on-hand, expensive to acquire or completely unavailable.
Furthermore, access data is also able to make recommendations for the many papers which never are cited. Its coverage is nearly complete.
As a minor result, we could confirm that a download expresses more interest than plain page views.

Throughout the whole thesis, we have been using future citation information as ground truth for relatedness. Hereby, we favored this kind of data, particularly because future citations are biased by past citations. This might not be the best way to measure relatedness, but it is the best we have had available for an offline evaluation.
Previous to the discovery of co-citation, co-reference has been widely used as a mean of relatedness. Our experiments confirm its inferiority.

Problematic with a lot of access data based analyzes is the fact that it is generally based on user behavior, observed on the server-side. A lot more confident data about user behavior could be learned by its measurement more nearby to the user. Even the examination of mouse movements might help to induce the activity of users in a better way, as eye-tracking studies showcase~\cite{granka04_eyetracking}. To assess the extent of possible improvements, the results of \cite{Koch06:clientlogs} remain to be seen.
Also, preprocessing of access logs has to be done with the help of expert knowledge about the most often (unwanted) patterns that can be found in the data.

Most existing work has concentrated on analysis to understand the data. Little research has been made to apply access data to specific, real problems.
Instead, the work of this thesis also comprises the creation of a real-world, large-scale recommendation system.
The massive datasets we have been using entailed multiple implementations for every processing step, until a feasible variant has been found. Despite the usage of Perl and Java thousands lines of code have been written. The need to work in external memory and for sparse and customized data structures increased the efforts.
As a positive side-effect, because most of it has been \mbox{implemented} as self-contained tools or as frameworks, much can be reused for different questions.

\section{Future work}

We have shown that access data is able to present recommendations which contain related documents. A still unanswered follow-up question is, if the recommendations significantly differ from those achieved on the basis of citation data, but also textual measures. We can only assume that link as well as access informations are also able to reveal less obvious relationships between documents which cannot be found by textual means.

To overcome the lack of a ground truth, we used unseen citation information and have been therefore restricted to evaluate on papers for which citations exist. Thus, we couldn't examine the quality on papers without citations. It would be interesting to see, if the utility of access data decreases for them, or more general, on papers with what properties does access data perform differently.
A preferable evaluation would be an online experiment, with real users involved, giving explicit or implicit feedback. The built research prototype is a good start for this. However, such experiments are very time-consuming, if the results shall be significant, i.e.\ based on a reasonable number of participants.

During this thesis, we also proposed importance measures which could improve recommendations. It seems reasonable to assume that authors tend to cite more important papers.
For our particular task, the combination of the recommendations based on different types of data into a final single ranking seems promising. Further information is contained in the social network of authors and their affiliations.
Also, more sophisticated measures could be derived from access data. Usage of the time spent on a page has been shown to be a clear indicator of interest of a user.
Then, the whole array of machine learning methods could be applied, for instance, to learn rankings or how to combine them, or use smoothing to achieve a higher coverage.

An unsolved problem for access data is the susceptibility to spam. This might become a problem if access data would be widely used. However, various download service provider in the internet use counter for the number of downloads to measure importance. Also, due to the little commercial interest involved in our task abuses become less likely.


%% file: chapters/app1.tex
\chapter{Figures \& Illustrations}
\label{sec:app:figs}

For the interested reader, this appendix provides additional figures, which would have prevented fluently reading of the text.

\begin{figure}[h]
{\setlength{\tabcolsep}{1pt} 
\newcommand{\evalfigwidth}{7.0cm} 
\newcommand{\evalfigwidthh}{7.0cm} 
\begin{tabular}[t]{m{\evalfigwidthh}m{\evalfigwidthh}}
\includegraphics[width=\evalfigwidth]{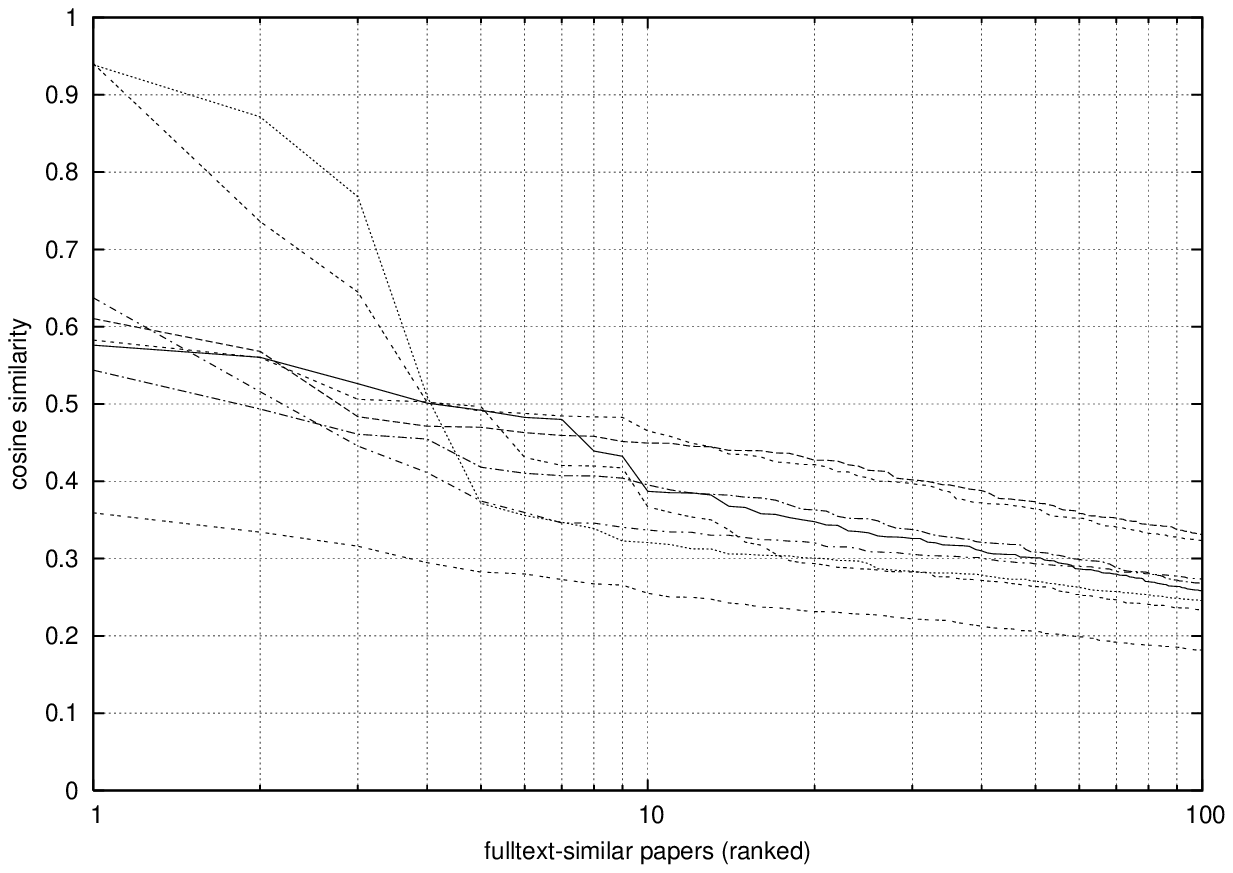} &
\includegraphics[width=\evalfigwidth]{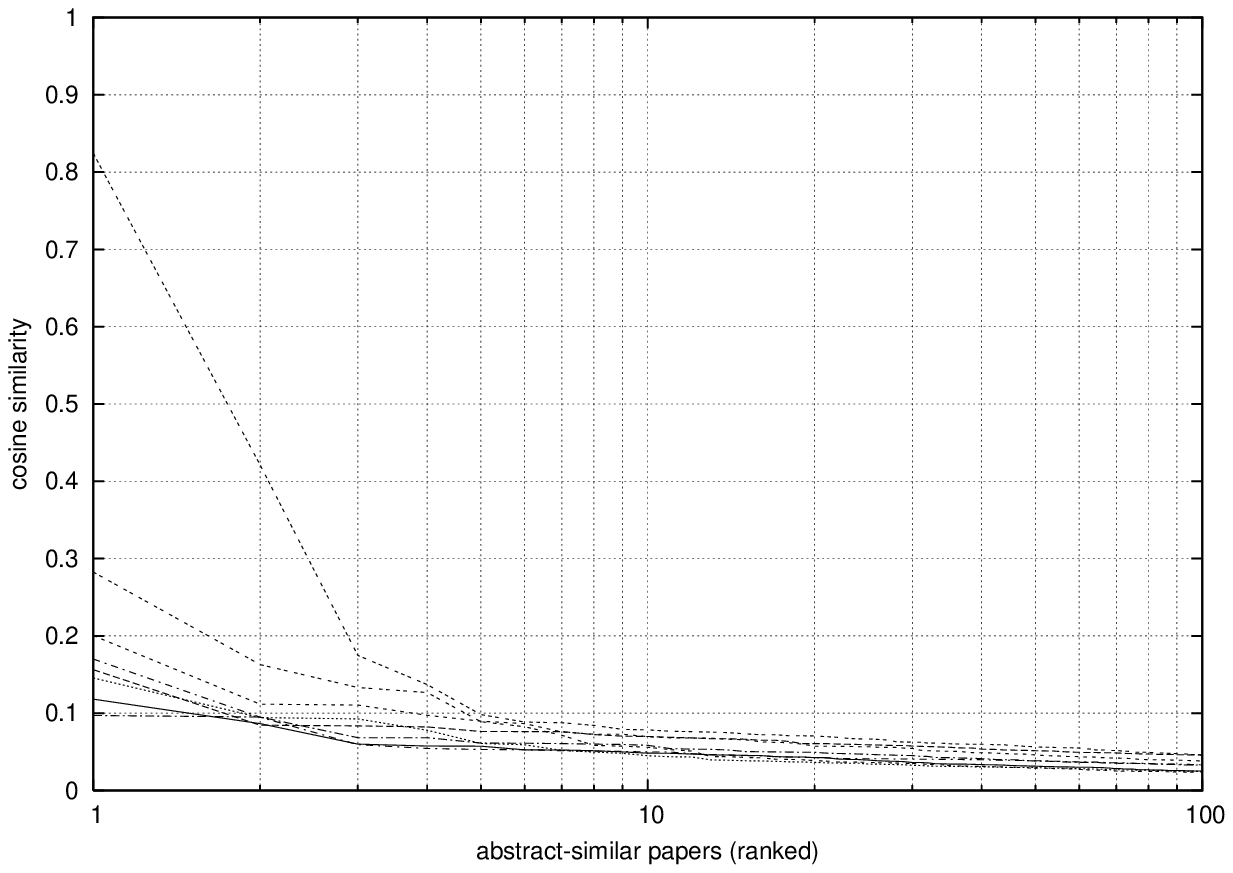} \\

\includegraphics[width=\evalfigwidth]{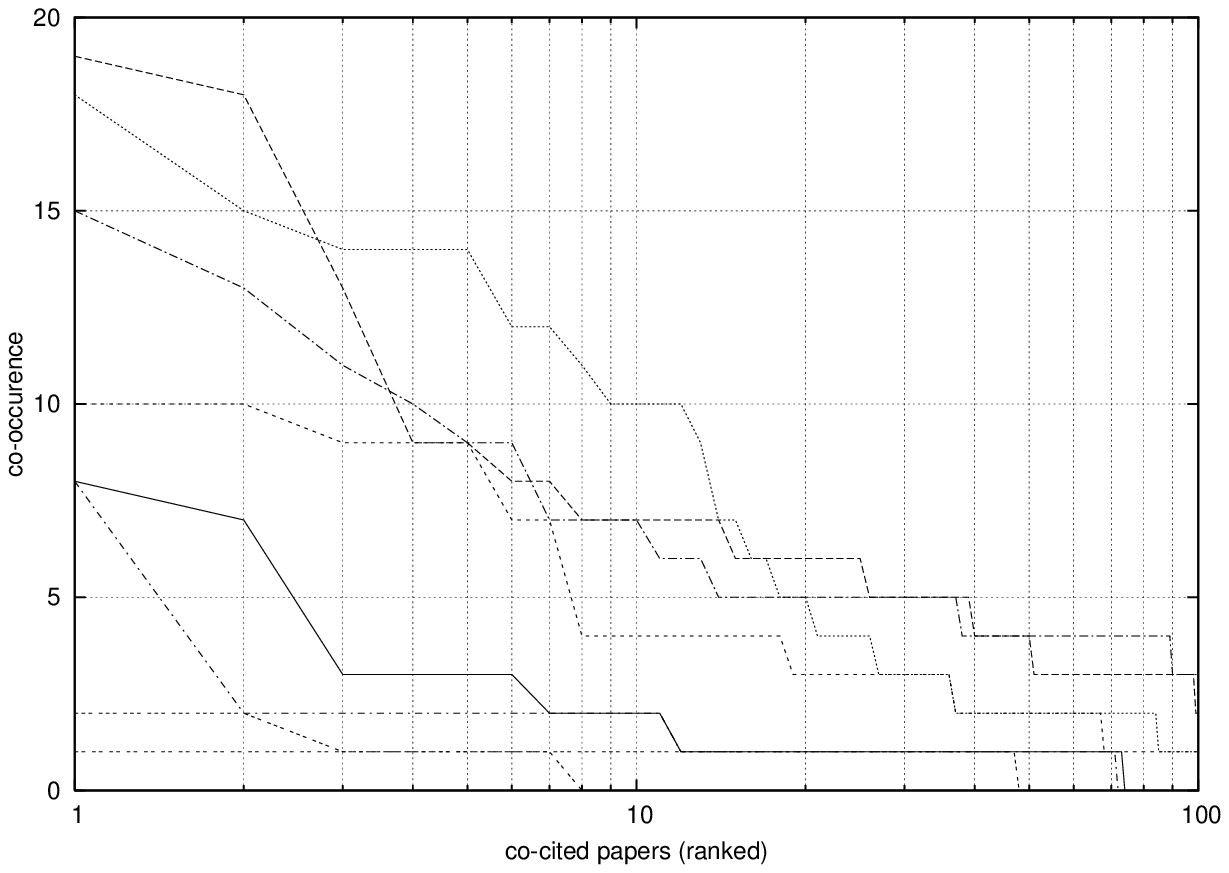} &
\includegraphics[width=\evalfigwidth]{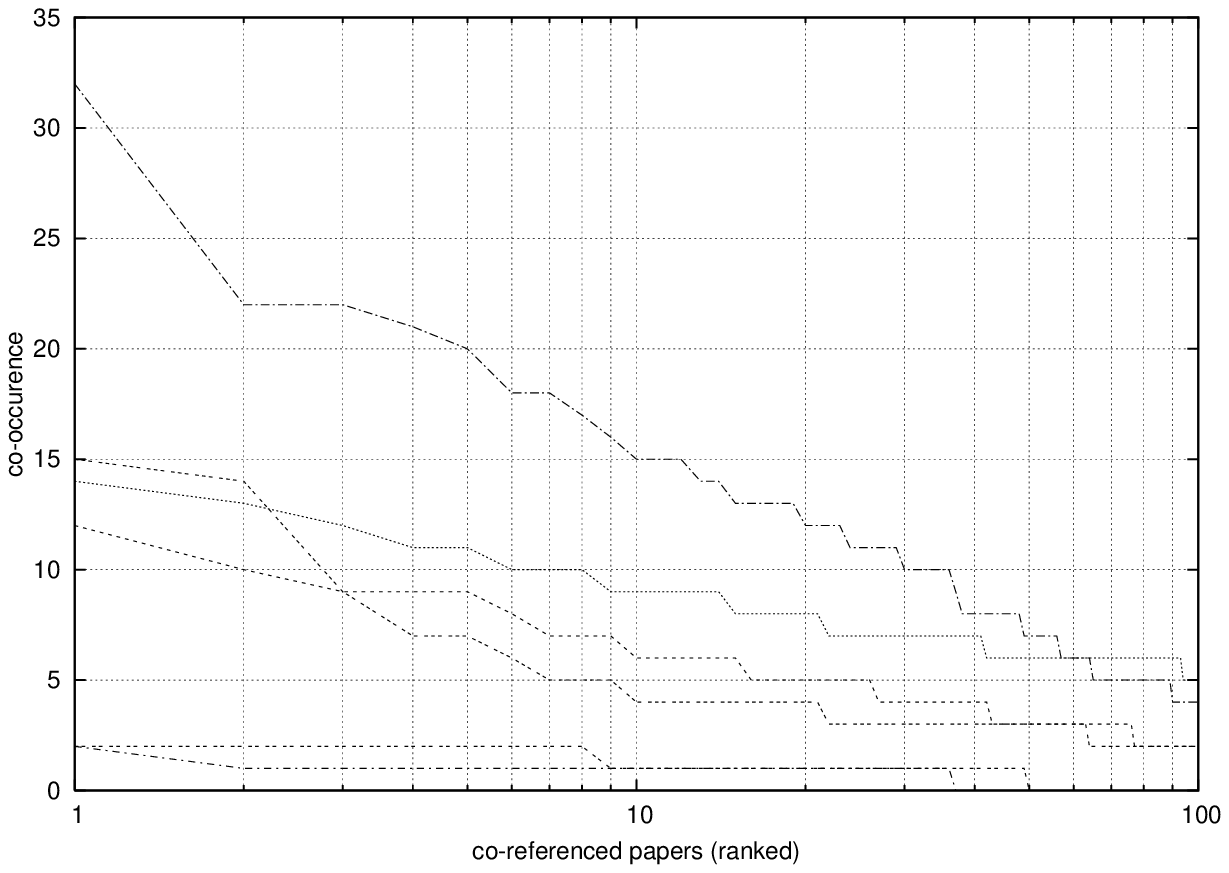} \\

\includegraphics[width=\evalfigwidth]{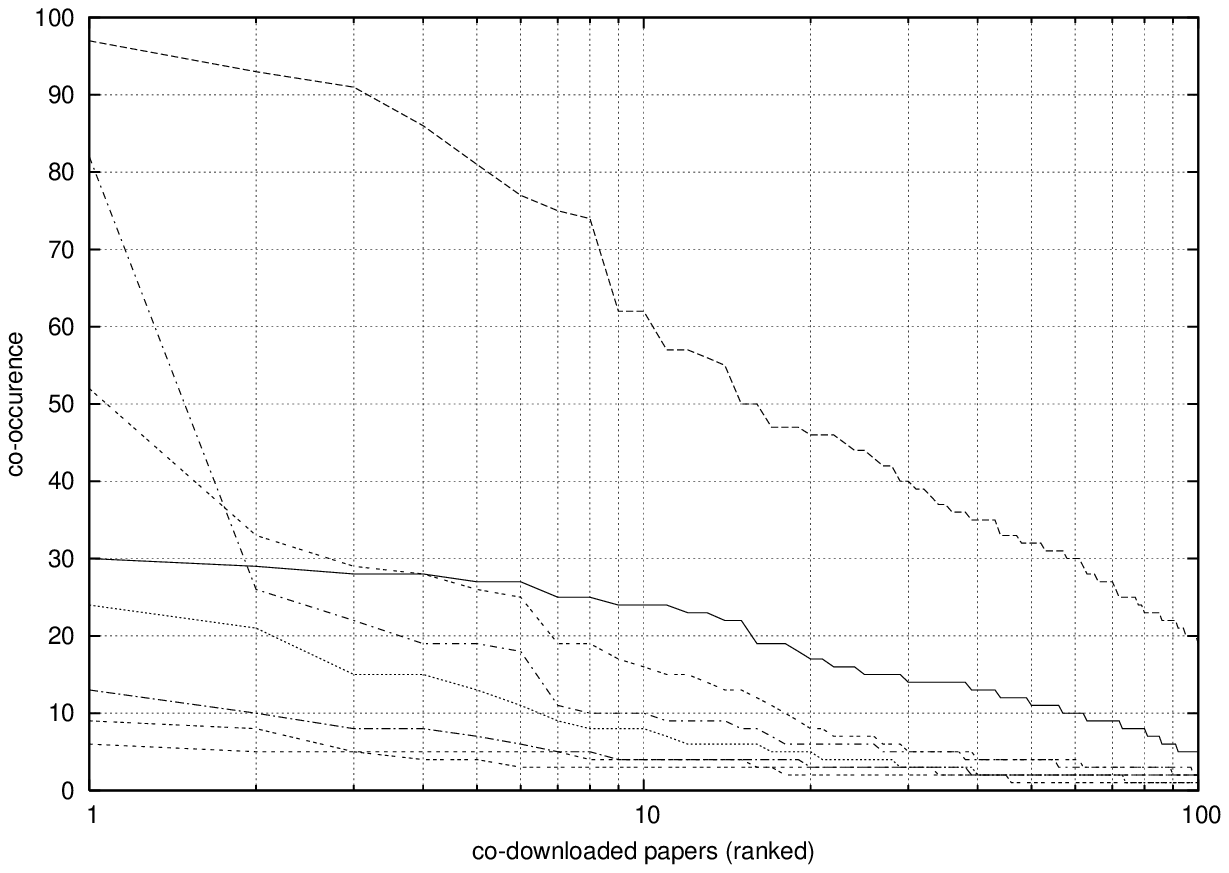} &
\includegraphics[width=\evalfigwidth]{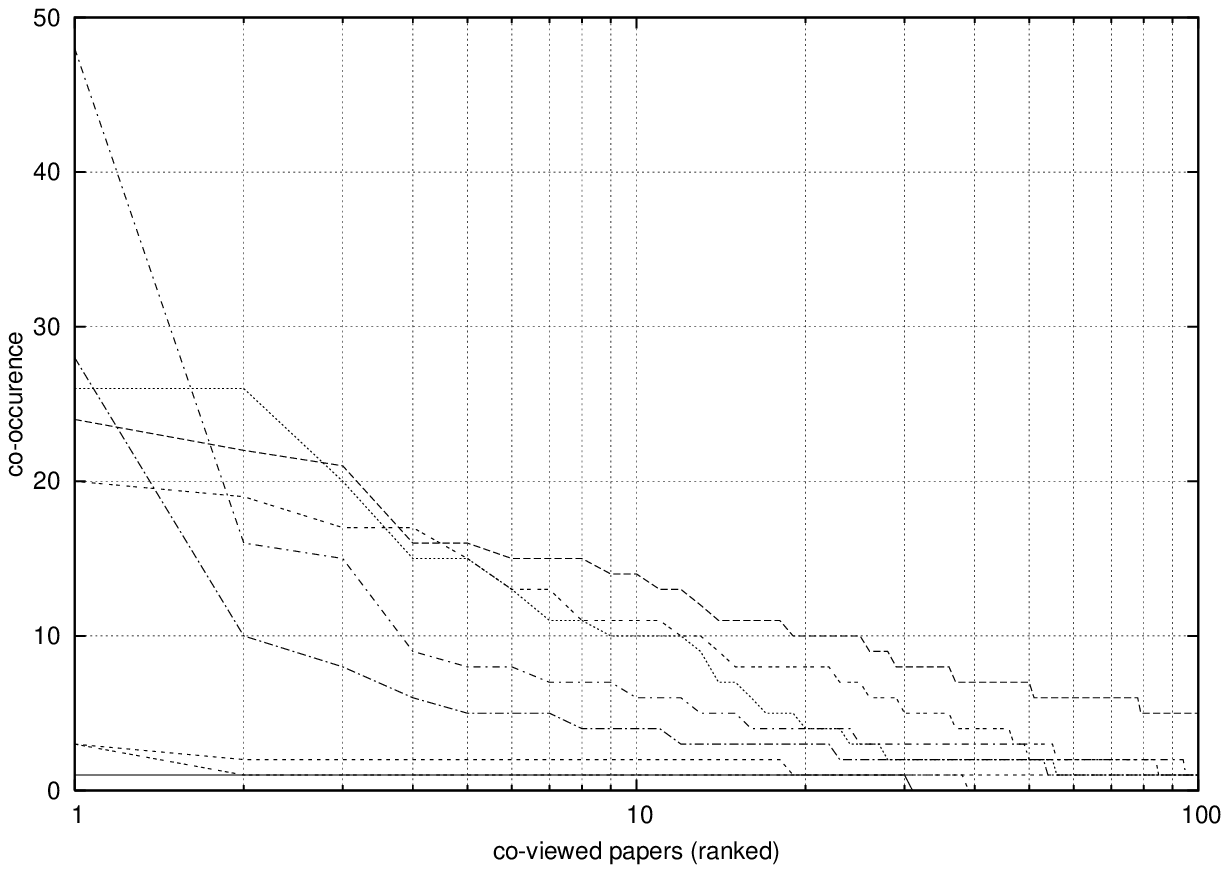} \\

\end{tabular}
}
\caption[Measures applied to 8 exemplarily chosen papers]{Measures applied to 8 exemplarily chosen papers (4 with high co-citation and 4 with a more usual low co-citation). For each of them the 100 most related have been ranked.}
\label{fig:measures:freq100}
\end{figure}


\begin{figure}
{\setlength{\tabcolsep}{1pt} 
\newcommand{\evalfigwidth}{2.3cm}
\newcommand{\evalfigwidthh}{2.5cm}
\begin{tabular}[t]{m{1.1cm}@{\ \ }|@{\ \ }m{\evalfigwidthh}m{\evalfigwidthh}m{\evalfigwidthh}m{\evalfigwidthh}m{\evalfigwidthh}l}
\hfill top-N & \centering \tfidf & \centering co-citation & \centering co-reference & \centering co-reference1 & \centering co-download & \\
\hline 
\\[-0.4cm] 
\hfill 1 & 
\includegraphics[width=\evalfigwidth]{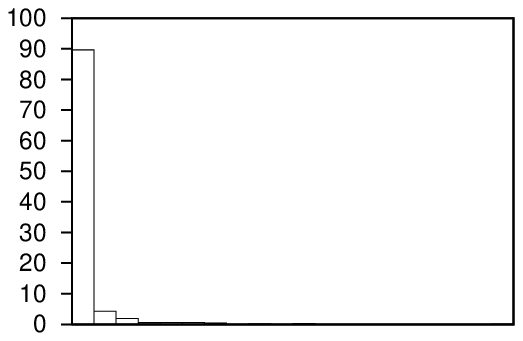} &
\includegraphics[width=\evalfigwidth]{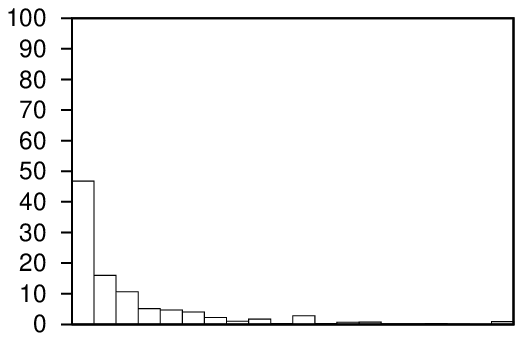} &
\includegraphics[width=\evalfigwidth]{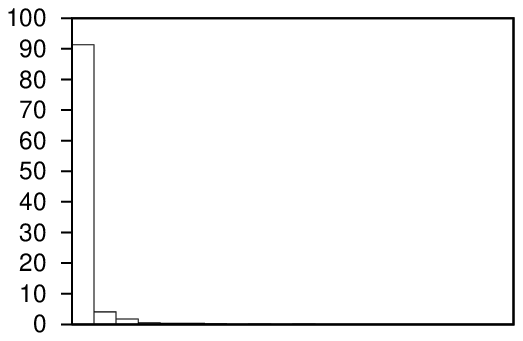} &
\includegraphics[width=\evalfigwidth]{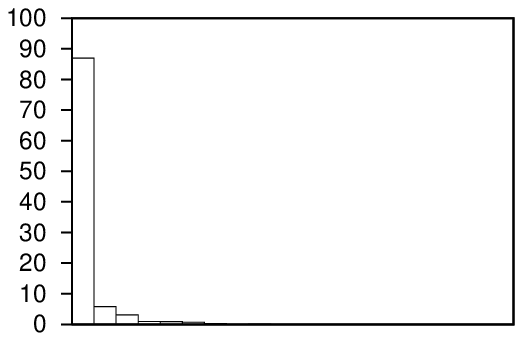} &
\includegraphics[width=\evalfigwidth]{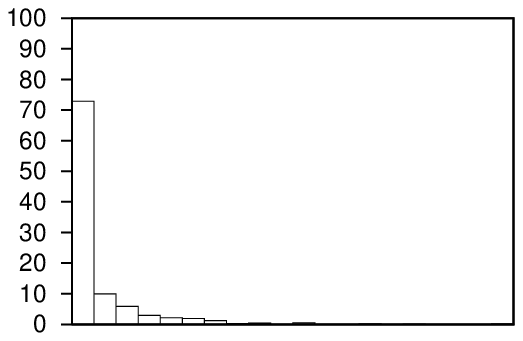} &
\\
\hfill 3 & 
\includegraphics[width=\evalfigwidth]{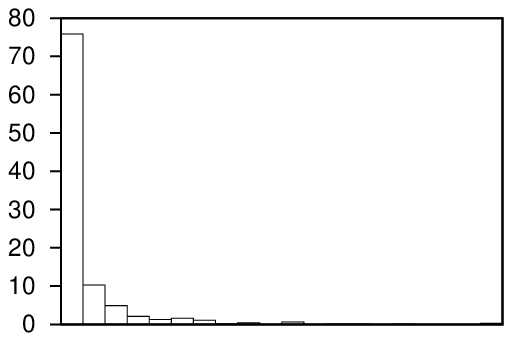} &
\includegraphics[width=\evalfigwidth]{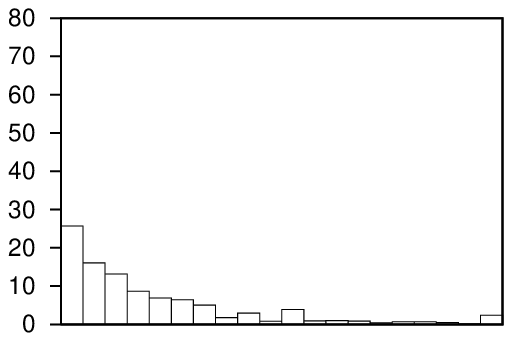} &
\includegraphics[width=\evalfigwidth]{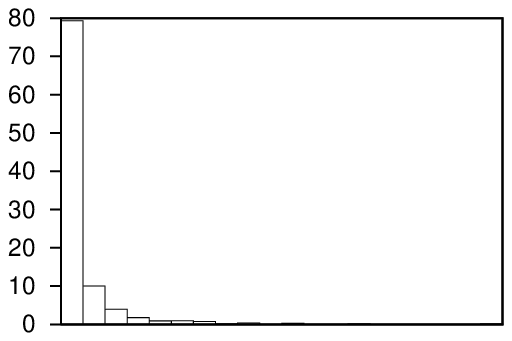} &
\includegraphics[width=\evalfigwidth]{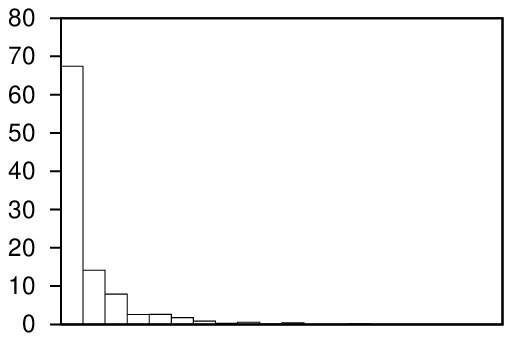} &
\includegraphics[width=\evalfigwidth]{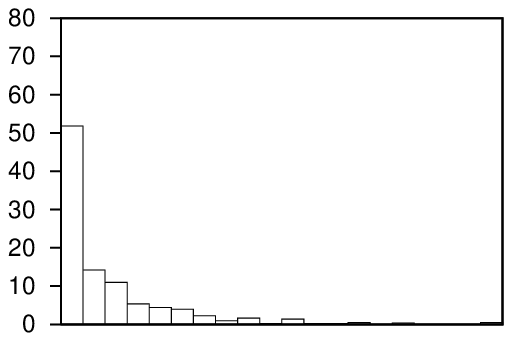} &
\\
\hfill 5 & 
\includegraphics[width=\evalfigwidth]{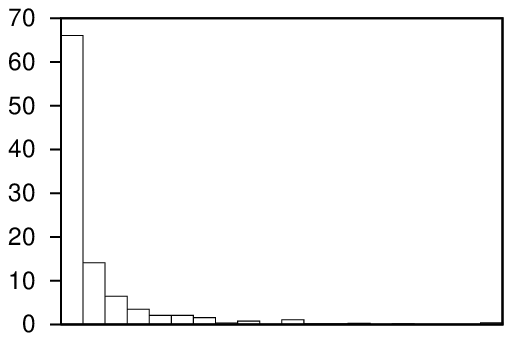} &
\includegraphics[width=\evalfigwidth]{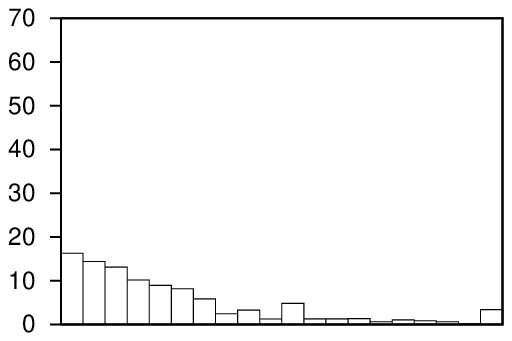} &
\includegraphics[width=\evalfigwidth]{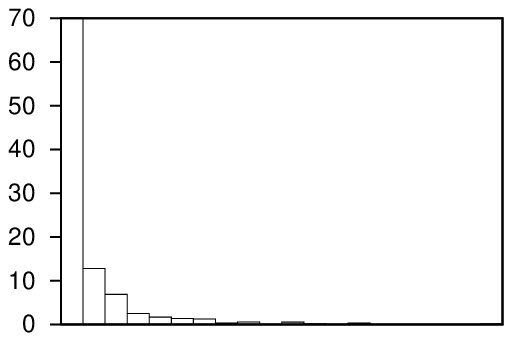} &
\includegraphics[width=\evalfigwidth]{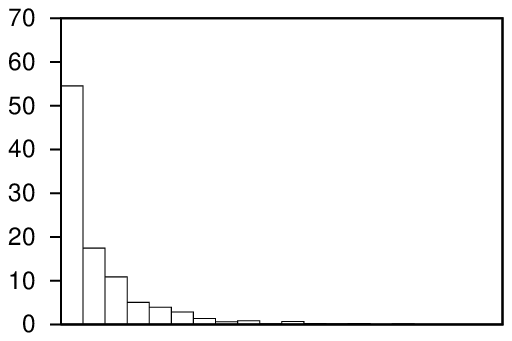} &
\includegraphics[width=\evalfigwidth]{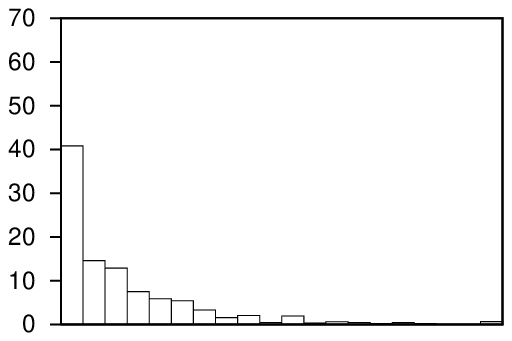} &
\\
\hfill 10 & 
\includegraphics[width=\evalfigwidth]{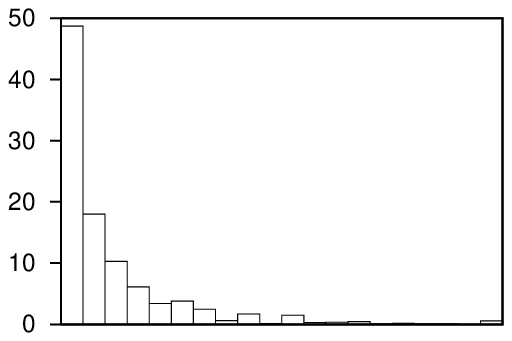} &
\includegraphics[width=\evalfigwidth]{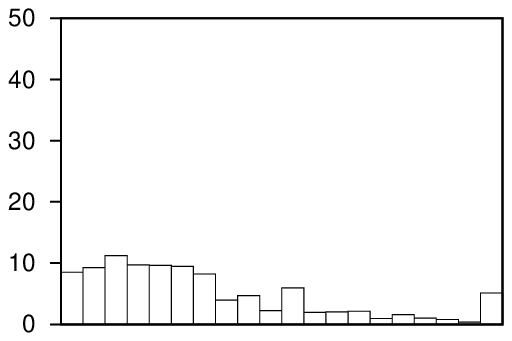} &
\includegraphics[width=\evalfigwidth]{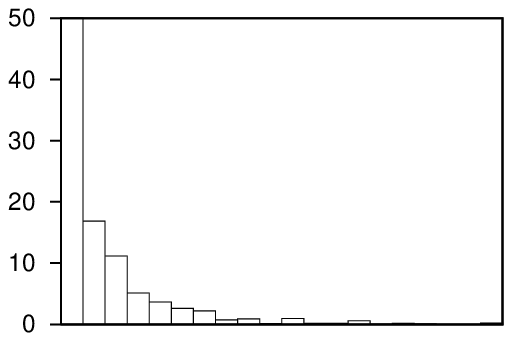} &
\includegraphics[width=\evalfigwidth]{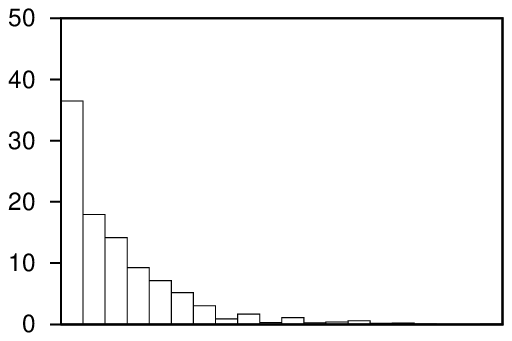} &
\includegraphics[width=\evalfigwidth]{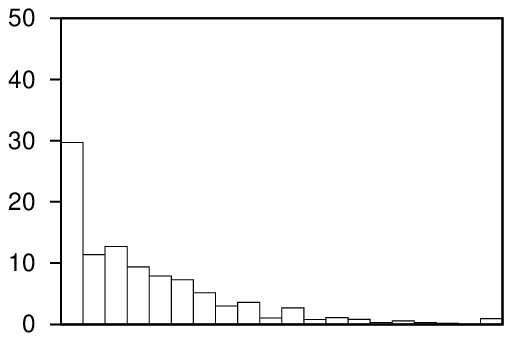} &
\\
\hfill 20 & 
\includegraphics[width=\evalfigwidth]{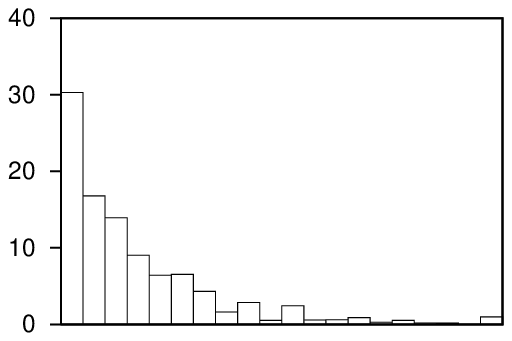} &
\includegraphics[width=\evalfigwidth]{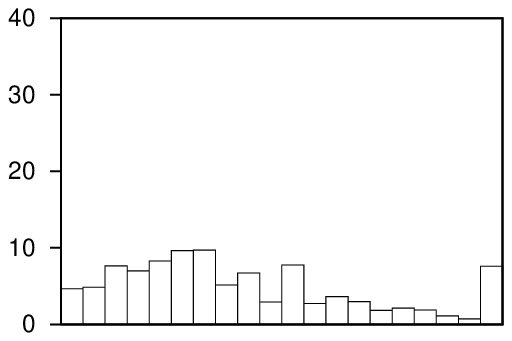} &
\includegraphics[width=\evalfigwidth]{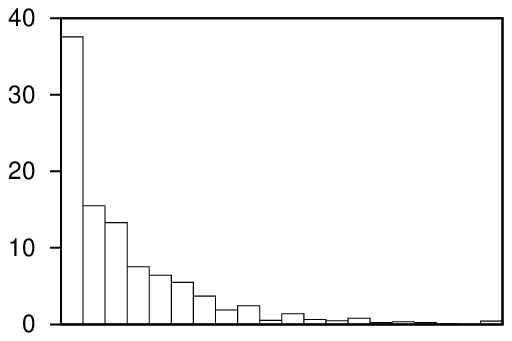} &
\includegraphics[width=\evalfigwidth]{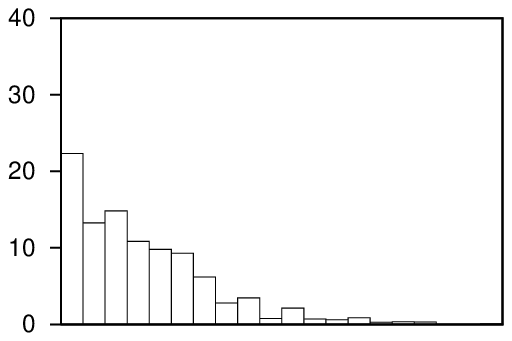} &
\includegraphics[width=\evalfigwidth]{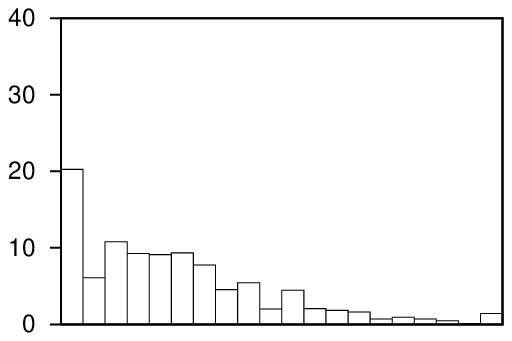} &
\\
\hfill 30 & 
\includegraphics[width=\evalfigwidth]{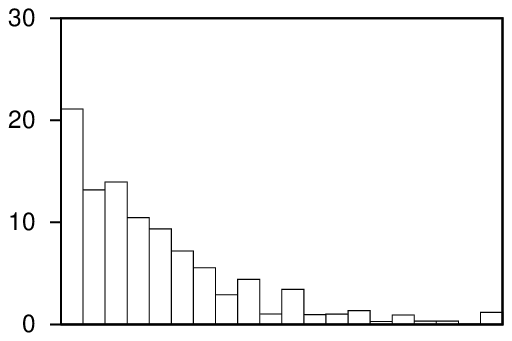} &
\includegraphics[width=\evalfigwidth]{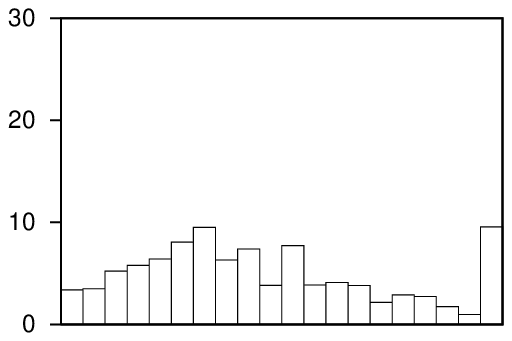} &
\includegraphics[width=\evalfigwidth]{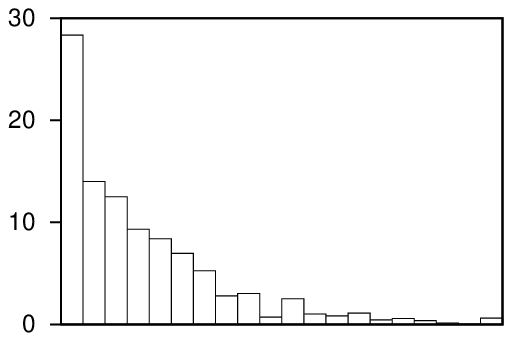} &
\includegraphics[width=\evalfigwidth]{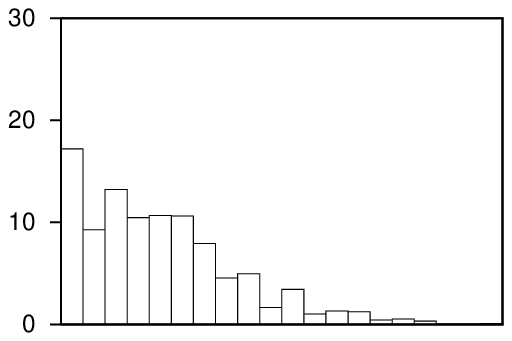} &
\includegraphics[width=\evalfigwidth]{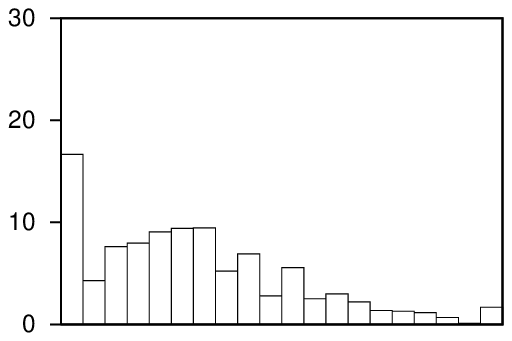} &
\\
\hfill 40 & 
\includegraphics[width=\evalfigwidth]{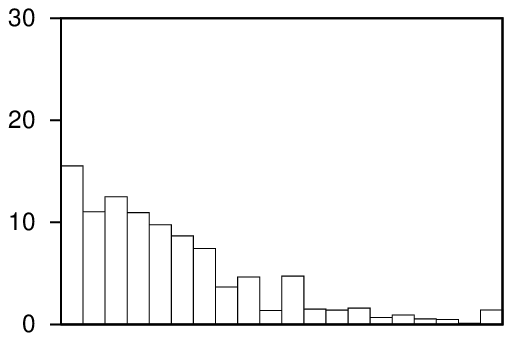} &
\includegraphics[width=\evalfigwidth]{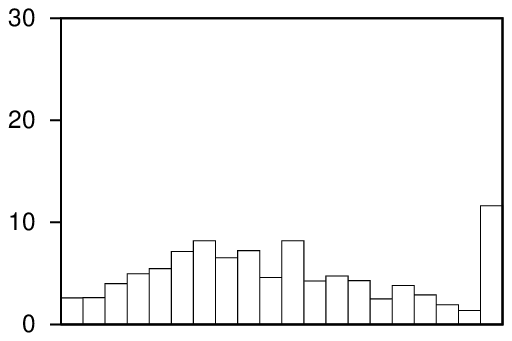} &
\includegraphics[width=\evalfigwidth]{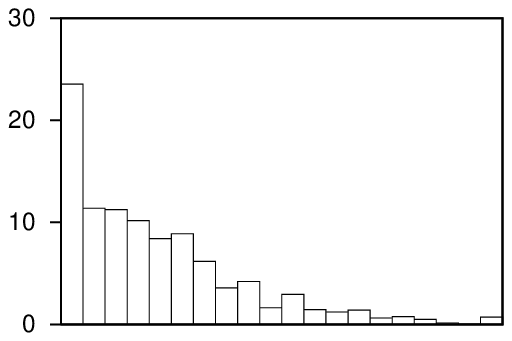} &
\includegraphics[width=\evalfigwidth]{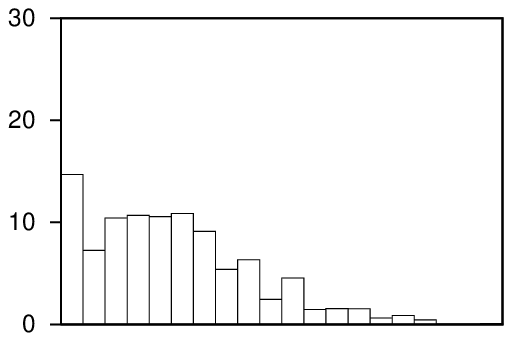} &
\includegraphics[width=\evalfigwidth]{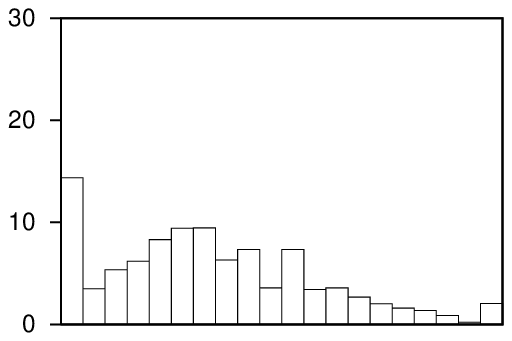} &
\\
\hfill 50 & 
\includegraphics[width=\evalfigwidth]{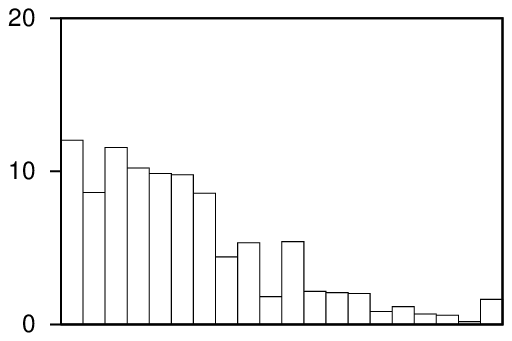} &
\includegraphics[width=\evalfigwidth]{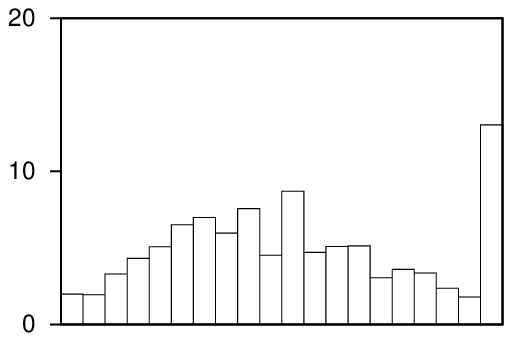} &
\includegraphics[width=\evalfigwidth]{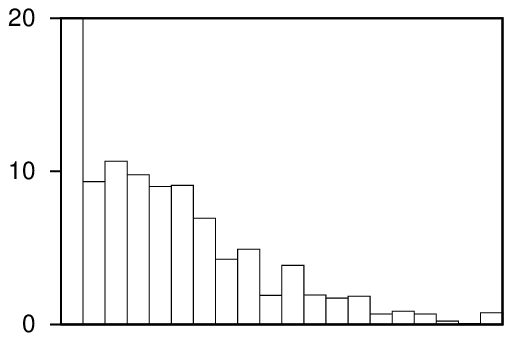} &
\includegraphics[width=\evalfigwidth]{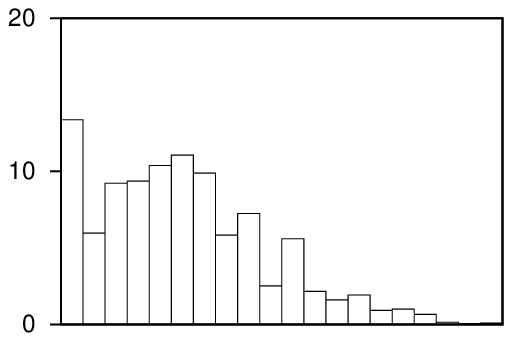} &
\includegraphics[width=\evalfigwidth]{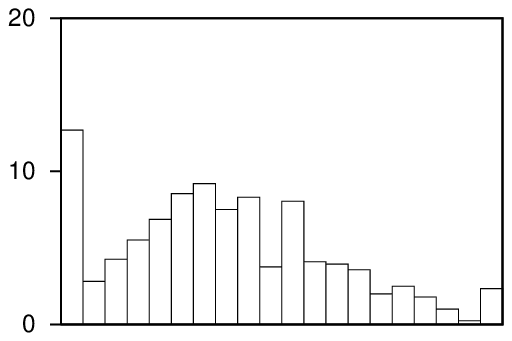} &
\\
\hfill 60 & 
\includegraphics[width=\evalfigwidth]{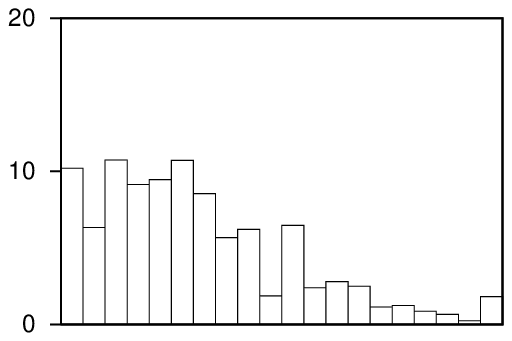} &
\includegraphics[width=\evalfigwidth]{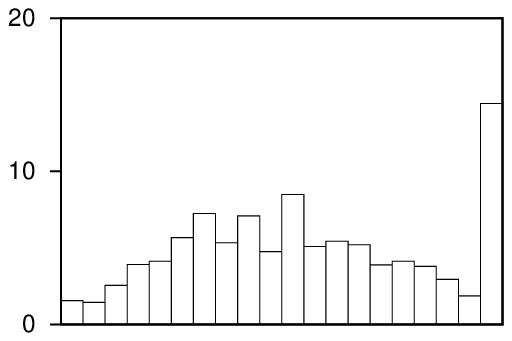} &
\includegraphics[width=\evalfigwidth]{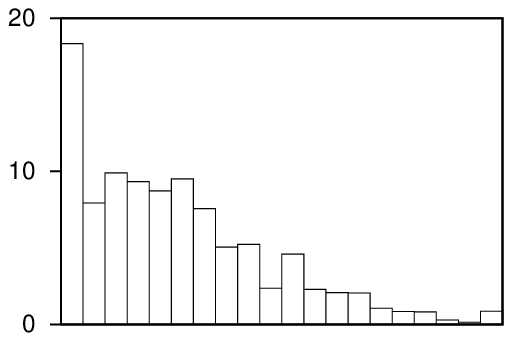} &
\includegraphics[width=\evalfigwidth]{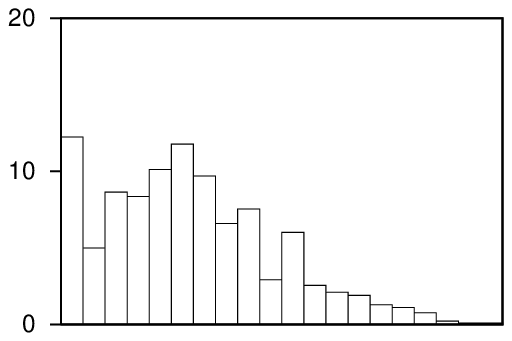} &
\includegraphics[width=\evalfigwidth]{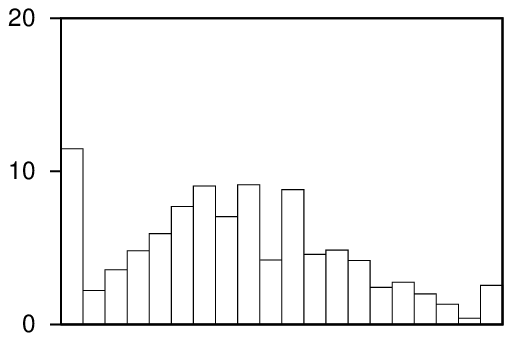} &
\\
\hfill 70 & 
\includegraphics[width=\evalfigwidth]{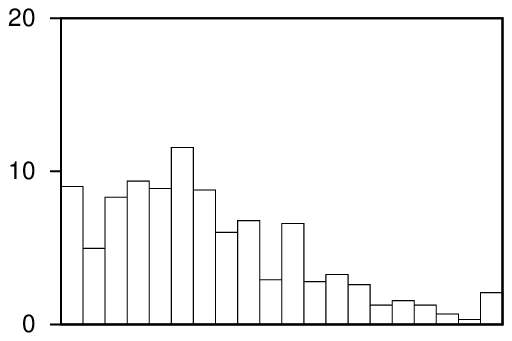} &
\includegraphics[width=\evalfigwidth]{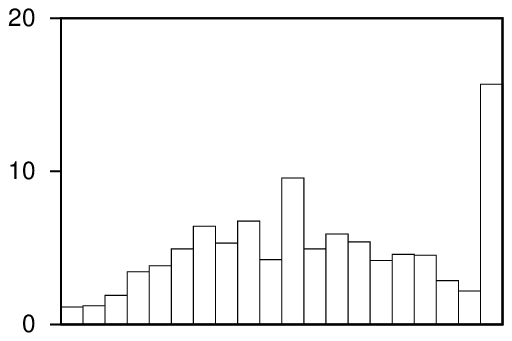} &
\includegraphics[width=\evalfigwidth]{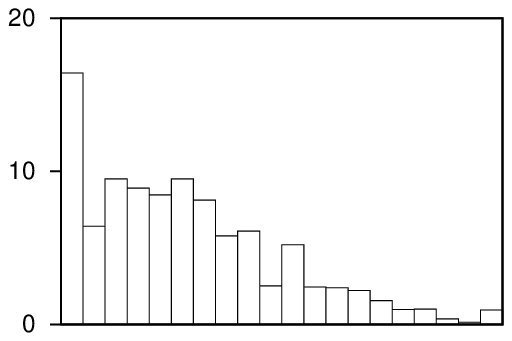} &
\includegraphics[width=\evalfigwidth]{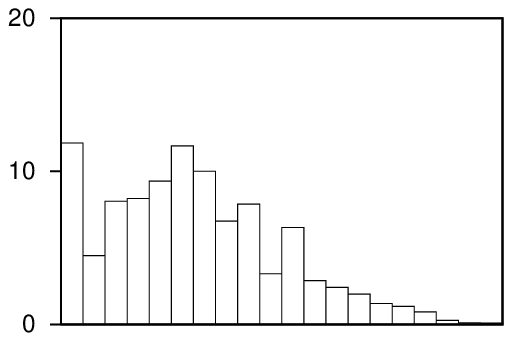} &
\includegraphics[width=\evalfigwidth]{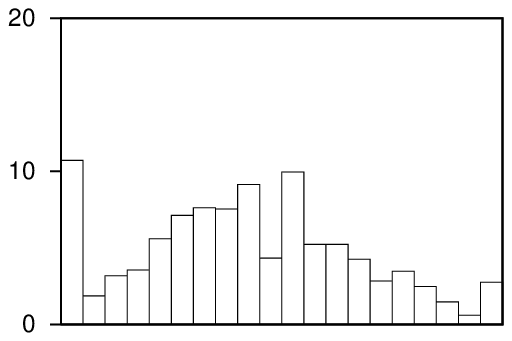} &
\\
\hfill 80 & 
\includegraphics[width=\evalfigwidth]{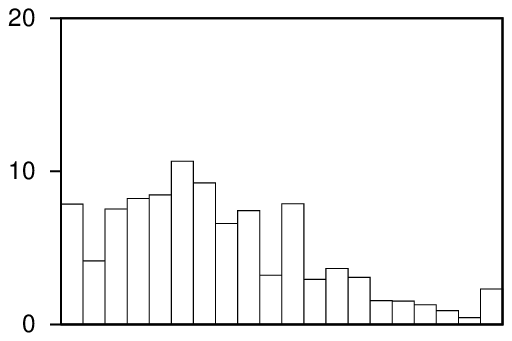} &
\includegraphics[width=\evalfigwidth]{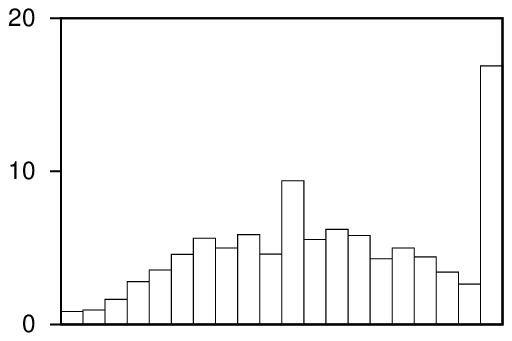} &
\includegraphics[width=\evalfigwidth]{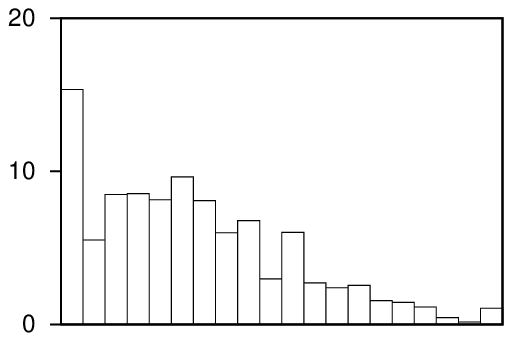} &
\includegraphics[width=\evalfigwidth]{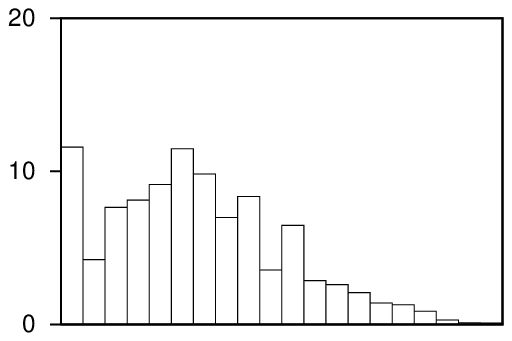} &
\includegraphics[width=\evalfigwidth]{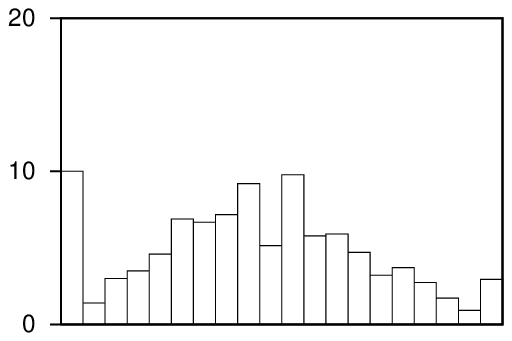} &
\\
\hfill 90 & 
\includegraphics[width=\evalfigwidth]{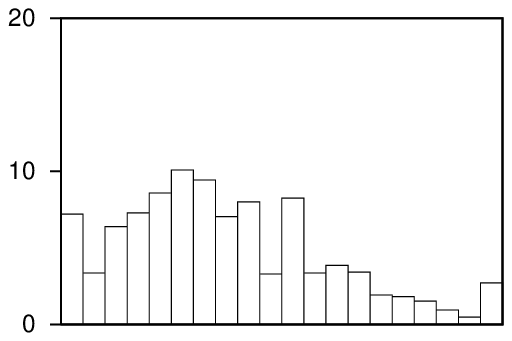} &
\includegraphics[width=\evalfigwidth]{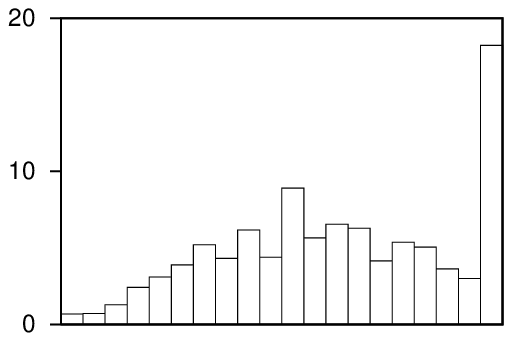} &
\includegraphics[width=\evalfigwidth]{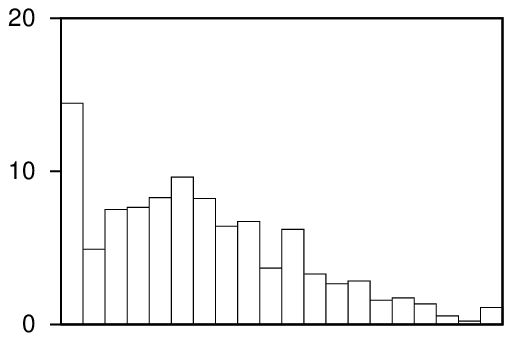} &
\includegraphics[width=\evalfigwidth]{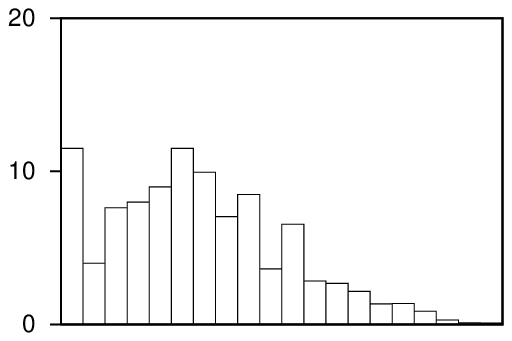} &
\includegraphics[width=\evalfigwidth]{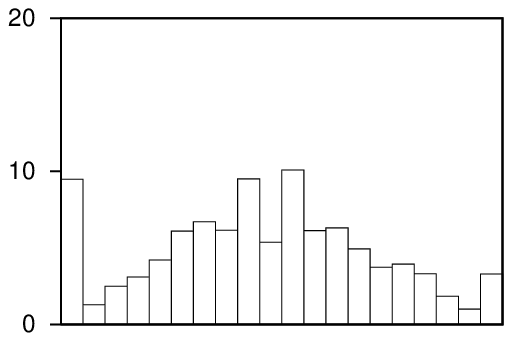} &
\\
\vspace{-0.3cm} \hfill 100 & 
\includegraphics[width=\evalfigwidth]{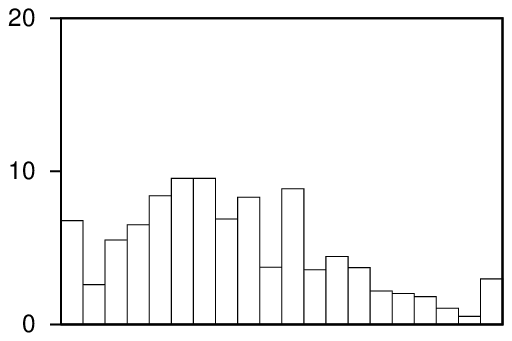} &
\includegraphics[width=\evalfigwidth]{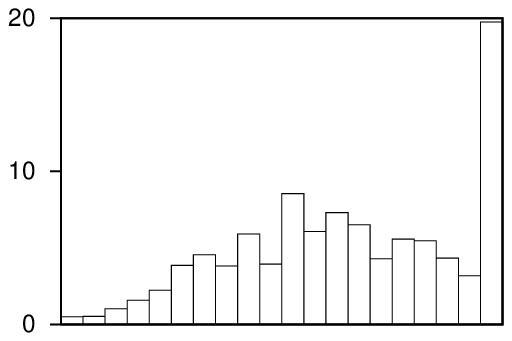} &
\includegraphics[width=\evalfigwidth]{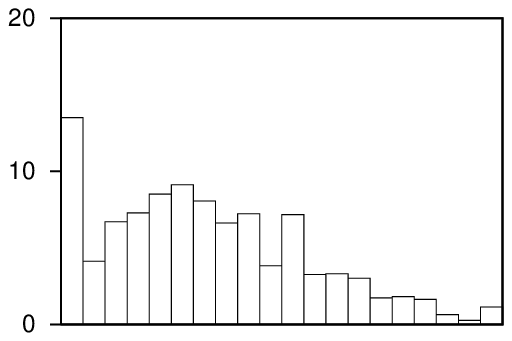} &
\includegraphics[width=\evalfigwidth]{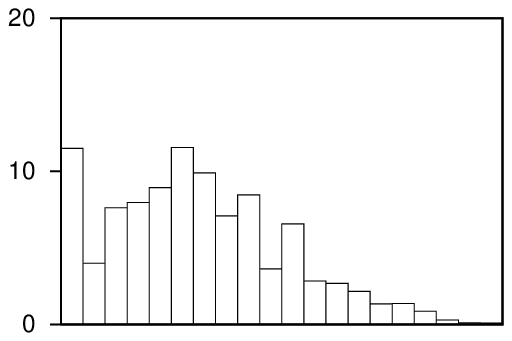} &
\includegraphics[width=\evalfigwidth]{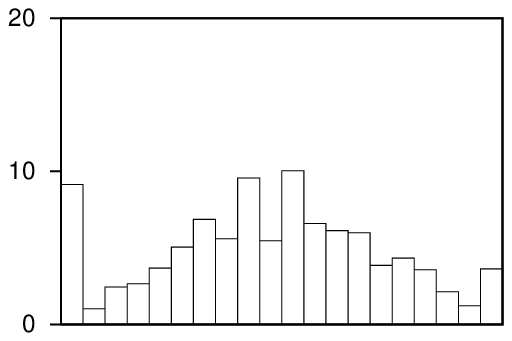} &
\\
\comment{
\hfill 150 & 
\includegraphics[width=\evalfigwidth]{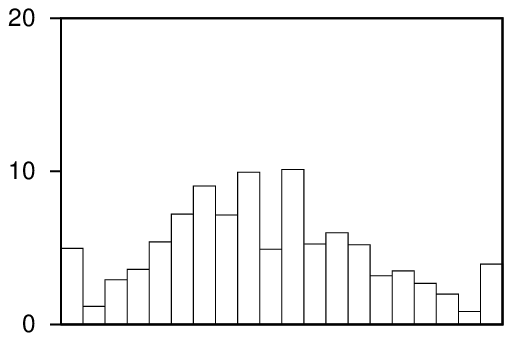} &
\includegraphics[width=\evalfigwidth]{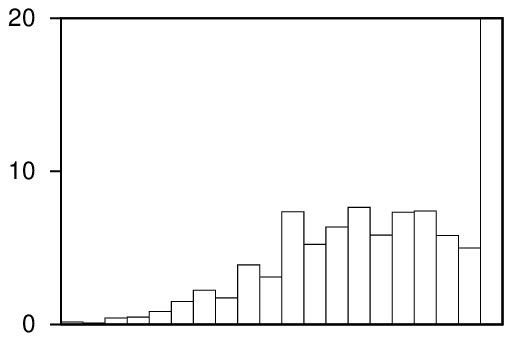} &
\includegraphics[width=\evalfigwidth]{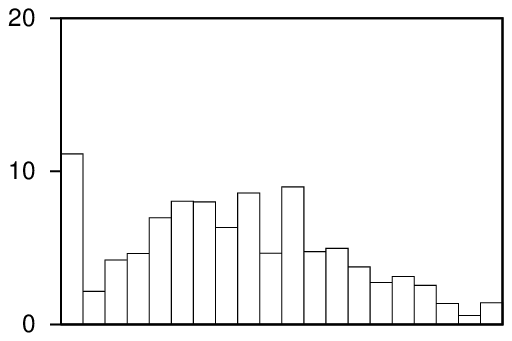} &
\includegraphics[width=\evalfigwidth]{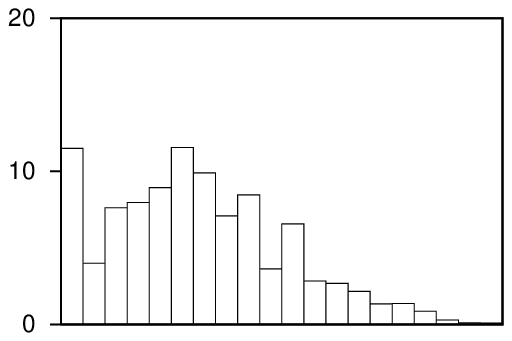} &
\includegraphics[width=\evalfigwidth]{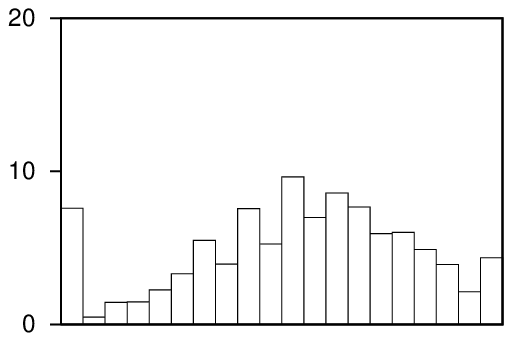} &
\\
\hfill 200 & 
\includegraphics[width=\evalfigwidth]{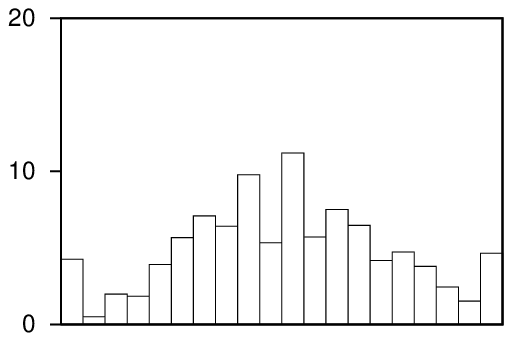} &
\includegraphics[width=\evalfigwidth]{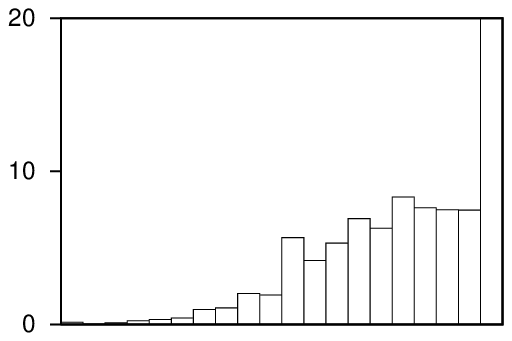} &
\includegraphics[width=\evalfigwidth]{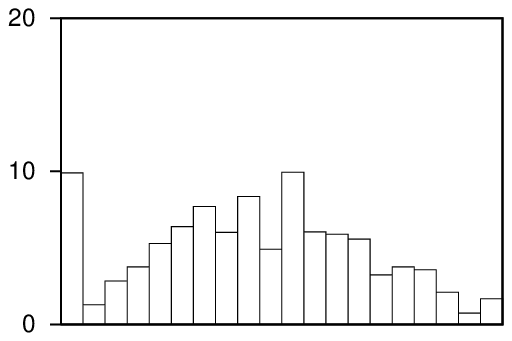} &
\includegraphics[width=\evalfigwidth]{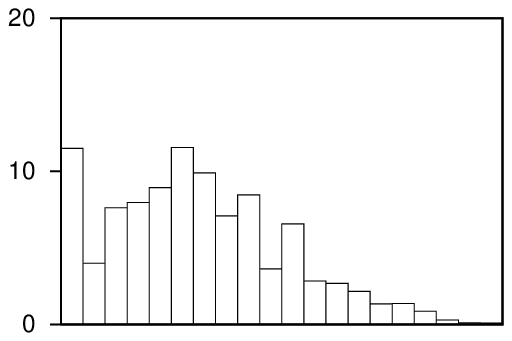} &
\includegraphics[width=\evalfigwidth]{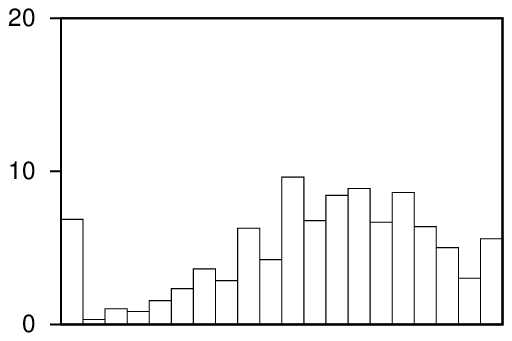} &
\\
\hfill 250 & 
\includegraphics[width=\evalfigwidth]{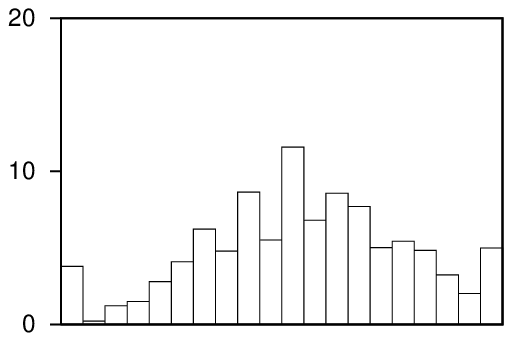} &
\includegraphics[width=\evalfigwidth]{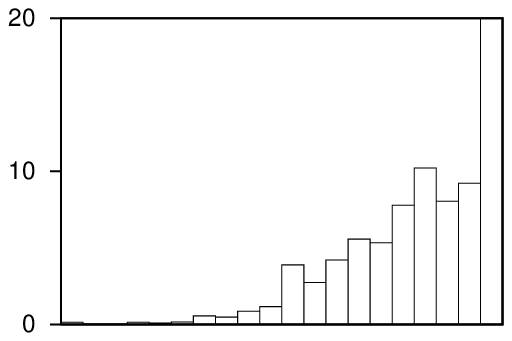} &
\includegraphics[width=\evalfigwidth]{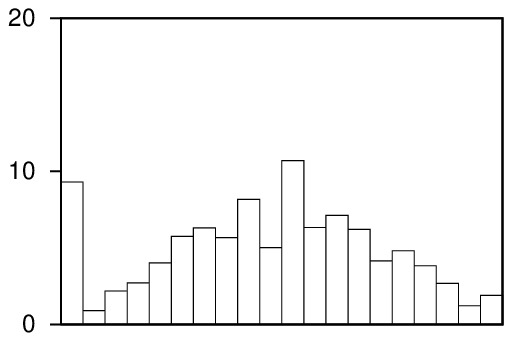} &
\includegraphics[width=\evalfigwidth]{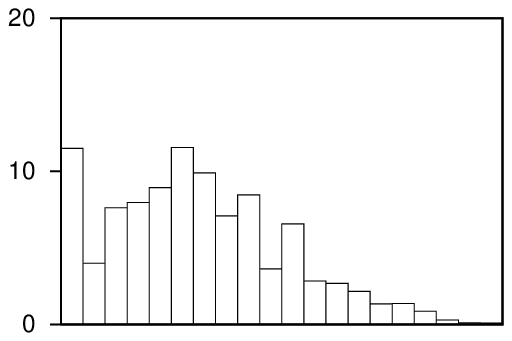} &
\includegraphics[width=\evalfigwidth]{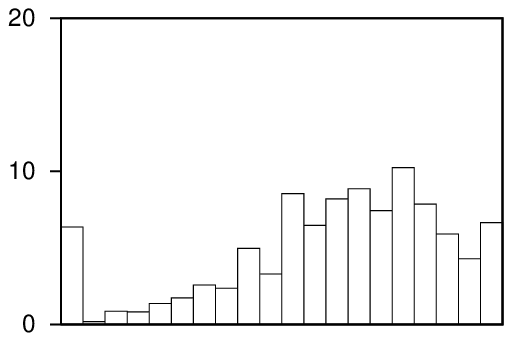} &
\\
\hfill 300 & 
\includegraphics[width=\evalfigwidth]{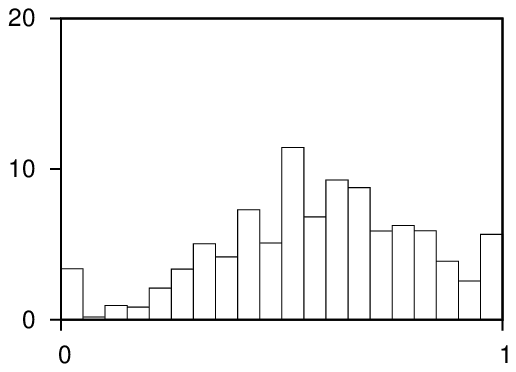} &
\includegraphics[width=\evalfigwidth]{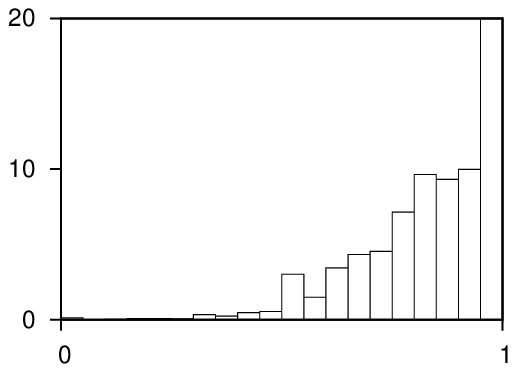} &
\includegraphics[width=\evalfigwidth]{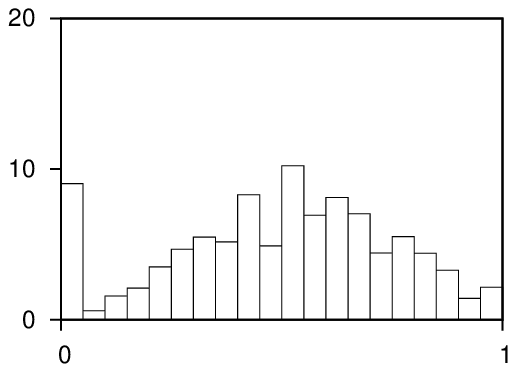} &
\includegraphics[width=\evalfigwidth]{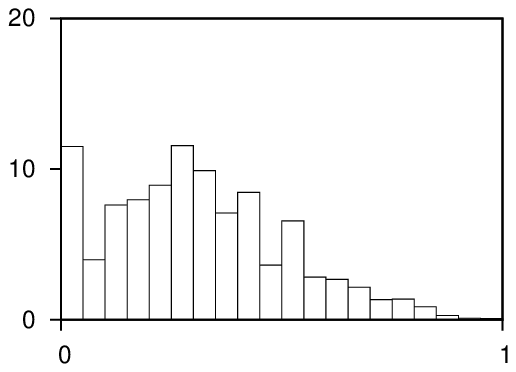} &
\includegraphics[width=\evalfigwidth]{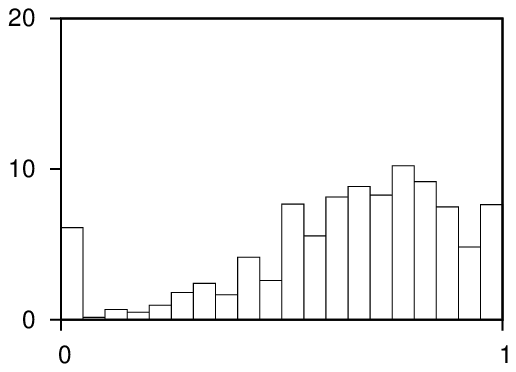} &
\\
}
\end{tabular}
}
  \caption[Distribution over fractions of predicted references]{Distribution over fractions of predicted references in top 1--100 (in percent).}
  \label{fig:eval_dist}
\end{figure}

\clearpage

\clearpage

\begin{figure}
{\centering
  \includegraphics[width=12cm]{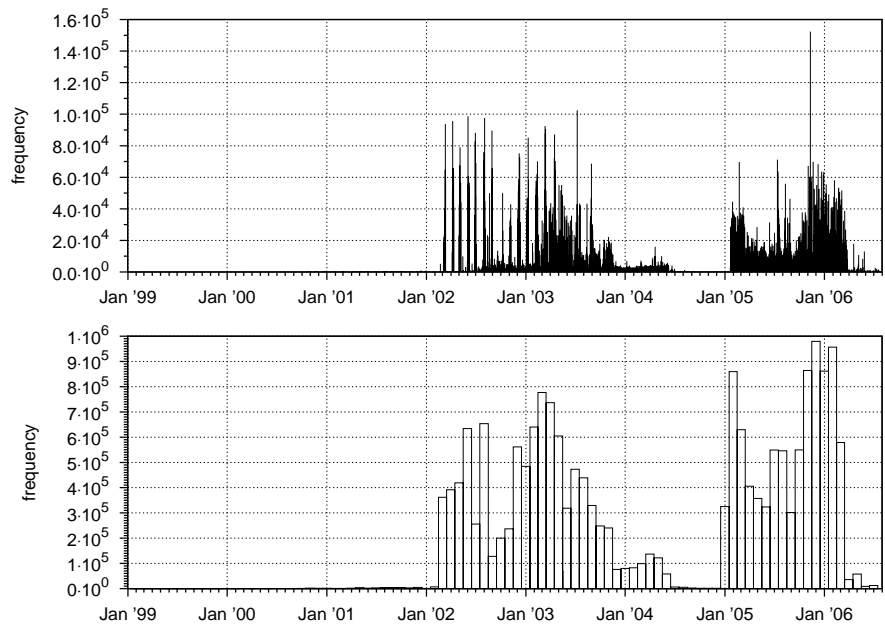}
  \caption[Activity of Google's robot]{Activity of Google's robot (daily and monthly)}
  \label{fig:googlebot}
}
The effect of filtering robots and automated scripts leads to a less spiky access distribution (see Figure~\ref{fig:accesses}).
\end{figure}

\begin{figure}
{\centering
  \includegraphics[width=12cm]{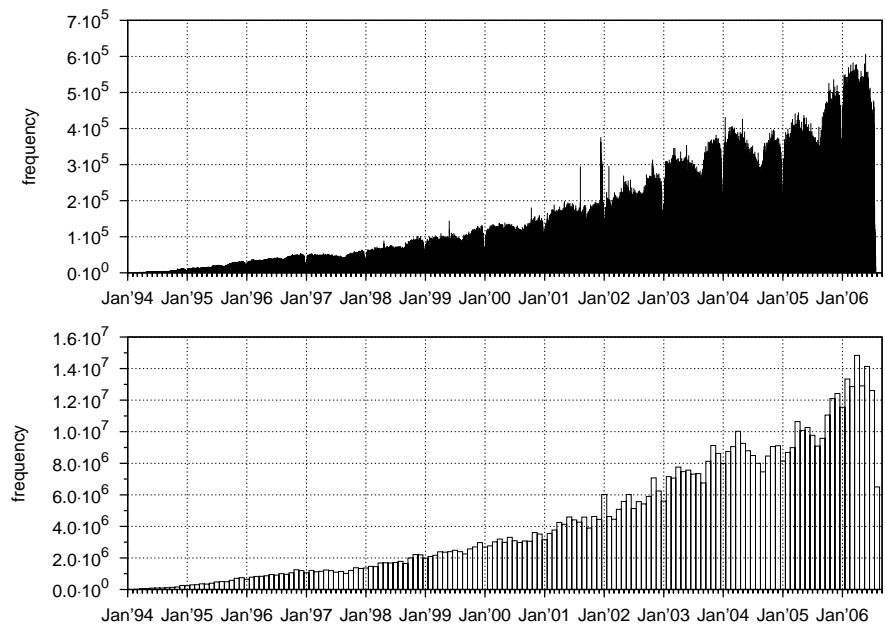}
  \caption[Accesses on arXiv.org]{Accesses on arXiv.org, after filtering (daily and monthly)}
  \label{fig:accesses}
}
Remark: In the daily accesses, one can see gaps before the turn of the year caused by holidays.
\end{figure}

\clearpage

{\newcommand{\widthconcur}{12cm}
\begin{figure}[h]
{\centering
  \includegraphics[width=\widthconcur]{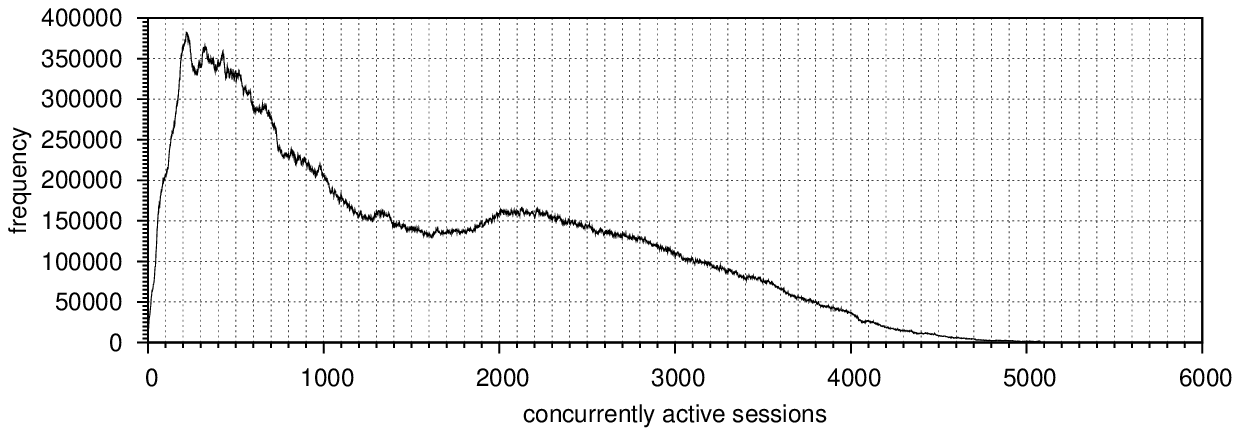}
  \caption[Concurrent sessions (30 min)]{Concurrent users/sessions in a sliding window of 30 minutes}
  \label{fig:sess:concur30}
}
\end{figure}
\begin{figure}[h]
{\centering
  \includegraphics[width=\widthconcur]{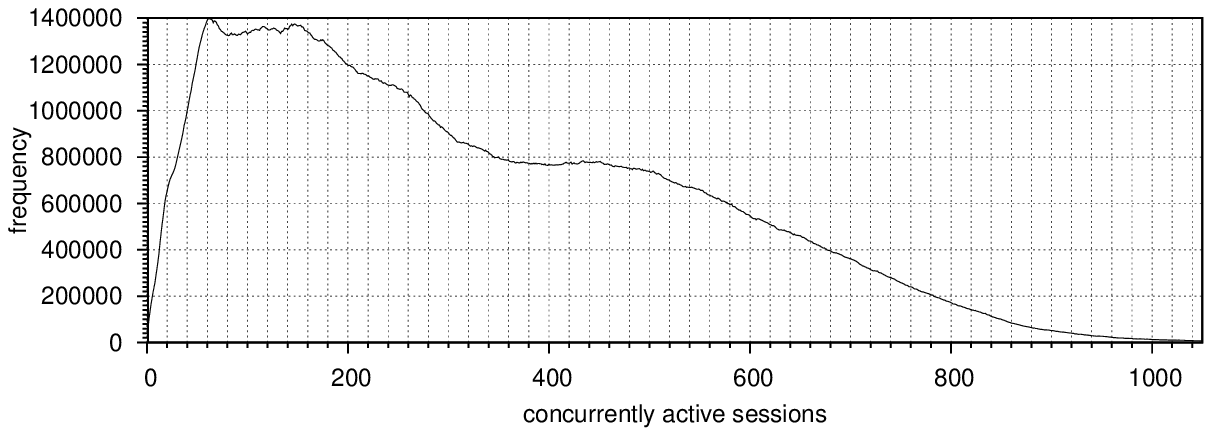}
  \caption[Concurrent sessions (5 min)]{Concurrent users/sessions in a sliding window of 5 minutes}
  \label{fig:sess:concur5}
}
\end{figure}
\begin{figure}[h]
{\centering
  \includegraphics[width=\widthconcur]{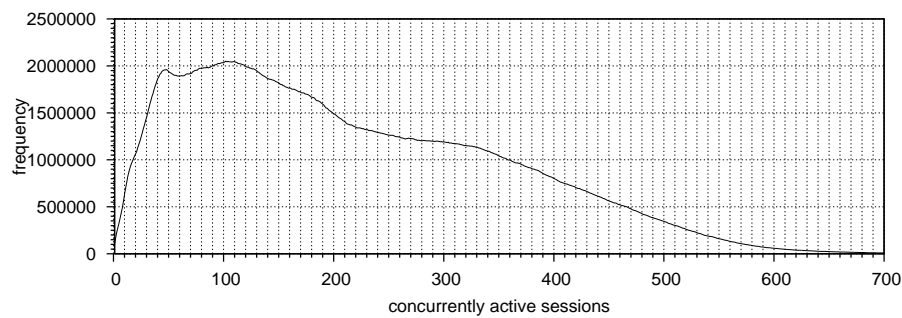}
  \caption[Concurrent sessions (3 min)]{Concurrent users/sessions in a sliding window of 3 minutes}
  \label{fig:sess:concur3}
}
\end{figure}
The tracking of sessions allows to retrospectively retrieve past access statistics of a website. Today, most websites show the number of currently active users, measured over the last five minutes. Figures~\ref{fig:sess:concur30}--\ref{fig:sess:concur3} reproduce the distribution over concurrently active users over different time periods. The bimodality provides evidence that there are times of either low or high traffic. With shorter considered time periods, the distribution of high traffic disappears in the dominating normal traffic distribution.
}

\clearpage 

\begin{figure}
{\centering
  \includegraphics{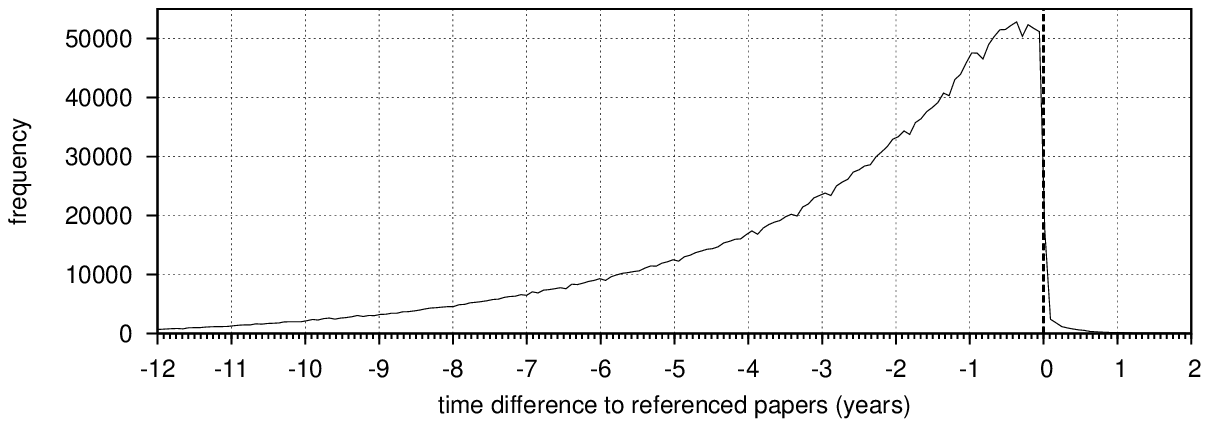}
  \caption{The distribution of the references of a paper over time.}
  \label{fig:citations:reference_dist}
}
The skewness of the curve is due to that it was calculated over all papers of \arxiv. Only recent papers can have references to the very early papers archived.
\end{figure}

\begin{figure}
{\centering
  \includegraphics{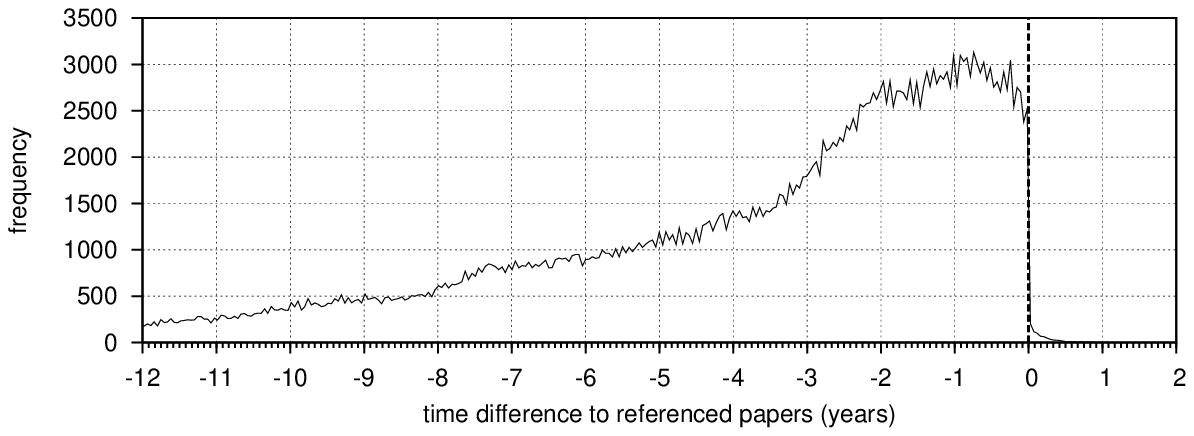}
  \caption{The distribution of the references of papers in 2005 over time.}
  \label{fig:citations:reference_dist2005}
}
\end{figure}

\begin{figure}
{\centering
  \includegraphics[width=0.8\textwidth]{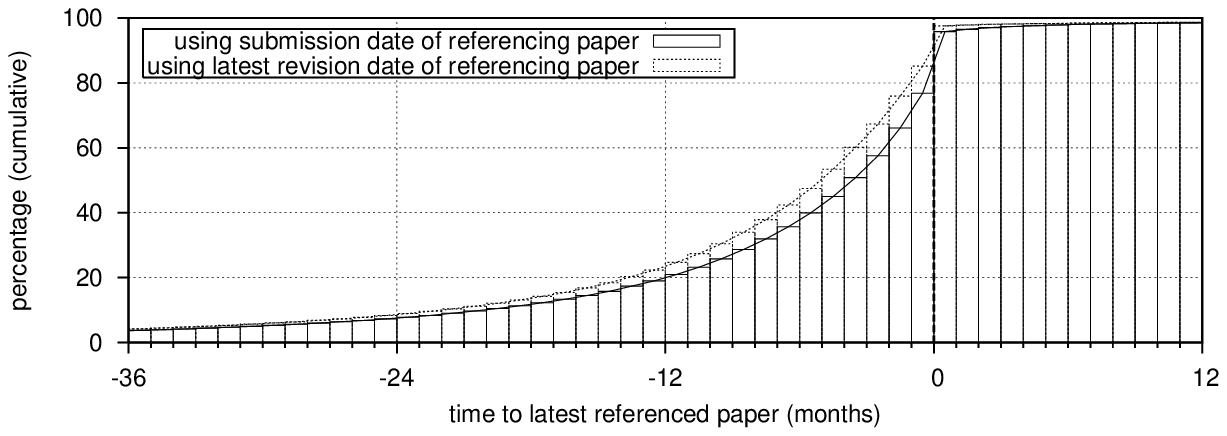}
  \caption[Time between paper publication and latest reference in it]{Time between paper publication and latest reference in it.
}
}
\end{figure}

\clearpage

\chapter{Tables \& Functions}

\label{preprocessing_re}
Regular expressions\index{Regular Expression} for finding the reference section in text documents, ordered by descending dependability. They are applied case-insensitive in order, the next expression is only applied if the previous fails or is ambiguous (i.e.\ matches multiple times in second half of text):

\begin{table}[h]
\begin{tabular}{r|l}
Coverage & Regular Expression \\
\hline
73.1\% & \begin{minipage}[t]{12cm}
\begin{verbatim}
"^\s*(\d{0,2}|[IVX]{0,4})\.?[ ]*REFERENCES\.?\s*$"
\end{verbatim}
\end{minipage} \\
11.1\% & \begin{minipage}[t]{12cm}
\begin{verbatim}
"^\s*(\d{0,2}|[IVX]{0,4})\.?[ ]*ACKNOWLEDGE?MENTS?\s*$"
\end{verbatim}
\end{minipage} \\
1.5\% & \begin{minipage}[t]{12cm}\hfuzz=5pt
\begin{verbatim}
"^\s*(\d{0,2}|[IVX]{0,4})\.?[ ]*(Bibliograph(y|ie))\s*$"
\end{verbatim}
\end{minipage} \\
5.1\% & \begin{minipage}[t]{12cm}
\singlespacing
\begin{verbatim}
// the following expression is typical before
// reference sections:
"work.{0,20}(partly)?.{0,20}support"
\end{verbatim}
\end{minipage} \\
5.9\% & \begin{minipage}[t]{12cm}
\begin{verbatim}
"\n"+  // matches on typical bibliography enumerations 
  "\s*(|\[)1( |\. |\] ).{10,700}\r?\n"+
  "("+
    "\s*(\1)2\2.{10,700}\r?\n"+
    "\s*(\1)3\2.{10,700}\r?\n"+
    "\s*(\1)4\2.{10,700}\r?\n"+
    "\s*(\1)5\2.{10,700}\r?\n"+
    "\s*(\1)6\2.{10,700}\r?\n"+
    "\s*(\1)7\2.{10,700}\r?\n"+
    "\s*(\1)8\2.{10,700}\r?\n"+
    "\s*(\1)9\2.{10,700}\r?\n"+
    "\s*(\1)10\2"+
  ")"
\end{verbatim}
\end{minipage} \\
 3.3\% & no recognized reference sections, mostly not existing \\
\hline
100.0\% \\
\end{tabular}
\caption{Detecting reference sections}
\label{tbl:regex}
\end{table}

\clearpage

\clearpage

\newcommand{\ALGSIZE}{\scriptsize}

\begin{function}
\caption{IncrementCoAccessesFromSession($S_k$)}
\ALGSIZE
\SetLine \linesnumbered \dontprintsemicolon
\setcounter{AlgoLine}{0}
\setcounter{algocf}{0}

\KwIn{time lag between sending of email alerts and reaction on them $t_{lag}$, session $S_k$}
\KwOut{Co-citation}

\ForEach{access to document $d_i$ in session $S_k$}{
    \ForEach{access to document $d_j$ in session $S_i \setminus \{d_i\}$}{
      \eIf{$inSameMonth(t(S_k)-t_{lag}, t(d_i))\ \&\&\ inSameMonth(t(d_i), t(d_j))$ $||$ $(t(S_i)-t_{lag} - t(d_j) \le 7 days\ \&\&\ | t(d_i) - t(d_j) |_1 \le 7 days)$)
      }{
        // ignore co-accesses, probably induced by (alert) lists\;
      }{
        incrementCoAccess($d_i$, $d_j$)
      }
    }
}

\label{alg:filtering}
\end{function}

\clearpage


%% file: mastersthesis.bbl
\begin{thebibliography}{10}

\bibitem{OConnell02:arxivhistory}
Heath~B. O'Connell.
\newblock Physicists thriving with paperless publishing.
\newblock {\em HEP Lib. Web.}, 6:3, 2002.
\newblock arXiv-id:~[physics/0007040].

\bibitem{Kurtz2004:OAI}
Michael~J. Kurtz.
\newblock Restrictive access policies cut readership of electronic research
  journal articles by a factor of two.
\newblock Harvard-Smithsonian Centre for Astrophysics, Cambridge, MA, 2004.

\bibitem{Linden03_Amazon}
Greg Linden, Brent Smith, and Jeremy York.
\newblock Amazon.com recommendations: {Item-to-Item} collaborative filtering.
\newblock {\em IEEE Internet Computing}, 7(1):76--80, 2003.

\bibitem{resnick94_grouplens_news}
P.~Resnick, N.~Iacovou, M.~Suchak, P.~Bergstorm, and J.~Riedl.
\newblock {GroupLens: An Open Architecture for Collaborative Filtering of
  Netnews}.
\newblock In {\em Proceedings of {ACM} 1994 Conference on Computer Supported
  Cooperative Work}, pages 175--186, Chapel Hill, North Carolina, 1994. ACM.

\bibitem{Miller03_movielens}
Bradley~N. Miller, Istvan Albert, Shyong~K. Lam, Joseph~A. Konstan, and John
  Riedl.
\newblock Movielens unplugged: Experiences with an occasionally connected
  recommender system.
\newblock In {\em {Proceedings of {ACM} 2003 Conference on Intelligent User
  Interfaces (IUI'03) (Accepted Poster)}}, Chapel Hill, North Carolina, 2003.
  ACM.

\bibitem{Shardanand95_social}
Upendra Shardanand and Patti Maes.
\newblock Social information filtering: Algorithms for automating ``{Word of
  Mouth}''.
\newblock In {\em Proceedings of {ACM} {CHI}'95 Conference on Human Factors in
  Computing Systems}, volume~1, pages 210--217, 1995.

\bibitem{goldberg01_eigentaste}
Ken Goldberg, Theresa Roeder, Dhruv Gupta, and Chris Perkins.
\newblock Eigentaste: {A} constant time collaborative filtering algorithm.
\newblock {\em Information Retrieval}, 4(2):133--151, 2001.

\bibitem{Schafer99_RecommenderSystems}
J.~Ben Schafer, Joseph Konstan, and John Riedl.
\newblock Recommender systems in e-commerce.
\newblock In {\em EC '99: Proceedings of the 1st ACM conference on Electronic
  commerce}, pages 158--166, New York, NY, USA, 1999. ACM Press.

\bibitem{Garfield79:citationindexing}
Eugene Garfield.
\newblock {\em {Citation Indexing: Its Theory and Application in Science,
  Technology, and Humanities}}.
\newblock John Wiley and Sons, New York, 1979.

\bibitem{McNee2002}
S.M. McNee, I.~Albert, D.~Cosley, P.~Gopalkrishnan, S.K. Lam, A.M. Rashid, J.A.
  Konstan, and J.~Riedl.
\newblock On the recommending of citations for research papers.
\newblock In {\em CSCW '02: Proceedings of the 2002 ACM conference on Computer
  supported cooperative work}, pages 116--125, New York, NY, USA, 2002. ACM
  Press.

\bibitem{McNee2005:Techlens}
J.A. Konstan, N.~Kapoor, S.M. McNee, and J.T. Butler.
\newblock Techlens: Exploring the use of recommenders to support users of
  digital libraries.
\newblock A Project Briefing at the Coalition for Networked Information, Fall
  2005 Task Force Meeting, Phoenix, AZ, December 2005.
\newblock [available at
  \url{http://www.grouplens.org/papers/pdf/CNI-TechLens-Final.pdf}].

\bibitem{KddCup2003}
\href{http://www.cs.cornell.edu/projects/kddcup/citation_prediction_task.html}{\texttt{http://www.cs.cornell.edu/projects/kddcup/citation\_prediction\_}} \href{http://www.cs.cornell.edu/projects/kddcup/citation_prediction_task.html}{\texttt{task.html}}.
\newblock [Accessed on 10/03/2006].

\bibitem{Brody2005:webusage}
Tim Brody, Stevan Harnad, and Leslie Carr.
\newblock Earlier web usage statistics as predictors of later citation impact.
\newblock {\em Journal of the American Society for Information Science and
  Technology (JASIST)}, 57(8):1060--1072, 2006.

\bibitem{Brody00_opcit}
{OpCit --- The Open Citation Project --- Reference Linking and Citation
  Analysis for Open Archives}.
\newblock \url{http://opcit.eprints.org/opcitresearch.shtml}, 1999--2002.
\newblock [Accessed on 08/12/2006].

\bibitem{Kurtz05_readcite}
Michael~J. Kurtz, Guenther Eichhorn, Alberto Accomazzi, Carolyn Grant, Markus
  Demleitner, Stephen~S. Murray, Nathalie Martimbeau, and Barbara Elwell.
\newblock The bibliometric properties of article readership information.
\newblock {\em Journal of the American Society for Information Science and
  Technology}, 56(2):111--128, 2005.

\bibitem{Woodruff2000}
A.~Woodruff, R.~Gossweiler, J.~Pitkow, E.~H. Chi, and S.~K. Card.
\newblock Enhancing a digital book with a reading recommender.
\newblock In {\em Proc. CHI 2000}, pages 153--160, 2000.

\bibitem{joachims97_web}
Thorsten Joachims, Dayne Freitag, and Tom~M. Mitchell.
\newblock {Web Watcher: A Tour Guide for the World Wide Web}.
\newblock In {\em {IJCAI} (1)}, pages 770--777, 1997.

\bibitem{Landauer97_bagofwords}
T.~K. Landauer, D.~Laham, R.~Rehder, and M.~E. Schreiner.
\newblock {How well can passage meaning be derived without using word order? A
  comparison of Latent Semantic Analysis and humans}.
\newblock In {\em 19th Annual Conference of the Cognitive Science Society},
  pages 412--417, Mahwah, NJ, 1997.

\bibitem{Salton89_IR}
Gerard Salton.
\newblock {\em Automatic text processing: the transformation, analysis, and
  retrieval of information by computer}.
\newblock Addison-Wesley Longman Publishing Co., Inc., Boston, MA, USA, 1989.

\bibitem{giles98_citeseer}
C.~Lee Giles, Kurt Bollacker, and Steve Lawrence.
\newblock Citeseer: An automatic citation indexing system.
\newblock In {\em Digital Libraries 98 - The Third {ACM} Conference on Digital
  Libraries}, pages 89--98, Pittsburgh, PA, June 23--26 1998. ACM Press.

\bibitem{Small73_cocitation}
H.~G. Small.
\newblock Co-citation in the scientific literature: A new measure of the
  relationship between two documents.
\newblock {\em Journal of American Society for Information Science},
  24(4):265--269, 1973.

\bibitem{Bharat2001_Hilltop}
Krishna Bharat and George~A. Mihaila.
\newblock Hilltop: A search engine based on expert documents.
\newblock \url{http://www.cs.toronto.edu/~georgem/hilltop}, 2001.

\bibitem{Gyongyi04_trustrank}
Zolt{\'a}n Gy{\"o}ngyi, Hector~G. Molina, and Jan Pedersen.
\newblock Combating web spam with {TrustRank}.
\newblock In {\em Proceedings of the Thirtieth International Conference on Very
  Large Data Bases (VLDB)}, pages 576--587, Toronto, Canada, August 2004.
  Morgan Kaufmann.

\bibitem{Benczur05_spamrank}
A.~Benczur, K.~Csalogany, T.~Sarlos, and M.~Uher.
\newblock Spamrank --- fully automatic link spam detection.
\newblock In {\em First International Workshop on Adversarial Information
  Retrieval on the Web}, 2005.

\bibitem{Page98_pagerank}
L.~Page, S.~Brin, R.~Motwani, and T.~Winograd.
\newblock {The PageRank citation ranking: Bringing order to the web}.
\newblock {\em Technical report, Stanford Digital Library Technologies
  Project}, 1998.

\bibitem{Brin98_google}
Sergey Brin and Lawrence Page.
\newblock The anatomy of a large-scale hypertextual web search engine.
\newblock In {\em WWW7: Proceedings of the seventh international conference on
  World Wide Web 7}, pages 107--117, Amsterdam, The Netherlands, The
  Netherlands, 1998. Elsevier Science Publishers B. V.

\bibitem{Kleinberg99_hits}
Jon~M. Kleinberg.
\newblock Authoritative sources in a hyperlinked environment.
\newblock {\em Journal of the ACM}, 46(5):604--632, 1999.

\bibitem{Motwani95_randomwalks}
R.~Motwani and P.~Raghavan.
\newblock {\em Randomized Algorithms}.
\newblock Cambridge University Press, New York, NY, USA, 1995.

\bibitem{Karypis04}
Mukund Deshpande and George Karypis.
\newblock {Item-based top-N recommendation algorithms}.
\newblock {\em ACM Trans. Inf. Syst.}, 22(1):143--177, 2004.

\bibitem{Hofmann99plsa}
Thomas Hofmann.
\newblock Probabilistic latent semantic analysis.
\newblock In {\em Proc. of Uncertainty in Artificial Intelligence, UAI'99},
  Stockholm, 1999.

\bibitem{Ziegler05divers}
Cai-Nicolas Ziegler, Sean~M. McNee, Joseph~A. Konstan, and Georg Lausen.
\newblock Improving recommendation lists through topic diversification.
\newblock In {\em WWW '05: Proceedings of the 14th international conference on
  World Wide Web}, pages 22--32, New York, NY, USA, 2005. ACM Press.

\bibitem{granka04_eyetracking}
L.~Granka, T.~Joachims, and G.~Gay.
\newblock {Eye-tracking analysis of user behavior in WWW search}.
\newblock In {\em Proceedings of SIGIR'04}. ACM Press, 2004.

\bibitem{Koch06:clientlogs}
David Koch.
\newblock Study of the discrepancy between client- and server side logging of
  clickstreams.
\newblock Master's thesis, Royal Technical Institute, Stockholm, 2006, in
  progress.

\end{thebibliography}
